\documentclass[fp,twocolumn]{jpsj3}
\usepackage{txfonts}
\newcommand{\lsim}
 {\ \raise.35ex\hbox{$<$}\kern-0.75em\lower.5ex\hbox{$\sim$}\ }
\newcommand{\gsim}
 {\ \raise.35ex\hbox{$>$}\kern-0.75em\lower.5ex\hbox{$\sim$}\ }
\def\journal #1#2#3#4{#1 {\bf #2}, #3 (#4)}

\def\SSC{Solid State Commun.}

\def\JPParis{J.~Phys.~(Paris)}

\def\JPCS{J.\ Phys.\ Chem.\ Solids}

\def\JPSJ{J.\ Phys.\ Soc.\ Jpn.}
\def\JPSCP{JPS Conf.\ Proc.}
\def\LTP{Low~Temp.~Phys.}

\def\NJP{New J.~Phys.}
\def\NC{Nat.~Commun.}
\def\NP{Nat.~Phys.}

\def\PP{Phys.~Proc.}

\def\PR{Phys.\ Rev.}

\def\PRB{Phys.\ Rev.\ B}

\def\PRL{Phys.\ Rev.\ Lett.}

\def\RMP{Rev.\ Mod.\ Phys.}
\def\SST{Supercond.~Sci.~Technol.}

\def\ZPB{Z.\ Phys.\ B}

\def\RPP{Rep.~Prog.~Phys.}
\title{Band-Renormalization Effects and Predominant Antiferromagnetic \\
Order in Two-Dimensional Hubbard Model} 

\author{ 
Ryo Sato\thanks{satoryo@cmpt.phys.tohoku.ac.jp} and Hisatoshi Yokoyama 
}
\inst{Department of Physics, Tohoku University, Sendai 980-8578, Japan } 
\abst{
Band renormalization effects (BRE) are comprehensively studied for 
a mixed state of $d_{x^2-y^2}$-wave superconducting ($d$-SC) and antiferromagnetic 
(AF) orders, in addition to simple $d$-SC, AF, and normal (paramagnetic) 
states, by applying a variational Monte Carlo method to a two-dimensional 
Hubbard ($t$-$t'$-$U$) model. 
In a weakly correlated regime ($U/t\lesssim 6$), BRE are negligible on all 
the states studied. 
As previously shown, the effective band of $d$-SC is greatly renormalized 
but the modifications of physical quantities, including energy 
improvement, are negligible. 
In contrast, BRE on the AF state considerably affects various features of 
the system. 
Because the energy is markedly improved for $t'/t<0$, the AF state occupies 
almost the whole underdoped regime in phase diagrams. 
A doped metallic AF state undergoes a kind of Lifshitz transition at 
$t'=t'_{\rm L}\sim -0.05t$ as $t'/t$ varies, irrespective of the values of 
$U/t$ and $\delta$ (doping rate). 
Pocket Fermi surfaces arise around $(\pi, 0)$ [$(\pi/2, \pi/2)$] for 
$t'>t'_{\rm L}$ [$t'<t'_{\rm L}$], which corresponds to the 
electron-hole asymmetry observed in angle-resolved photoemission spectroscopy (ARPES) spectra. 
The coexistent state of the two orders is possible basically for 
$t'>t'_{\rm L}$, because the existence of Fermi surfaces near 
$(\pi, 0)$ is a requisite for the electron scattering of ${\bf q}=(\pi, \pi)$. 
Actually, the coexistent state appears mainly for $t'_{\rm L}/t<t'/t\lesssim0.2$ 
in the mixed state. 
Nevertheless, the AF and coexisting states become unstable toward phase 
separation for $-0.05\lesssim t'/t\lesssim 0.2$ but become stable at other values 
of $t'/t$ owing to the energy reduction by the diagonal hopping of 
doped holes. 
We show that this instability does not directly correlate with the strength of 
$d$-SC. 
} 

\begin{document} 
\maketitle

\section{Introduction\label{sec:intro}}
To clarify the physics of cuprate superconductors 
(SCs),\cite{Bednorz,Exp-JPSJ} 
we have to know the fundamental properties of the $t$-$J$ and Hubbard models 
on a square lattice 
with an extension in the kinetic part ($t$-$t'$ and $t$-$t'$-$t''$, 
etc.) as basic models.\cite{theory} 
In this paper, we mainly focus on  the following subjects in the Hubbard 
($t$-$t'$-$U$) model: 
\par
(A) 
The primary subject is the ground-state phase diagram in the model-parameter 
space. 
Although a typical view to date is that the antiferromagnetic 
(AF) order arising at half filling rapidly vanishes on doping holes and the 
$d_{x^2-y^2}$-wave superconductivity ($d$-SC) appears,\cite{theory,Y2013} 
in accordance with the behavior of cuprates, in recent studies using 
advanced techniques it was argued that AF orders or inhomogeneous phases 
prevail in wider ranges of $\delta$ (doping rate).\cite{Misawa,Otsuki,DMET} 
\par
(B) 
In phase diagrams of cuprates, the areas of superconducting (SC) and AF 
phases are in proximity. 
In the SC phase, appreciable AF correlation or short-range AF orders are 
observed,\cite{AF-SRO} but the coexistence of two long-range orders has 
not been detected except for in multilayered systems. 
In theory, it is still unclear in what parameter range the two long-range orders coexist and why they are coexisting or mutually exclusive.
\par
(C) 
Another subject is whether or not homogeneous states are stable against 
phase separation. 
Actually, signs of inhomogeneous electronic states or phase separation are 
often noticed in cuprates such as a stripe structure of charge and spin and 
a mosaic distribution of the gap magnitude. 
Theoretically, it is again unclear as to the ranges of $U/t$, $t'/t$, and 
$\delta$ and the cause of the state becoming unstable toward phase 
separation. 
\par

So far, these subjects have been addressed by many researchers with 
a variety of methods, in particular, dynamical mean field theories 
(DMFTs) with some extensions \cite{Capone,CDMFT,Aichhorn,Kancharla,Otsuki,DMET} 
and variational Monte Carlo (VMC) 
methods\cite{Lee-mix,Himeda-mix,Ivanov,Shih-mix,Pathak,GL,Koba-mix,
Koba-ISS14,Y2013,Misawa} 
are useful tools to quantitatively treat strong local correlations. 
One also needs to consider the effects of antiferromagnetism (AF) because it is 
crucial even for subject (C). 
The results regarding (A)-(C) of the above studies do not seem 
unified but are rather scattered at first glance. 
Although inconsistencies exist among them, we feel that 
the main source of confusion resides in insufficient consideration of the difference 
in the diagonal hopping term ($t'$). 
In most of the above studies, $t'/t$ (and $t''/t$) was 
set to specific values, say $0$ and/or $-0.3$, but we are apt to 
read the results associated with (A)-(C) without care in while also considering the value of $t'/t$. 
If we arrange the results by specifying the value of $t'/t$, they are 
often consistent beyond our expectation, as shown later 
in Table~\ref{table:Summary-AF} for some results obtained by the VMC method. 
This also applies to many results of DMFT. 
From this point of view, the results of recent studies with high 
accuracy\cite{Misawa,Otsuki,DMET} are consistent. 
In fact, a small number of studies have considered the difference 
in the features of (A)-(C) between the cases of $t'/t=0$ and other cases, 
although they were not sufficiently elaborate or 
analytic.\cite{Y2013,Koba-ISS14,Shih-mix,DMET} 
\par

To study (A)-(C) in an ordinary VMC framework, one has to use a mixed 
state which represents the AF and SC orders simultaneously. 
The properties associated with (B) have been studied for the 
$t$-$J$-type \cite{Lee-mix,Himeda-mix,Ivanov,Shih-mix,Pathak} and 
Hubbard\cite{GL,Koba-mix,Pathak,Koba-ISS14} models.
In addition, it is crucial to take account of the effects of band 
renormalization (BR) owing to strong correlations in the one-body 
part of the wave function. 
To date, band renormalization effects (BRE) have been introduced into 
$d$-SC states\cite{Himeda-BR,Shih-BR,Koba-old,Wata-t-J,Tocchio} or the 
$d$-SC part of mixed states.\cite{Liu,Wata-Org,Koba-mix,Pathak,Koba-ISS14} 
Because BRE were disregarded in the AF part in these studies, an AF order 
does not arise for $t'/t\sim -0.3$, or it vanishes rapidly with doping for 
$t'/t\sim 0$. 
Such features are inconsistent with recent research.\cite{Misawa,Otsuki,DMET} 
Unexpectedly, BRE have not been introduced into normal (paramagnetic) and 
AF states and the AF part of mixed states,\cite{note-Misawa} probably 
because optimization is technically bothersome, as mentioned in 
Sect.~\ref{sec:VMC} and the Appendices. 
\par

In this paper, we study ground-state properties of the 
Hubbard ($t$-$t'$-$U$) model by applying a VMC method with BRE of up to 
fifth-neighbor hopping to a mixed state $\Psi_{\rm mix}$ in addition to 
normal (paramagnetic), pure $d$-SC, and pure AF states. 
In $\Psi_{\rm mix}$, we renormalize the energy dispersions $\varepsilon_{\bf k}^{\rm SC}$ and 
$\varepsilon_{\bf k}^{\rm AF}$ independently. 
This parametrization is a key to finding correct features of a mixed 
state. 
The present results are quantitatively consistent with those in recent 
research.\cite{Misawa,Otsuki,DMET}
As the merits of the present study, we stress the following points: 
(a) We systematically study the dependence on the model parameters, in 
particular, $t'/t$ and $\delta$. 
(b) We clarify the physics underlying the properties of $\Psi_{\rm mix}$ 
(or the Hubbard model) by comparing various levels of wave functions. 
Through these merits, we will acquire a more enlightened view of subjects 
(A)-(C). 
\par

This paper is organized as follows.
In Sect.~\ref{sec:form}, we explain the model and method used in this study. 
In Sect.~\ref{sec:d-wave}, we discuss the results of BRE on the $d$-SC state. 
In Sect.~\ref{sec:normal}, the results of BRE on the normal (or paramagnetic) 
state are presented. 
In Sect.~\ref{sec:AF}, we consider the BRE on an AF state, referring to a 
Lifshitz transition arising at $t'/t\sim -0.05$. 
In Sect.~\ref{sec:mix}, we study BRE on a mixed state of $d$-SC and AF 
orders, and discuss prerequisites for the appearance of $d$-SC. 
In Sect.~\ref{sec:summary}, we recapitulate the main results and make 
additional comments. 
In Appendices\ref{sec:normal-A} and \ref{sec:AF-A}, details of the calculations 
and analyses of the normal and AF states are described, respectively. 
The preliminary results referred to in this paper were presented in three 
preceding publications.\cite{proceedings1,proceedings2,proceedings3}
\par

\section{Formulation\label{sec:form}}
After introducing the model in Sect.~\ref{sec:model}, in 
Sect.~\ref{sec:wf} we describe the setup of trial wave functions, which is the core of 
variation theory. 
In Sect.~\ref{sec:VMC}, we comment on a way of computing expectation 
values with the present wave functions. 
\par

\subsection{Hubbard model\label{sec:model}}
%
\begin{figure}
\begin{center} 
\vskip -0mm
\hskip -0mm
\includegraphics[width=7.5cm,clip]{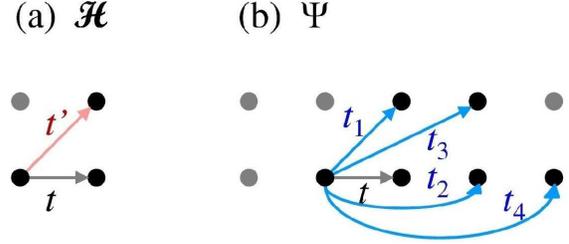}
\end{center} 
\vskip -0mm
\caption{
(a) Hopping processes in Hamiltonian [Eqs.~(\ref{eq:Hamil}) and 
(\ref{eq:bareband})] and 
(b) those corresponding to band-adjusting parameters $t_\eta$ 
($\eta=1$--$4$) in trial wave functions 
[Eqs.~(\ref{eq:gamma})-(\ref{eq:epsilon})]. 
In both figure, $t$ is the unit. 
}
\label{fig:bp} 
\vskip -5mm
\end{figure}
With cuprate SCs in mind, we consider the Hubbard model ($U\ge 0$) 
on a square lattice with diagonal hopping:
\begin{eqnarray}
{\cal H}&=&{\cal H}_{\rm kin}+{\cal H}_U \nonumber \\
        &=&-\sum_{(i,j),\sigma}t_{ij}\left( 
        c^\dagger_{i\sigma}c_{j\sigma} + \mbox{H.c.}\right)+U\sum_j n_{j\uparrow}n_{j\downarrow}, 
\label{eq:Hamil}
\end{eqnarray} 
where $c_{j\sigma}$ annihilates an electron of spin $\sigma$ at site $j$, 
$n_{j\sigma}=c^\dag_{j\sigma}c_{j\sigma}$, and $(i,j)$ indicates the sum 
of pairs on sites $i$ and $j$. 
In this work, the hopping integral $t_{ij}$ is $t$ for nearest neighbors 
($\ge 0$), $t'$ for diagonal neighbors, and $0$ otherwise 
(${\cal H}_{\rm kin}={\cal H}_t+{\cal H}_{t'}$) [Fig.~\ref{fig:bp}(a)]. 
The bare energy dispersion becomes
\begin{equation}
\tilde\varepsilon_{\bf k}=
-2t\left(\cos k_x+\cos k_y\right)-4t'\cos k_x\cos k_y. 
\label{eq:bareband}
\end{equation}
As we will see, the diagonal hopping term ${\cal H}_{t'}$ plays a crucial 
role in the present theme. 
We use $t$ and the lattice spacing as the units of energy and length, 
respectively. 
\par

\subsection{Trial wave functions\label{sec:wf}} 
Because our interest here is to grasp the nature of BRE rather than 
obtain accurate numerical values, we employ forms of trial functions that 
capture the essence of physics but are as simple as possible. 
As many-body trial states, we use a Jastrow type, $\Psi={\cal P}\Phi$, 
where ${\cal P}$ is a two-body correlation factor 
(projector) and $\Phi$ is a one-body (mean-field-type) wave function. 
We use a simple form of ${\cal P}$ common to all trial states, 
${\cal P}={\cal P}_{\rm G}{\cal P}_Q$, where ${\cal P}_{\rm G}$ is 
the well-known onsite Gutzwiller projector 
${\cal P}_{\rm G}=\prod_j[1-(1-g)n_{j\uparrow}n_{j\downarrow}$]\cite{Gutz} 
and ${\cal P}_Q$ is the nearest-neighbor doublon-holon (D-H) binding 
factor,\cite{D-H1,D-H2,Y2013}
\begin{equation}
{\cal P}_Q=\prod_j\left[1
-\zeta_{\rm d}d_j\prod_\tau\left(1-h_{j+\tau}\right)
-\zeta_{\rm h}h_j\prod_\tau\left(1-d_{j+\tau}\right)
\right],
\label{eq:D-H}
\end{equation}
where $d_j=n_{j\uparrow}n_{j\downarrow}$, 
$h_j=(1-n_{j\uparrow})(1-n_{j\downarrow})$, and $\tau$ runs over the 
nearest-neighbor sites of site $j$. 
As shown before,\cite{YOT,YTOT} the D-H binding effect included in 
${\cal P}_Q$ is crucial for properly treating Mott physics. 
The projector ${\cal P}$ has three variational parameters, $g$, 
$\zeta_{\rm d}$, and $\zeta_{\rm h}$, which trigger BR in $\Phi$. 
\par

\begin{table}
\caption{
Elements modified by band renormalization in one-body part for finite 
systems (indicated by circles). 
The two elements merge for $L\rightarrow\infty$. 
}
\begin{center}
\begin{tabular}{l|cccc}
\hline
Modified elements 
& $\Phi_{\rm N}$ & $\Phi_d$ & $\Phi_{\rm AF}$ & $\Phi_{\rm mix}$  
\\
\hline
$\{{\bf k}\}_{\rm occ}$ or Fermi surface
& $\bigcirc$ & $-$        & $\bigcirc$ & $\bigcirc$ 
\\
Direct modification of $\varepsilon_{\bf k}$  
& $-$        & $\bigcirc$ & $\bigcirc$ & $\bigcirc$ 
\\
\hline
\end{tabular}
\end{center}
\vskip -9mm
\label{table:BR}
\end{table}
We turn to the one-body part $\Phi$, which is the main point for BRE. 
We start with the normal (paramagnetic) state. 
Let $\{{\bf k}\}_{\rm occ}$ denote the set of {\bf k} points occupied by 
electrons in $\Phi$ according to 
$\varepsilon_{\bf k}\le\varepsilon_{{\bf k}_{\rm F}}$ 
(or symbolically ${\bf k}\in{\bf k}_{\rm F}$). 
Then, the one-body normal state we use (a Fermi sea) is written as 
\begin{equation}
\Phi_{\rm N}=
\prod_{\{\bf k\}_{\rm occ},~\sigma}c^\dag_{{\bf k},\sigma}|0\rangle. 
\label{eq:FS}
\end{equation}
If $\{{\bf k}\}_{\rm occ}$ is determined according to the bare band 
dispersion $\tilde\varepsilon_{\bf k}$ in Eq.~(\ref{eq:bareband}), 
$\Phi_{\rm N}$ is the exact ground state of ${\cal H}$ for $U=0$. 
When the interaction is introduced, $\varepsilon_{\bf k}$ will be modified 
by its self-energy. 
In the framework of many-body variation theory, $\varepsilon_{\bf k}$ should 
be optimized along with the other parameters so as to reduce the total 
energy $E=\langle{\cal H}\rangle/N_{\rm s}$ ($N_{\rm s}$: number of sites). 
Note that in $\Phi_{\rm N}$ [Eq.~(\ref{eq:FS})], $\varepsilon_{\bf k}$ 
does not explicitly appear but has the effect of determining $\{{\bf k}\}_{\rm occ}$ 
or the Fermi surface (see Table~\ref{table:BR} for comparison). 
Namely, the operation of BR for $\Phi_{\rm N}$ is simply reduced to the 
choice of $\{{\bf k}\}_{\rm occ}$. 
To obtain full BRE, we need to find the $\{{\bf k}\}_{\rm occ}$ that yields 
the lowest $E/t$ among all the $\{{\bf k}\}_{\rm occ}$, but the number of 
choices of $\{{\bf k}\}_{\rm occ}$ grows exponentially 
[roughly as $_{N_{\rm s}/4}C_{N/8}$ ($N$: number of electrons)] as the 
system size grows. 
In this work, we optimize $E/t$ within the $\{{\bf k}\}_{\rm occ}$ that are 
generated by a tight-binding form of $\varepsilon_{\bf k}$ with diagonal 
transfer: 
\begin{equation}
\varepsilon_{\bf k}^{\rm N}=
-2t\left(\cos k_x+\cos k_y\right)-4t_1\cos k_x\cos k_y, 
\label{eq:disp-N}
\end{equation}
where $t_1$ is varied. 
This form of $\varepsilon_{\bf k}$ has often been used for $d$-SC states 
in previous studies\cite{Himeda-BR,Shih-BR,Koba-old,Wata-t-J,Tocchio} and 
also seems reasonable as a first setting for $\Phi_{\rm N}$. 
Details of optimizing $\Psi_{\rm N}={\cal P}\Phi_{\rm N}$ are described in 
Appendix\ref{sec:normal-A}.
The ordered states $\Phi_d$, $\Phi_{\rm AF}$, and $\Phi_{\rm mix}$ introduced 
below are reduced to $\Phi_{\rm N}$ in the limit of $\Delta_{\rm AF}$ and/or 
$\Delta_d\rightarrow 0$. 
\par
We move on to the mixed state of AF and $d$-SC orders of a fixed electron 
number, $\Phi_{\rm mix}$. 
This state is written as a $d$-wave BCS state composed of AF 
quasiparticles:\cite{GL} 
\begin{equation}
\Phi_{\rm mix} =\left(\sum_{\bf k}\phi({\bf k})~
a_{{\bf k}\uparrow}^\dagger a_{{\bf -k}\downarrow}^\dagger
\right)^\frac{N}{2}|0\rangle, 
\label{eq:Phi_mix}
\end{equation}
with
\begin{equation}
\phi({\bf k})
=\frac{\Delta_{\bf k}}{\varepsilon^{\rm SC}_{\bf k}-\mu+
\sqrt{(\varepsilon^{\rm SC}_{\bf k}-\mu)^2+\Delta_{\bf k}^2}}. 
\label{eq:BCSDelta}
\end{equation}
Here, $\mu$ is a variational parameter, which is reduced to the chemical 
potential for $U/t\rightarrow 0$, and a $d_{x^2-y^2}$-wave gap is assumed as 
\begin{equation}
\Delta_{\bf k}=\Delta_d(\cos k_x-\cos k_y), 
\label{eq:gap}
\end{equation} 
with $\Delta_d$ being a $d$-wave pairing gap parameter. 
As the AF quasiparticles in Eq.~(\ref{eq:Phi_mix}), we employ a form of an AF
Hartree-Fock solution at half filling with $t'/t=0$: 
\begin{eqnarray}
a^\dag_{{\bf k},\sigma}=\alpha_{\bf k} c^\dag_{{\bf k},\sigma}+
\mbox{sgn}(\sigma)\ \beta_{\bf k} c^\dag_{{\bf k}+{\bf Q},\sigma}, 
\label{eq:QP1} \\
a^\dag_{{\bf k}+{\bf Q},\sigma}=-\mbox{sgn}(\sigma)\ \beta_{\bf k} 
c^\dag_{{\bf k},\sigma}+\alpha_{\bf k} c^\dag_{{\bf k}+{\bf Q},\sigma}, 
\label{eq:QP2} 
\end{eqnarray}
where ${\bf Q}$ is the AF nesting vector $(\pi,\pi)$, 
$\mbox{sgn}(\sigma)=1$ ($-1$) for $\sigma=\uparrow$ ($\downarrow$), and
\begin{equation}
\alpha_{\bf k}\ (\beta_{\bf k})=\frac{1}{\sqrt{2}}
\sqrt{1-(+)\frac{\varepsilon^{\rm AF}_{\bf k}} 
       {\left(\varepsilon^{\rm AF}_{\bf k}\right)^2+\Delta_{\bf AF}^2}}. 
\label{eq:alpha-beta} 
\end{equation}
Here, $\Delta_{\rm AF}$ corresponds to the AF gap parameter in the sense of 
mean-field theory. 

To introduce BRE into $\Phi_{\rm mix}$, we extend the band dispersions 
$\varepsilon^{\rm SC}_{\bf k}$ in Eq.~(\ref{eq:BCSDelta}) and 
$\varepsilon^{\rm AF}_{\bf k}$ in Eq.~(\ref{eq:alpha-beta}) independently
by including tight-binding hopping terms up to three-step processes shown 
in Fig. \ref{fig:bp}(b), 
\begin{equation}
\varepsilon^\Lambda_{\bf k}=\gamma_{\bf k}
  +\varepsilon^\Lambda_1({\bf k})+\varepsilon^\Lambda_2({\bf k}) 
  +\varepsilon^\Lambda_3({\bf k})+\varepsilon^\Lambda_4({\bf k}). 
\label{eq:ep-k}
\end{equation}
with $\Lambda=$~SC or AF and 
\begin{eqnarray}
&&\gamma_{\bf k}=-2t(\cos k_x+\cos k_y), 
\label{eq:gamma}
\\
&&\varepsilon^\Lambda_1({\bf k})=-4t^\Lambda_1\cos k_x\cos k_y, \\
&&\varepsilon^\Lambda_2({\bf k})=-2t^\Lambda_2(\cos 2k_x+\cos 2k_y), \\
&&\varepsilon^\Lambda_3({\bf k})=
    -4t^\Lambda_3(\cos 2k_x\cos k_y+\cos k_x\cos 2k_y), \quad\qquad \\
&&\varepsilon^\Lambda_4({\bf k})=-2t^\Lambda_4(\cos 3k_x+\cos 3k_y). 
\label{eq:epsilon}
\end{eqnarray}
Here, the eight band-adjusting parameters $t^\Lambda_\eta/t$ 
($\Lambda=\mbox{SC}$ or AF, $\eta=1$--4) are independent of $t'/t$ in 
${\cal H}$ and are optimized along with the other variational parameters 
($g$, $\zeta_{\rm d}$, $\zeta_{\rm h}$, $\Delta_d$, $\mu$, $\Delta_{\rm AF}$). 
Note that the ${\bf k}$ points used in Eqs.~(\ref{eq:QP1}) and (\ref{eq:QP2}) 
belong to $\{{\bf k}\}_{\rm occ}$ determined by 
$\varepsilon^{\rm AF}_{\bf k}$ (not $\gamma_{\bf k}$).\cite{note-HF} 
As a result, if $\{{\bf k}\}_{\rm occ}$ includes ${\bf k}$ points outside 
the folded AF Brillouin zone, $\phi({\bf k})$ for the corresponding 
${\bf k}$ in the sum in Eq.~(\ref{eq:Phi_mix}) is doubled, and 
$\phi({\bf k})$ for ${\bf k}$ ($\notin {\bf k}_{\rm F}$) inside the AF 
Brillouin zone becomes null. 
In $\Phi_{\rm mix}$, $\varepsilon^{\rm SC}_{\bf k}$ and 
$\varepsilon^{\rm AF}_{\bf k}$ are explicitly renormalized, and the weight 
of $\phi({\bf k})$ is also modified by $\{{\bf k}\}_{\rm occ}$ determined 
by $\varepsilon^{\rm AF}_{\bf k}$, as summarized in Table \ref{table:BR}.  
\par 

A pure one-body AF state $\Phi_{\rm AF}$ is given by the 
$\Delta_d\rightarrow 0$ limit of $\Phi_{\rm mix}$ as 
\begin{equation}
\Phi_{\rm AF}=\prod_{\{{\bf k}\}_{\rm occ},~\sigma}
a^\dag_{{\bf k},\sigma}|0\rangle, 
\label{eq:AF}
\end{equation}
where the AF quasiparticles are given by Eqs.~(\ref{eq:QP1}) and 
(\ref{eq:QP2}) and $\{{\bf k}\}_{\rm occ}$ is determined by 
$\varepsilon^{\rm AF}_{\bf k}$ in Eq.~(\ref{eq:ep-k}).
There are five variational parameters ($t^{\rm AF}_\eta$, $\Delta_{\rm AF}$) 
in $\Phi_{\rm AF}$. 
A pure $d_{x^2-y^2}$-wave singlet pairing (BCS) state of a fixed electron 
number\cite{Lhuillier} is given by the $\Delta_{\rm AF}\rightarrow 0$ limit 
of $\Phi_{\rm mix}$ as
\begin{equation}
\Phi_d =\left(\sum_{\bf k}\phi({\bf k})~
c_{{\bf k}\uparrow}^\dagger c_{{\bf -k}\downarrow}^\dagger
\right)^\frac{N}{2}|0\rangle, 
\label{eq:Phi_d}
\end{equation}
with $\phi({\bf k})$ given by Eq.~(\ref{eq:BCSDelta}). 
There are six variational parameters ($t^{\rm SC}_\eta$, $\Delta_d$, $\mu$) in 
$\Phi_d$. 

\subsection{Variational Monte Carlo calculations\label{sec:VMC}} 
In general, it is impossible to accurately calculate variational expectation 
values of a many-body wave function $\langle{\cal O}\rangle$, with ${\cal O}$ 
being an operator, by analytical means. 
Instead, in many cases, the expectation values can be accurately numerically estimated using VMC 
methods\cite{VMC1,VMC2,VMC3,VMC4}. 
Recently, many parameters (up to more than $10^6$) in 
$\langle{\cal H}\rangle$ have been efficiently optimized by newly introduced 
algorithms.\cite{VMCopt} 
In the present cases, however, we cannot adopt ordinary optimization 
schemes using derivatives of energy because $E(\{\gamma\})$ is constant 
($\Psi_{\rm N}$) or nearly constant ($\Psi_{\rm AF}$ and $\Psi_{\rm mix}$) 
as a function of the band parameters ($t_\eta^\Lambda$) in the parameter set 
$\{\gamma\}$ and has irregularly distributed discontinuities. 
To address BR in $\Psi_{\rm N}$, we combine a VMC method with the 
extrapolation scheme described in Appendix\ref{sec:normal-A}. 
For $\Psi_{\rm AF}$ and $\Psi_{\rm mix}$, we repeat a primitive 
linear optimization method in this 
study until optimization becomes successful, 
although better ways are applicable. 
Details are described in Appendix\ref{sec:AF-A}. 
For $\Psi_d$, ordinary optimization algorithms are applicable unless 
$\Delta_d$ approaches zero. 
For $\Delta_d\sim 0$, a difficulty similar to that for $\Psi_{\rm AF}$ manifests 
itself. 
\par

We calculate physical quantities using more than $2.5\times 10^5$ samples. 
The accuracy of the total energy of $10^{-4}t$ is preserved, similarly to in previous 
studies. 
It is laborious to accurately converge $\Delta_{\rm AF}$ (or $\Delta_d$) 
and the band parameters to specific values because there is redundancy 
among these parameters. 
However, this affects the calculations of physical quantities only slightly 
in most cases.  
\par

We use systems of $N_{\rm s}=L\times L$ sites with $L=10$--$18$ under 
periodic-antiperiodic boundary conditions. 
The closed-shell condition is not satisfied because we allow 
$\{{\bf k}\}_{\rm occ}$ to be optimized automatically, although the total 
momentum is preserved at zero. 
In this paper, we often consider rough system-size dependence for 
$\delta\sim 0.08$ using $L=10$, 12, 14, 16, and 18 with $N=92$, 132, 180, 
236, and 296 ($\delta=0.08$, 0.0833, 0.0816, 0.0781, and 0.0864), 
respectively. 
\par

\section{BRE on Pure $d$-Wave Pairing State\label{sec:d-wave}}
In this section, we discuss BRE on the pure $d$-wave pairing state without 
an AF order, $\Psi_d={\cal P}\Phi_d$. 
In Sect.~\ref{sec:d-BRE}, we confirm that there is a large BRE in $\Psi_d$, 
as found in previous 
studies.\cite{Himeda-BR,Shih-BR,Koba-old,Wata-t-J,Tocchio} 
In Sect.~\ref{sec:d-ene}, however, we show that the improvement in energy is 
unexpectedly small. 
In Sect.~\ref{sec:d-PQ}, we also find that the modification of relevant physical 
quantities is negligible. 
\par

\subsection{Large BRE for (doped) Mott insulators\label{sec:d-BRE}}
%
\begin{figure}
\begin{center}
\hskip 0mm
\includegraphics[width=9.5cm,clip]{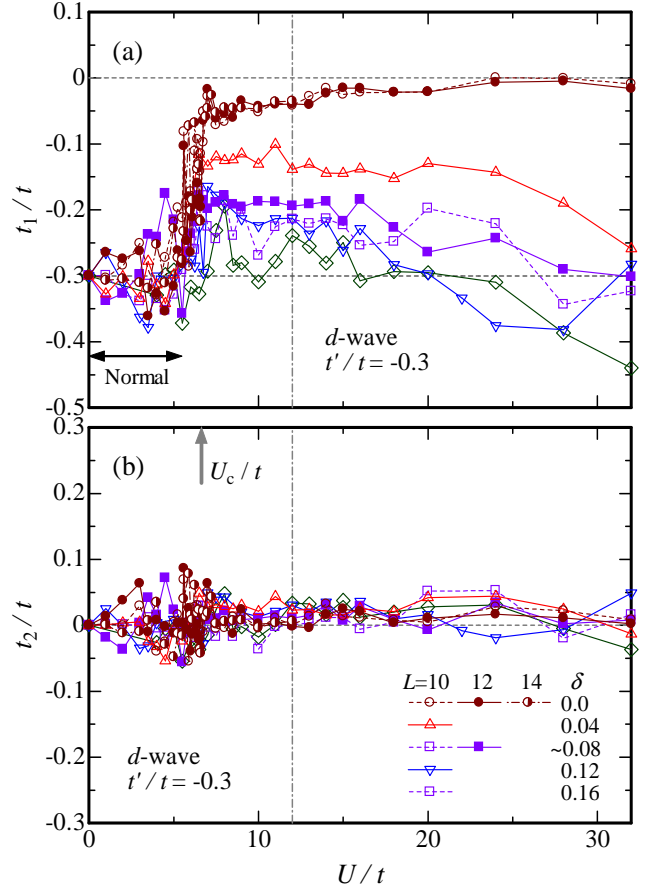} 
\end{center} 
\vskip -60mm
\caption{(Color online) 
Optimized band parameters (a) $t_1$ and (b) $t_2$ of $d$-wave singlet 
pairing state (BR2) as functions of $U/t$ for several doping rates. 
In (a), the area where $\Psi_d$ is reduced to $\Psi_{\rm N}$ is shown by 
an arrow labeled `Normal'.
In (b), the Mott transition point at half filling is indicated by a gray arrow.
}
\label{fig:para-td-t2-jpsj} 
\end{figure}
\begin{figure}
\begin{center}
\vskip -4mm
\hskip 0mm
\includegraphics[width=9.5cm,clip]{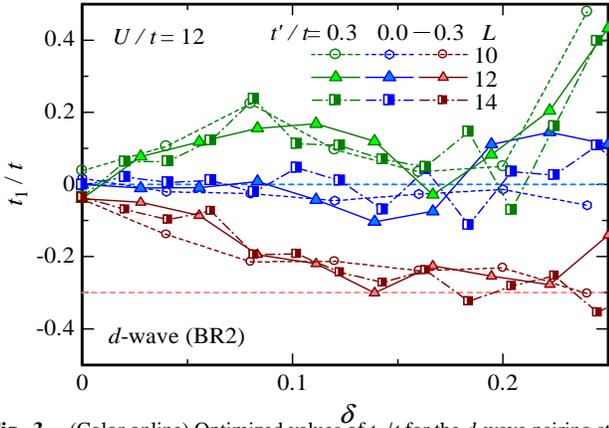} 
\end{center} 
\vskip -34mm
\caption{(Color online)
Optimized values of $t_1/t$ for the $d$-wave pairing state (BR2) plotted as functions of doping rate for three values of $t'/t$ in the regime 
of a doped Mott insulator ($U/t=12$). 
}
\label{fig:para-td-vsn-jpsj} 
\end{figure}

First, we attempted to optimize $\Psi_d$ only with two band parameters 
$t_1$ and $t_2$ by putting $t_3=t_4=0$ for simplicity. 
We abbreviate this two-band-parameter optimization to BR2. 
\par 

In Fig.~\ref{fig:para-td-t2-jpsj}(a), we show the optimized values of 
$t_1/t$ as functions of $U/t$, fixing the model parameter at $t'/t=-0.3$. 
For $U/t\lesssim 5$, $t_1$ preserves the bare value $t_1\sim t'$ irrespective 
of $\delta$; no substantial BR exists. 
This is because $\Delta_d$ is very small in this range of $U/t$ and 
the state is reduced to the normal state, as in the cases without 
BRE.\cite{YOT,Y2013} 
$\Psi_{\rm N}$ also shows no substantial BR in this range of $U/t$
as shown in Sect.~\ref{sec:normal}.
The relatively large statistical fluctuation in the case of $\Delta_d\sim 0$ 
stems from the same difficulty as in $\Psi_{\rm N}$ in optimizing the band 
parameters [see Appendix\ref{sec:normal-A}]. 
On the other hand, at $U=U_{\rm c}\sim 6.5t$\cite{YOT} ($U_{\rm c}/t$: 
Mott transition point in $\Psi_d$), $t_1/t$ abruptly increases, 
in particular, $t_1/t$ approaches 0 at $\delta=0$. 
As previously pointed out,\cite{Himeda-BR,Y2013} this BR of $\Psi_d$ occurs 
so that the quasi-Fermi surface overlaps with or approaches antinodal points 
[$(\pi,0)$, etc.], where the van Hove singularity exists for $|t'/t|\le 0.5$ 
and the $d$-wave gap becomes maximum. 
Furthermore, the elastic electron scattering of ${\bf q}={\bf Q}$ connects these 
points with opposite signs of $\Delta_d$. 
Restoration of the nesting condition, which is the principal cause of BRE 
for the AF state, seems a subordinate aspect for $\Psi_d$. 
As $\delta$ increases, $t_1$ slowly approaches the value of $t'$ for the 
same reason (see Fig.~\ref{fig:para-td-vsn-jpsj}). 
In contrast to $t_1/t$, the optimized $t_2/t$ remains almost zero (the bare 
value) for all $U/t$ and $\delta$ for $t'/t=-0.3$, as shown in 
Fig.~\ref{fig:para-td-t2-jpsj}(b). 
\par

\begin{table}
\caption{
Rough estimate of coefficients in fitting function Eq.~(\ref{eq:t1t})
for $U/t=12$ estimated from data for $L=10$-$14$. 
}
\begin{center}
\begin{tabular}{l|ccccc}
\hline
$\delta$   & $0.0$  & $0.04$ & $\sim0.08$ & $0.12$ & $0.16$  
\\
\hline
$\alpha_+$ & $0.14$ & $0.41$ & $0.55$     & $0.25$ & $0.05$
\\
$\alpha_-$ & $0.14$ & $0.41$ & $0.67$     & $0.90$ & $0.95$
\\
\hline
\end{tabular}
\end{center}
\vskip -9mm
\label{table:Coeff}
\end{table}

Next, we look at the $t'/t$ dependence of $t_1/t$ and $t_2/t$ for $U>U_{\rm c}$. 
We find that the optimized $t_1/t$ is roughly fitted by separate linear functions 
of $t'/t$ for the hole- and electron-doped cases:  
\begin{equation}
t_1/t=\alpha_\pm(\delta)\times t'/t, 
\label{eq:t1t}
\end{equation}
where $\alpha_+(\delta)$ [$\alpha_-(\delta)$] is the coefficient for 
$t'/t>0$ [$t'/t<0$] at a fixed $\delta$. 
If $t_\eta$ ($\eta\ge 2$) is ineffective (we actually see it shortly), BRE 
are nonexistent for $\alpha_\pm=1$ and, inversely, 
$\varepsilon_{\bf k}^{\rm SC}$ is renormalized to the case of $t'=0$ for 
$\alpha_\pm=0$. 
The values of $\alpha_\pm$ depend on $U/t$ only slightly and are shown 
for $U/t=12$ in Table~\ref{table:Coeff}. 
Although the magnitudes of BR exhibit opposite tendencies between $\alpha_+$ 
and $\alpha_-$ for $\delta\gtrsim 0.08$, $\alpha_\pm$ is always positive. 
As a result of this positiveness, the convexity ($t'/t>0$) or concavity ($t'/t<0$) 
of the bare Fermi surface near $(\pi/2, \pi/2)$ is preserved in the 
renormalized quasi-Fermi surface of $\varepsilon_{\rm k}^{\rm SC}$. 
As a result, the locus of a hot spot ---the intersection of a (quasi-) Fermi 
surface and the AF Brillouin zone boundary, where scattering of 
${\bf q}=(\pi, \pi)$ takes place---\cite{note-hotspot} is near $(\pi, 0)$ 
for $t'/t<0$ but approaches $(\pi/2, \pi/2)$ to some extent for 
$t'/t>0$.\cite{Hirashima,Wata-t-J,nonmonotonic} 
As we will see in Sect.~\ref{sec:Co}, the loci of hot spots become a 
condition that a coexistent state arises. 
\par

In contrast to $t_1/t$, $t_2/t$ is again found to be almost zero for any 
$t'/t$ and $\delta$. 
The effect of $t_3$ and $t_4$ is considered using $\Psi_d$ with four band 
parameters $t_1$--$t_4$ in $\varepsilon_{\bf k}$ [Eq.~(\ref{eq:ep-k})] (BR4). 
The behavior of $t_1/t$ and $t_2/t$ for BR4 is basically similar to 
that for BR2 mentioned above. 
We found that both the optimized $t_3$ and $t_4$ have small positive values 
($t_3/t\lesssim 0.11$, $t_4/t\lesssim 0.095$, at largest at half filling) 
almost independent of $t'/t$. 
These values decrease as $\delta$ increases and almost vanish for 
$\delta\gtrsim 0.1$. 
As we will see in Sect.~\ref{sec:d-ene}, the effects of $t_3$ and $t_4$ 
on energy and other quantities are also slight. 
\par

To summarize, BRE on $\Psi_d$ are large for $U\gtrsim U_{\rm c}$, 
$\delta\sim 0$ and large $|t'/t|$.  
If these conditions are satisfied, the effective band tends to 
the bare band of a square lattice 
($\varepsilon_{\bf k}\rightarrow\gamma_{\bf k}$ or 
$|t_1/t|\rightarrow 0$). 
This feature of BRE on the $d$-wave pairing state has already been pointed 
out in previous 
studies.\cite{Himeda-BR,Shih-BR,Liu,Wata-Org,Koba-old,Wata-t-J,Tocchio} 
\par

\subsection{Slight improvement in energy by BRE\label{sec:d-ene}}
%
\begin{figure}
\begin{center}
\vskip 0mm
\hskip -0mm
\includegraphics[width=11.5cm,clip]{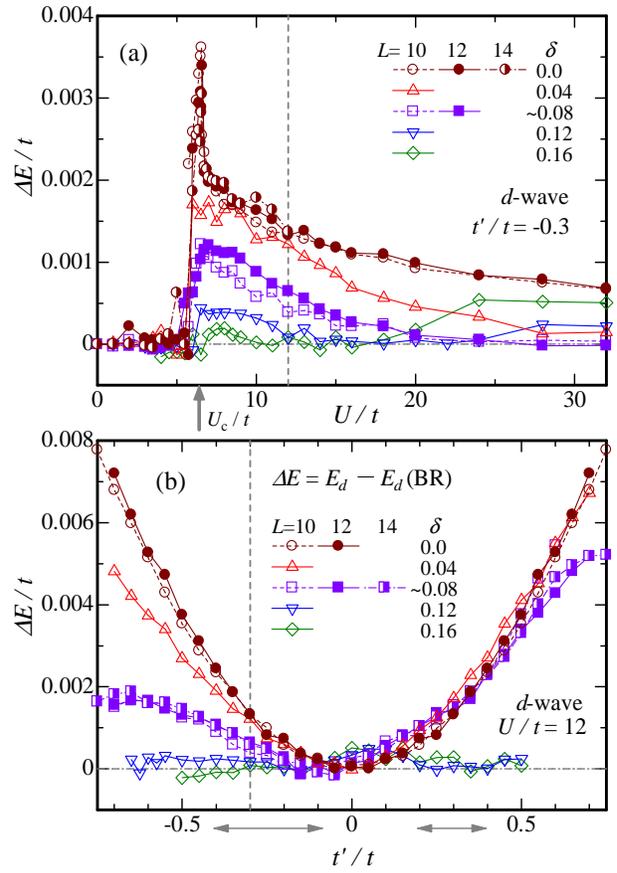} 
\end{center} 
\vskip -1mm
\caption{(Color online)
Energy improvement [Eq.~(\ref{eq:DelE})] owing to BRE for the $d$-wave 
pairing state (BR2) shown for some values of $\delta$ and $L$, 
(a) as functions of correlation strength with $t'/t=-0.3$ and
(b) as functions of $t'/t$ with $U/t=12$. 
The Mott transition point at half filling is indicated by a thick gray arrow 
in (a). 
In (b), plausible areas of $t'/t$ for hole-doped ($t'/t<0$) and 
electron-doped ($t'/t>0$) cuprates are indicated with gray arrows. 
}
\label{fig:DelE-br-jpsj}
\end{figure}
\begin{figure}
\begin{center}
\hskip -0mm
\includegraphics[width=8.5cm,clip]{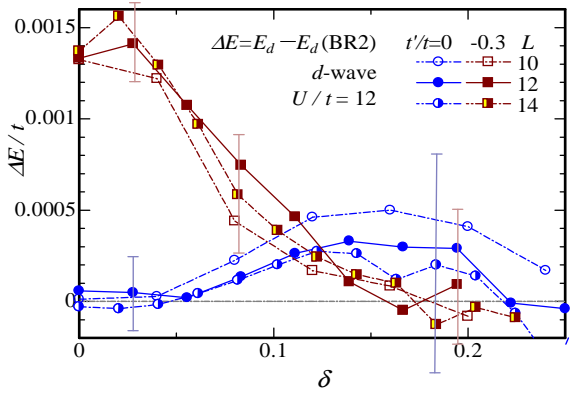} 
\end{center} 
\vskip -34mm
\caption{(Color online)
Energy improvement [Eq.~(\ref{eq:DelE})] owing to BRE for the $d$-wave 
pairing state (BR2) plotted as functions of doping rate for two values 
of $t'/t$ in a strongly correlated regime ($U/t=12$).  
Broad statistical errors are indicated by bars for some data points. 
}
\label{fig:DelE-br-vsn} 
\end{figure}
Here and in some later sections, we consider the improvement in the total energy 
per site owing to BRE, represented as 
\begin{equation}
\Delta E=E_\Lambda-E_\Lambda(\mbox{BR}),
\qquad  (\Lambda=d, \mbox{N, or AF})
\label{eq:DelE}
\end{equation} 
where $E_d$ [$E_d(\mbox{BR})$] is the energy of $\Psi_d$ without [with] 
BRE; $\Delta E/t\ge 0$ holds except for statistical errors. 
In Fig.~\ref{fig:DelE-br-jpsj}(a), the $U/t$ dependence of $\Delta E/t$ is shown 
for some values of $\delta$ for $t'/t=-0.3$. 
The regime of finite $\Delta E/t$ for $U>U_{\rm c}$ corresponds to that of 
the finite BR of $t_1/t$ shown in Fig.~\ref{fig:para-td-t2-jpsj}(a). 
As $\delta$ increases, both the magnitude of BR and $\Delta E/t$ decrease 
and almost vanish in the overdoped regime ($\delta\gtrsim 0.15$). 
Figure \ref{fig:DelE-br-jpsj}(b) shows the $t'/t$ dependence of $\Delta E/t$ 
for $U/t=12$, which mostly corresponds to the degree of BR of $t_1/t$ 
given by Eq.~(\ref{eq:t1t}) with $\alpha_\pm$ in Table~\ref{table:Coeff}. 
The exception for $t'/t>0$ and large $\delta$ is caused by the vanishing of hot 
spots, which BRE alone cannot control. 
Shown in Fig.~\ref{fig:DelE-br-vsn} is the $\delta$ dependence of $\Delta E/t$, 
which again corresponds to the degree of BRE on $t_1/t$ shown in 
Fig.~\ref{fig:para-td-vsn-jpsj}. 
\par

\begin{table}
\caption{
Examples of total energy per site shown for a specific case 
($t'/t=-0.3$, $U/t=12$, $L=10$) for comparison among four states 
with different BR levels and three doping rates. 
The brackets denote errors in the last digits. 
}
\label{table:E-comp}
\begin{center}
\begin{tabular}{l|l|lll}
\hline
State 
& Condition & \multicolumn{3}{c}{$E/t$} 
\\
\cline{3-5}
& \quad of $\varepsilon_{\bf k}$ 
& \multicolumn{1}{c}{$\delta=0.0$} 
& \multicolumn{1}{c}{$0.04$} 
& \multicolumn{1}{c}{$0.08$} 
\\
\hline\hline
Normal
& no BR  & $-0.1855(2)$  & $-0.3230(2)$ & $-0.4259(2)$
\\                       
& BR     & $-0.2660(1)$  & $-0.3360(2)$ & $-0.4310(1)$
\\                       
\hline                   
$d$-wave 
& no BR  & $-0.3222(2)$  & $-0.3816(4)$ & $-0.4602(2)$
\\                      
& BR2    & $-0.3235(2)$  & $-0.3827(1)$ & $-0.4606(3)$
\\                      
& BR4    & $-0.3241(2)$  & $-0.3828(4)$ & $-0.4606(10)$
\\
\hline
AF                       
& no BR  & $-0.1879(2)$  & $-0.3288(2)$ & $-0.4259(2)$
\\                       
& BR4    & $-0.35319(2)$ & $-0.4201(3)$ & $-0.4881(1)$ 
\\
\hline
Mixed 
& BR 4+4 & $-0.3559(2)$  & $-0.4211(2)$ & $-0.4915(2)$
\\                      
\hline
\end{tabular}
\end{center}
\vskip -9mm
\label{table:E-d-wave}
\end{table}
Now we are aware that the energy is basically improved according to the 
degree of BRE on $t_1/t$ for every model parameter. 
Nevertheless, what we should notice here is that the magnitude of $\Delta E/t$ 
is unexpectedly small. 
The precision (statistical error) of the energy in the present VMC calculations for $\Psi_d$ 
is on the order of $10^{-4}t$ as shown by bars in 
Fig.~\ref{fig:DelE-br-vsn}, while the maximum value of $\Delta E/t$ is only $\sim10^{-3}t$ (only slightly larger than the errors). 
In Table \ref{table:E-comp}, $E/t$ for $\Psi_d$ is compared among the cases of 
without BR, BR2, and BR4 for typical model parameters. 
We also find that the difference between BR2 and BR4 is very small. 
What is more, the difference in $\Psi_d$ is an order (two orders) of magnitude 
smaller than that in $\Psi_{\rm N}$ ($\Psi_{\rm AF}$) for any $\delta$ 
presented. 
This difference is visually perceived in Fig.~\ref{fig:DelE-vsn-a03-comp}. 
\par

\subsection{Small modification of quantities by BRE\label{sec:d-PQ}}
%
\begin{figure}
\begin{center}
\vskip 4mm
\hskip -0mm
\includegraphics[width=9.5cm,clip]{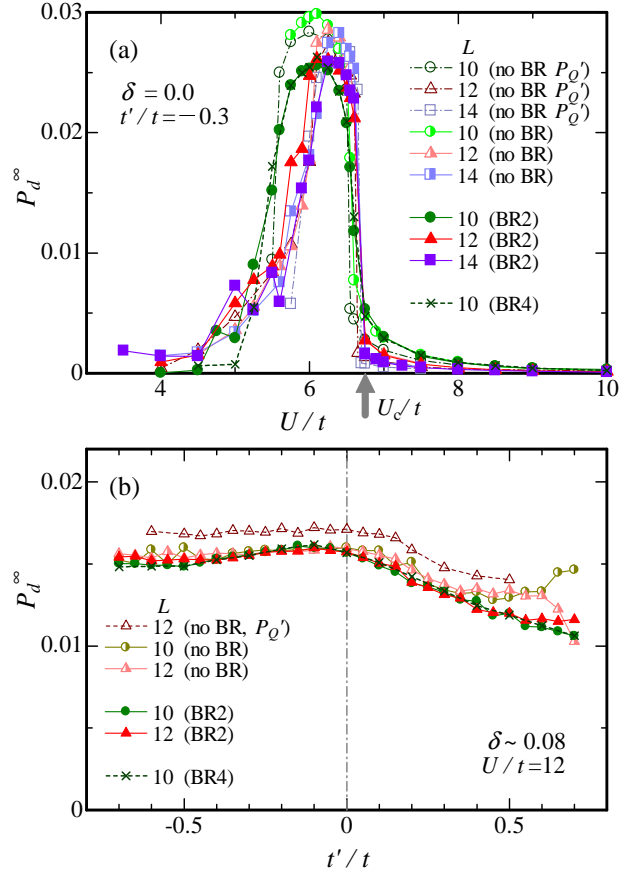} 
\end{center} 
\vskip -55mm
\caption{(Color online)
Behavior of the $d$-wave SC correlation functions in $\Psi_d$ compared 
between BR cases and no-BR cases 
(a) at half filling and $t'/t=-0.3$ as functions of $U/t$ and 
(b) for $\delta\sim 0.08$ and $U/t=12$ as a function of $t'/t$. 
On the horizontal axis in (a), the Mott transition point is indicated by 
a thick gray arrow. 
The data for ``no BR ${\cal P}_Q'$" are adopted from Ref.~\citen{Y2013}, 
in which a similar but somewhat different D-H factor is used. 
}
\label{fig:pd-jpsj}
\end{figure}
First, we consider a $d$-wave pairing correlation function, 
\begin{equation}
P_d({\bf r})=\frac{1}{N_{\rm s}}
\sum_{i}\sum_{\tau,\tau'=\hat {\bf x},\hat {\bf y}}
(-1)^{1-\delta(\tau,\tau')}
\left\langle{\Delta _\tau^\dag({\bf R}_i)\Delta_{\tau'}
({\bf R}_i+{\bf r})}\right\rangle, 
\label{eq:pd}
\end{equation}
where $\hat{\bf x}$ ($\hat{\bf y}$) denotes the lattice vector in the 
$x$ ($y$) direction, $\delta(\tau,\tau')$ indicates the Kronecker delta, and 
$\Delta_\tau^\dag({\bf R}_i)$ is the creation operator of a nearest-neighbor 
singlet pair at site ${\bf R}_i$, 
\begin{equation}
\Delta_\tau^\dag({\bf R}_i)=
(c_{{i}\uparrow}^\dag c_{{i}+\tau\downarrow}^\dag+ 
 c_{{i}+\tau\uparrow}^\dag c_{{i}\downarrow}^\dag)
 /{\sqrt 2}. 
\label{eq:singlet}
\end{equation}
If $P_d({\bf r})$ remains finite for 
$|{\bf r}|\rightarrow\infty$ ($P_d^\infty$), a $d$-wave off-diagonal 
long-range order exists; $P_d^\infty$ roughly represents the square of 
the SC gap. 
For $\Psi_d$, we estimate $P_d^\infty$ in the same way as discussed in 
Appendix~C in Ref.~\citen{Y2013}. 
As an example, in Fig.~\ref{fig:pd-jpsj}(a), we show $P_d^\infty$ at 
half filling for some levels of BR (and ${\cal P}$) for $L=10$-$14$. 
As discussed in Ref.~\citen{Y2013}, $P_d^\infty$ is negligible for 
small values of $U/t$. 
As $U/t$ increases, $P_d^\infty$ abruptly increases at $U/t\sim 5$, 
exhibits a sharp peak near the Mott transition point $U_{\rm c}/t\sim 6.5$, 
and vanishes in the Mott insulator regime $U>U_{\rm c}$. 
Although the peak value of $P_d^\infty$ tends to be slightly decreased by BRE, 
the behavior does not vary as a whole. 
For $\delta>0$, the area where $P_d^\infty$ is sizable extends to large 
values of $U/t$, but the modification of $P_d^\infty$ by BRE remains 
small (not shown). 
The modification of $P_d^\infty$ is also small when $t'/t$ is varied, as shown 
in Fig.~\ref{fig:pd-jpsj}(b). 
Furthermore, the difference between BR2 and BR4 is negligible.
\par

\begin{figure}
\begin{center}
\hskip -0mm
\includegraphics[width=9.5cm,clip]{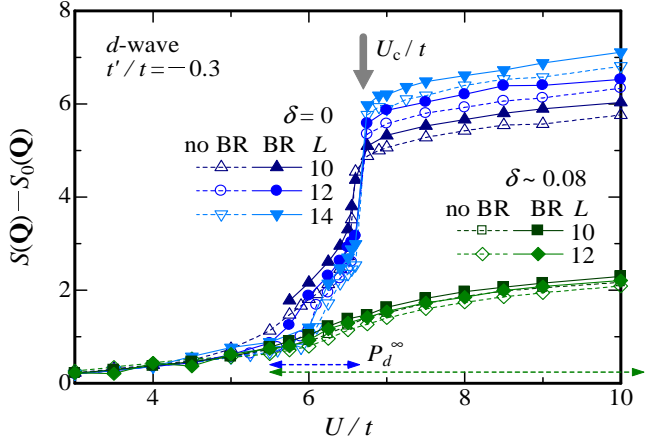} 
\end{center} 
\vskip -34mm
\caption{(Color online)
Spin structure factor at ${\bf Q}=(\pi,\pi)$ measured from the bare ($U=0$) 
value $S_0({\bf Q})$ ($=1$) compared between 
BR2 and no-BR cases for $\delta=0$ and $\sim 0.08$ for $t'/t=-0.3$ as 
functions of $U/t$.
A few cases with different values of $L$ are shown. 
The Mott transition point at half filling are indicated by a thick gray arrow. 
Near the horizontal axis, the areas where the $d$-wave correlation function 
$P_d^\infty$ becomes sizable are indicated by dashed arrows for 
$\delta=0$ (blue) and $\sim0.08$ (green). 
See Fig.~\ref{fig:pd-jpsj} for $\delta=0$. 
}
\label{fig:sq-Q-vsU-jpsj} 
\end{figure}

Next, we look at the ${\bf q}={\bf Q}$ element of the spin structure factor 
\begin{equation} 
S({\bf q})=\frac{1}{N_{\rm s}}\sum_{ij}{e^{i{\bf q}
\cdot({\bf R}_i-{\bf R}_j)} 
\left\langle{S_{i}^zS_{j}^z}\right\rangle}. 
\label{eq:sq}
\end{equation} 
The $U/t$ dependence of $S({\bf Q})$ is shown in Fig.~\ref{fig:sq-Q-vsU-jpsj} 
for $\delta=0$ and $\sim 0.08$. 
As previous studies pointed out, an increase in $S({\bf Q})$ is necessary for 
an increase in $P_d^\infty$ because the electron scattering of ${\bf Q}$ 
yields an attractive force for pairing. 
Anyway, the modification of $S({\bf Q})$ by BRE is also small and only 
quantitative even at half filling. 
\par

\begin{figure}
\begin{center}
\hskip -0mm
\includegraphics[width=10.5cm,clip]{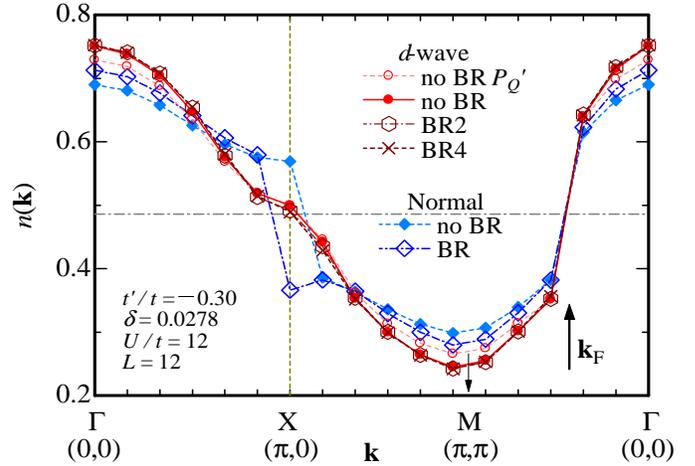} 
\end{center} 
\vskip -0mm
\caption{(Color online)
Comparison of momentum distribution function of $d$-wave pairing state 
(red and brown) among various levels of BR (and ${\cal P}_Q$) in the regime 
of doped Mott insulators ($U/t=12$) along the path 
$(0,0)\rightarrow(\pi,0)\rightarrow(\pi,\pi)\rightarrow(0,0)$. 
The Fermi surface in the nodal ($\Gamma$-M) direction is indicated by 
an arrow labeled ${\bf k}_{\rm F}$. 
For comparison, we add $n({\bf k})$ for the normal state (blue) with and 
without BRE as discussed in Sect.~\ref{sec:normal}. 
}
\label{fig:nk-jpsj} 
\end{figure}
Finally, we discuss the momentum distribution function 
\begin{equation}
n({\bf k})=
\frac{1}{2}\sum_\sigma\langle c^\dag_{\bf k\sigma}c_{\bf k\sigma}\rangle. 
\label{eq:nk}
\end{equation}
It seems that $n({\bf k})$ sensitively reflects the variation of the effective 
band $\varepsilon_{\bf k}$, which is considerably renormalized depending on the case. 
Figure \ref{fig:nk-jpsj} depicts $n({\bf k})$ in such cases with red 
(without BRE) and brown (with BRE) symbols for $\Psi_d$ and blue symbols for 
the normal state $\Psi_{\rm N}$ (see Sect.~\ref{sec:normal}). 
In accordance with the above expectation, the locus of ${\bf k}_{\rm F}$ 
(discontinuity) near the X point in $\Psi_{\rm N}$ is shifted to a neighbor 
${\bf k}$ point by BRE. 
Nevertheless, in $\Psi_d$, the modification by BRE is very small for the gap 
behavior in the antinodal area (${\bf k}\sim$X) as well as for the 
discontinuity in the nodal direction [${\bf k}\sim(\pi/2,\pi/2)$], 
despite the large BRE 
($t_1/t\sim -0.05$ for $t'/t=-0.3$ in Fig.~\ref{fig:para-td-vsn-jpsj}). 
This is probably because the choice of $\{\bf k\}_{\rm occ}$, which is 
controlled by $\varepsilon_{\bf k}$ in $\Psi_{\rm N}$ and $\Psi_{\rm AF}$, 
is unnecessary in $\Psi_d$ as shown in Table \ref{table:BR}. 
\par

In summary, the BR of $\varepsilon_{\bf k}^{\rm SC}$ itself is large 
($t_1/t\rightarrow 0$) for $U\gtrsim U_{\rm c}$, large $|t'/t|$, 
and $\delta\sim 0$, as previous studies elucidated. 
Notwithstanding, BRE on relevant quantities as well as on the energy for $\Psi_d$ 
are very small and insignificant compared with those on the normal and AF states 
discussed below. 
\par

\section{BRE on Normal (Paramagnetic) State\label{sec:normal}}
In this section, we discuss BRE on the normal or paramagnetic state 
(projected Fermi sea), 
\begin{equation}
\Psi_{\rm N}={\cal P}\Phi_{\rm N}={\cal P}
\prod_{\{{\bf k}_{\rm occ}\},\sigma}c^\dag_{{\bf k},\sigma}|0\rangle. 
\label{eq:normal}
\end{equation}
We cannot apply ordinary optimization procedures to $\Psi_{\rm N}$ that 
use the gradients of $E/t$ with respect to band parameters because $E/t$ for 
$\Psi_{\rm N}$ with a finite $N$ is constant in a certain area of 
the band-parameter space. 
Hence, we must resort to a different way of optimizing $\Psi_{\rm N}$, 
which is described in Appendix\ref{sec:normal-A}. 
Here, we focus on the features of the optimized $\Psi_{\rm N}$. 
\par 

Before discussing BRE, we briefly review some aspects of $\Psi_{\rm N}$ 
without BRE ($t_1=t'$).\cite{YOT,Y2013} 
At half filling, a Mott transition occurs at $U_{\rm c}/t\sim 8.5$ for 
$t'/t=0$; $U_{\rm c}/t$ increases as $|t'/t|$ increases: 
$U_{\rm c}/t\sim 11.2$ for $|t'/t|=0.3$. 
Although the Mott transition does not exist for $\delta>0$, the nature of 
$\Psi_{\rm N}$ markedly changes at $U\sim U_{\rm c}$. 
For $U\gtrsim U_{\rm c}$, $\Psi_{\rm N}$ is not a simple metal but takes on 
a typical feature of Mott physics (D-H binding effect) as a doped Mott 
insulator. 
\par

\begin{figure}
\begin{center}
\hskip -0mm
\includegraphics[width=10.5cm,clip]{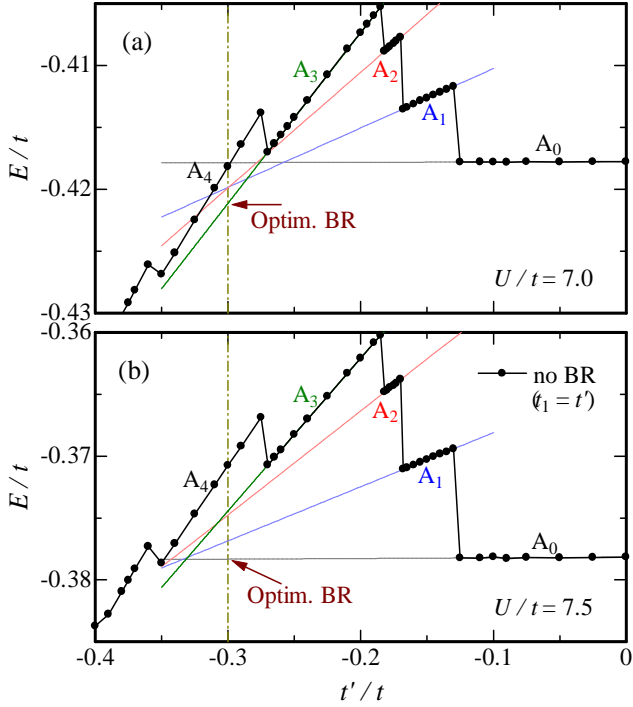} 
\end{center} 
\vskip -0mm
\caption{(Color online) 
Total energy of $\Psi_{\rm N}$ at half filling compared between 
the cases of (a) $U/t=7.0$ and (b) $7.5$ as a function of $t'/t$. 
Solid circles indicate $E/t$ without BRE, namely $t_1=t'$. 
A$_\ell$ ($\ell$: integer) indicates the area of $t'/t$ corresponding to 
$\{{\bf k}_\ell\}_{\rm occ}$. 
The optimized energy owing to BRE is given by the lowest value among all 
extrapolated lines. 
For $t'/t=-0.3$, the optimized value of $E/t$ is indicated by an arrow 
in each panel. 
A detailed explanation of the optimization is given 
in Appendix\ref{sec:normal-A}. 
}
\label{fig:evsa-jpsj}
\end{figure}

Now, we consider BRE. 
Because BRE are inefficient or weak for $t'/t\sim 0$, similarly 
to the case of $\Psi_d$, we first consider the moderate case $t'/t=-0.3$. 
We start with half-filled cases. 
Similarly to in $\Psi_d$, the energy reduction by BRE is zero or very small 
for $U/t\lesssim 6$, even if the optimized $t_1$ (accurately, the area including 
$t_1$) somewhat shifts from $t'$ (the area including $t'$). 
As shown in Fig.~\ref{fig:evsa-jpsj} for $L=14$, the optimized energy 
indicated by an arrow is 
given by $\{{\bf k}_3\}_{\rm occ}$ (${\rm A}_3=[-0.27,-0.19]$) for $U/t=7.0$, 
while it is given by $\{{\bf k}_0\}_{\rm occ}$ (${\rm A}_0=[-0.125,0.125]$) for $U/t=7.5$. 
Namely, the optimized band parameter $t_1$ rapidly varies from 
$\sim t'$ ($\in{\rm A}_3$) to $\sim 0$ ($\in{\rm A}_0$) between $U/t=7.0$ 
and $7.5$ in this case, and the nesting condition is restored. 
For $U/t\gtrsim 7$, the optimized $\{{\bf k}\}_{\rm occ}$ remains 
equal to $\{{\bf k}_0\}_{\rm occ}$, or the optimized value of $t_1/t$ remains 
$\sim 0$ ($\varepsilon^{\rm N}_{\bf k}=\gamma_{\bf k}$). 
Also the renormalized state becomes identical to the normal state without BR 
of the simple square lattice, whose behavior is reviewed above.\cite{YOT} 
Owing to BRE, the Mott transition point for $t'/t=-0.3$ shifts from 
$U_{\rm c}/t\sim 11.2$ to $\sim 8.5$. 
In fact, the optimal energy at $U/t=8.5$ for $|t'/t|\lesssim 0.5$ ($L=12$) 
is given by $\{{\bf k}_0\}_{\rm occ}$; thus, if BRE are introduced, 
the properties of the Mott transitions and Mott insulators for 
$|t'/t|\lesssim 0.5$ are reduced to those for the simple square-lattice 
case ($t'=0$) without BRE. 
\par

\begin{figure}
\begin{center}
\vskip 0mm
\hskip -0mm
\includegraphics[width=11.5cm,clip]{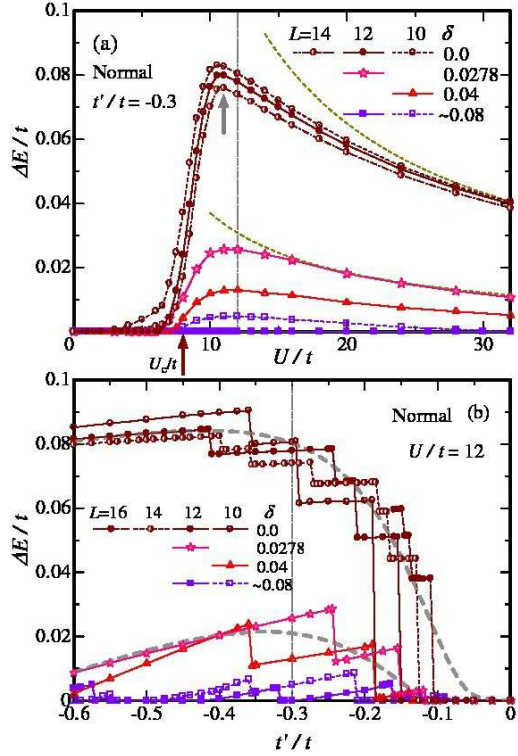} 
\end{center} 
\vskip -60mm
\caption{(Color online)
Energy improvement [Eq.~(\ref{eq:DelE})] owing to BRE for the normal state 
shown for some doping rates and values of $L$, 
(a) as functions of correlation strength with $t'/t=-0.3$, and
(b) as a function of $t'/t$ ($<0$) with $U/t=12$. 
In (a), the Mott transition points at half filling are indicated by thick 
arrows (brown for BR case, gray for no-BR case). 
Guide lines proportional to $t/U$ are added (dashed lines). 
In (b), data for each $\delta$ are well fitted by 
$\Delta E/t=-(\alpha/x)\exp(\beta/x)+\gamma x$ with $x=t'/t$ and $\alpha$, 
$\beta$, and $\gamma$ being positive constants, as shown with gray dashed 
lines. 
}
\label{fig:DelE-N-jpsj}
\end{figure}

In Fig.~\ref{fig:DelE-N-jpsj}(a), we show the energy reduction owing 
to BRE [Eq.~(\ref{eq:DelE})] for $t'/t=-0.3$ as a function of $U/t$. 
At half filling, as $U/t$ increases, $\Delta E/t$ abruptly increases at 
$U/t\sim 7$ owing to the reason mentioned above, roughly as 
$\Delta E/t=\alpha\exp(-\beta t/U)$ with $\alpha$ and $\beta$ being 
positive constants. 
Then, $\Delta E/t$ exhibits a peak at $U/t\sim 11$, which corresponds 
to $U_{\rm c}/t$ for the case without BRE, then slowly decreases 
(proportionally to $t/U$ for $U/t\rightarrow\infty$). 
As the doping rate increases from $\delta=0$, the overall feature 
of the $U/t$ dependence is preserved but the magnitude rapidly decreases. 
For all the doping rates shown, $\Delta E/t$ is negligible for a weakly 
correlated regime ($U/t\lesssim 7$), meaning that appreciable BRE are 
also a characteristic of strong correlation for the normal state. 
In Fig.~\ref{fig:DelE-N-jpsj}(b), the $t'/t$ dependence ($t'/t<0$) of 
$\Delta E/t$ is shown in the regime of Mott physics ($U/t=12$) 
for some doping rates. 
The BRE are largest at $|t'/t|\sim 0.3$-$0.4$ and slight for 
$|t'/t|\sim 0$. 
\par

\begin{figure}
\begin{center}
\vskip -2mm
\hskip -0mm
\includegraphics[width=8.5cm,clip]{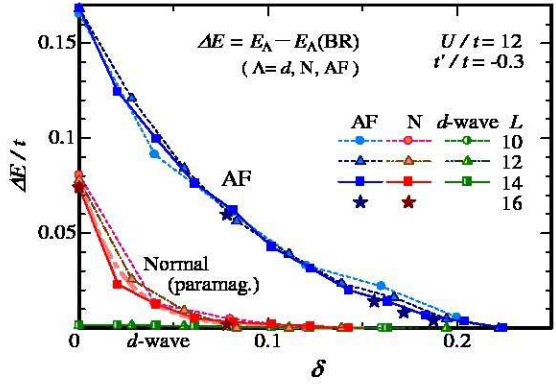} 
\end{center} 
\vskip -40mm
\caption{(Color online)
Energy improvement [Eq.~(\ref{eq:DelE})] by BRE compared among 
the normal, AF (Sect.~\ref{sec:AF}), and $d$-wave pairing 
(Sect.~\ref{sec:d-wave}) states as functions of doping rate. 
Data for some system sizes are plotted. 
The thick pink dash-dotted line is a curve fitted using 
Eq.~(\ref{eq:DE-fitted}) with $\Psi_{\rm N}$ for all values of $L$. 
}
\label{fig:DelE-vsn-a03-comp}
\end{figure}
We turn to the doping dependence of $\Delta E/t$. 
Shown in Fig.~\ref{fig:DelE-vsn-a03-comp} is  $\Delta E/t$ for $U/t=12$ 
and $t'/t=-0.3$; these values are marked with vertical gray lines in 
Fig.~\ref{fig:DelE-N-jpsj}. 
As $\delta$ increases, $\Delta E/t$ rapidly decreases as 
\begin{equation}
\Delta E/t \propto \alpha\exp(-\delta/\delta_{\rm N}), 
\label{eq:DE-fitted}
\end{equation}
with $\delta_{\rm N}\sim 0.022$ ($\alpha$ : positive constant) 
in this case, as shown with a thick dash-dotted line 
in Fig.~\ref{fig:DelE-vsn-a03-comp}. 
Thus, BRE substantially vanishes for $\delta\gtrsim 0.1$. 
However, it should be emphasized that $\Delta E$ for $\Psi_{\rm N}$ is 
much larger than that for $\Psi_d$ near half filling, as shown in 
Fig.~\ref{fig:DelE-vsn-a03-comp}, and the BRE on $\Psi_{\rm N}$ are never 
negligible. 
\par

Finally, we analyze $\Delta E$ by dividing it into three components: 
$\Delta E=\Delta E_t+\Delta E_{t'}+\Delta E_U$. 
We find that $\Delta E_t$ is positive and becomes large for a large 
$|t'/t|$, while $\Delta E_U$ is negative and its magnitude is 
relatively small. 
Namely, the effective band is transformed so as to gain kinetic 
energy $E_t$ at the cost of the interaction energy $E_U$. 
This corresponds to a general tendency for a state in a strongly 
correlated regime to undergo a transition to reduce the kinetic 
energy.\cite{YTOT,YOT} 
This feature applies to $\Psi_{\rm AF}$ and $\Psi_{\rm mix}$. 
For $\Delta E_{t'}$, the magnitude is small as compared with those of the 
other two components, except for $t'/t\sim -0.1$. 
\par

\section{BRE on Pure Antiferromagnetic State\label{sec:AF}}
In this section, we consider the features of BRE on the AF state without 
a SC order, 
\begin{equation}
\Psi_{\rm AF}={\cal P}\Phi_{\rm AF}. 
\label{eq:psiAF}
\end{equation}
In Sect.~\ref{sec:AF-para}, we discuss the optimized parameters. 
In Sect.~\ref{sec:AF-optim}, a large improvement in energy due to BRE is 
revealed. 
In Sect.~\ref{sec:Lifshitz}, topics associated with the Lifshitz transition are considered.
Details of optimizing $\Psi_{\rm AF}$ are given in Appendix\ref{sec:AF-A}.

\subsection{Optimized band parameters\label{sec:AF-para}}
%
\begin{figure}
\begin{center}
\vskip -2mm
\hskip -0mm
\includegraphics[width=9.5cm,clip]{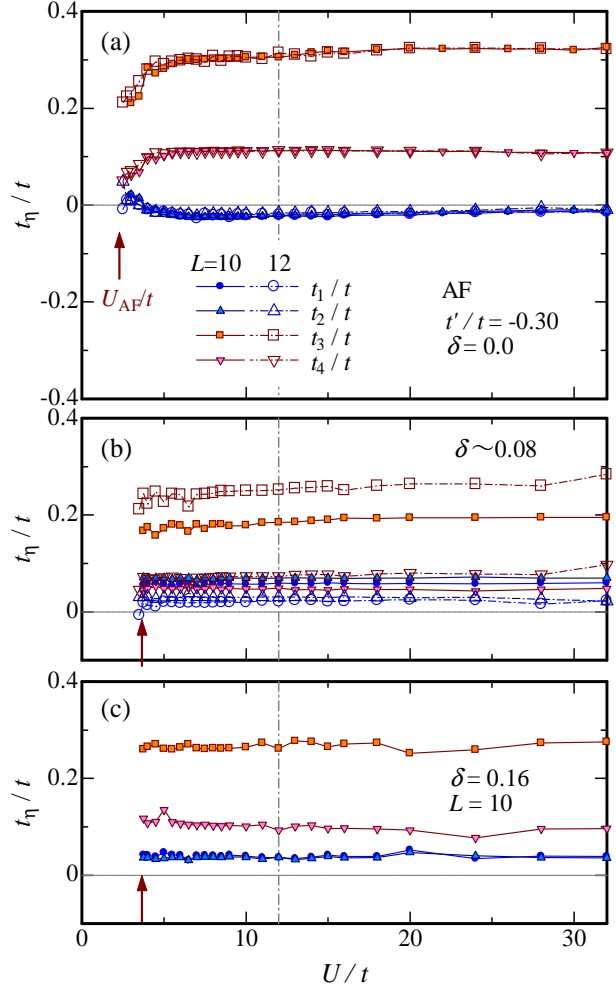} 
\end{center} 
\vskip -0mm
\caption{(Color online)
Optimized band parameters in pure AF state for $t'/t=-0.3$ plotted 
as functions of correlation strength. 
The doping rate is different among the three panels. 
The legends displayed in (a) are common to (b) and (c).
The arrow in each panel indicates the AF transition point. 
}
\label{fig:para-t-all-jpsj}
\end{figure}
We start by clarifying the features of the optimized band parameters in 
$\Psi_{\rm AF}$, for which we always use $t_\eta$ ($\eta=1$--$4$) (BR4). 
In Fig.~\ref{fig:para-t-all-jpsj}, the $U/t$ dependence of the optimized value 
of $t_\eta$ is shown for $t'/t=-0.3$. 
For a small $U/t$ ($<U_{\rm AF}/t\sim 2.75$--$3.5$ for $t'/t=-0.3$), no AF 
order exists and $\Psi_{\rm AF}$ is reduced to $\Psi_{\rm N}$. 
At $U=U_{\rm AF}$ (AF transition point), $\Delta_{\rm AF}$ and the sublattice 
magnetization (AF order parameter)
\begin{equation}
m=\frac{2}{N_{\rm s}}\sum_j
\left| e^{i{\bf Q}\cdot{\bf r}_j}\langle S^{z}_j\rangle \right| 
\label{eq:mag}
\end{equation}
suddenly become finite (not shown), probably as a first-order transition, 
for $t'/t\ne 0$. 
For $U>U_{\rm AF}$, marked BRE appears and $t_\eta$ becomes almost 
constant as a function of $U/t$. 
Although $t_\eta$ is almost invariant as a function of $U/t$, it varies with $\delta$ 
to some extent, at least for $t'/t=-0.3$, as seen in 
Fig.~\ref{fig:para-t-all-jpsj}. 
In fact, this feature depends on $t'/t$, as described in the next paragraph. 
Anyway, we find that $\varepsilon_{\bf k}^{\rm AF}$ is renormalized so as to 
restore the nesting condition, irrespective of $\delta$. 
\par

\begin{figure}
\begin{center}
\vskip -2mm
\hskip -0mm
\includegraphics[width=10.0cm,clip]{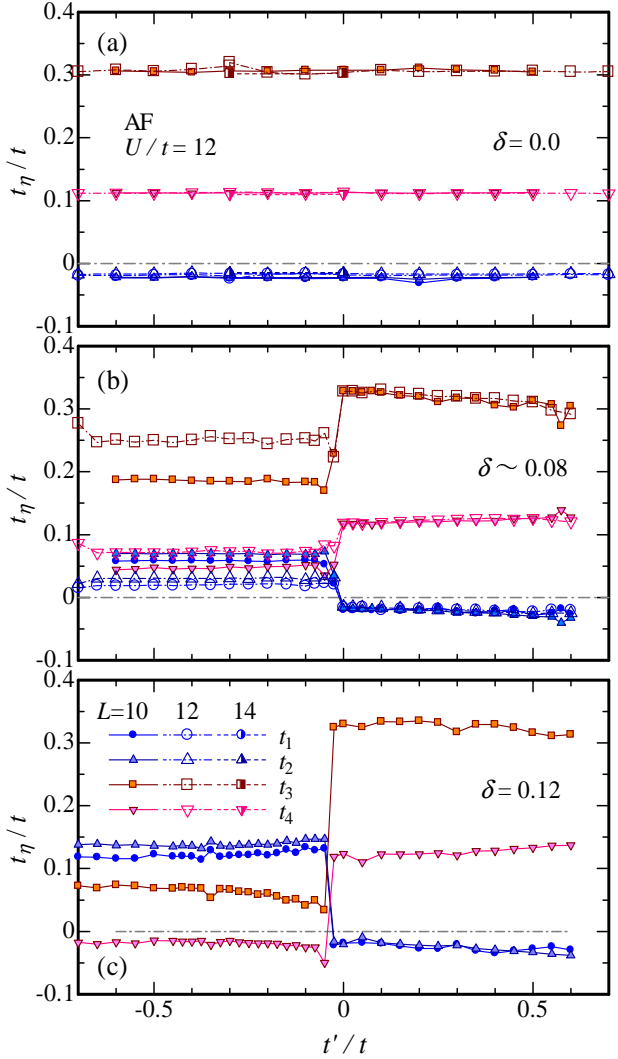} 
\end{center} 
\vskip -0mm
\caption{(Color online)
Optimized band parameters in pure AF state for $U/t=12$ plotted 
as functions of $t'/t$. 
The doping rate is different among the three panels. 
In (a), (b), and (c), data for $L=10$--$14$, $10$ and $12$, and $10$ are plotted respectively. 
The legends displayed in (c) are common to (a) and (b). 
The arrows in (b) and (c) indicate the values of $t'_{\rm L}/t$. 
}
\label{fig:para-t-all-vsa}
\end{figure}
Shown in Fig.~\ref{fig:para-t-all-vsa} is the $t'/t$ dependence of the optimized 
$t_\eta/t$ for $U/t=12$. 
At half filling [(a)], the renormalized values of $t_\eta/t$ and the other 
variational parameters (not shown) are constant with respect to $t'/t$. 
The optimized AF state is independent of $t'/t$; this feature is common 
to all values of $U/t$ ($>U_{\rm AF}/t$). 
In contrast, for $\delta>0$ [(b) and (c)], $t_\eta/t$ discontinuously 
changes at $t'=t'_{\rm L}\sim -0.05t$, and the other parameters (not shown) 
also exhibit singular behaviors (a cusp or discontinuity) there. 
Checking various cases, we find that the value of $t'_{\rm L}/t$ subtly varies as the model parameters ($\delta$, 
$L$) vary but is necessarily situated in the range $-0.1<t'_{\rm L}/t<0$. 
Thus, in doped cases, the AF phase is divided into two subphases 
according to whether $t'>t'_{\rm L}$ [type (i)] or $t'<t'_{\rm L}$ 
[type (ii)]. 
In each subphase, the $t'/t$ dependence of $t'_\eta/t$ is weak. 
However, the $\delta$ dependence of $t_\eta$ is weak 
in the type-(i) AF, whereas $t'_\eta$ changes markedly as $\delta$ increases 
in the type-(ii) AF. 
Thus, the effective band $\varepsilon_{\bf k}^{\rm AF}$ will be distinct between 
the two subphases. 
As we will discuss in Sects.~\ref{sec:Lifshitz} and \ref{sec:mix}, 
this transition is regarded as a Lifshitz transition in the AF phase.  
\par 

Finally, let us compare the optimized $\varepsilon_{\bf k}^{\rm SC}$ and 
$\varepsilon_{\bf k}^{\rm AF}$. 
$t_1/t$, which is the sole effective band parameter in $d$-SC, behaves as a 
linear function of $t'/t$ with a positive coefficient [Eq.~(\ref{eq:t1t}) 
and Table~\ref{table:Coeff}], indicating that BRE are mild, and a trace of 
the bare band remains [see Fig.~\ref{fig:schematic}(a) later].    
On the other hand, for the AF part, BRE are prominent in that $t_\eta/t$ is 
almost independent of $t'/t$, and for $t'<t'_{\rm L}$, the sign of 
$t_1/t$ becomes opposite that of $t'/t$ 
[Figs.~\ref{fig:para-t-all-vsa}(b) and \ref{fig:para-t-all-vsa}(c)].  
Generally, the optimized forms of $\varepsilon_{\bf k}^{\rm SC}$ and 
$\varepsilon_{\bf k}^{\rm AF}$ are distinct, particularly, in the case 
of $t'<t'_{\rm L}$. 
This applies to the mixed state. 
\par

\subsection{Large energy reduction by BRE\label{sec:AF-optim}}
As shown in a previous VMC study without BRE,\cite{Y2013} the energy of the AF 
state is not lowered with respect to the paramagnetic state even 
at half filling for $|t'/t|>0.35$--$0.41$ (depending on $L$) and $U/t=12$ 
(see Fig.~\ref{fig:evsa-comp} later). 
For $\delta>0$ and $t'/t<0$, this boundary value of $|t'/t|$ tends to 
decrease; for example, for $t'/t=-0.3$, the optimized AF gap 
$\Delta_{\rm AF}$ substantially vanishes for $\delta\gtrsim 0.03$. 
However, as discussed in preceding reports,\cite{proceedings1,proceedings2} 
the AF state $\Psi_{\rm AF}$ with BRE [Eq.~(\ref{eq:psiAF})] is stabilized 
with respect to $\Psi_{\rm N}$ up to $\delta\sim 0.16$ ($0.21$) 
for $t'/t=0$ ($\pm 0.3$). 
First, we look at this great improvement by BRE more systematically. 
\par 

\begin{figure}
\begin{center}
\vskip -2mm
\hskip -0mm
\includegraphics[width=9.5cm,clip]{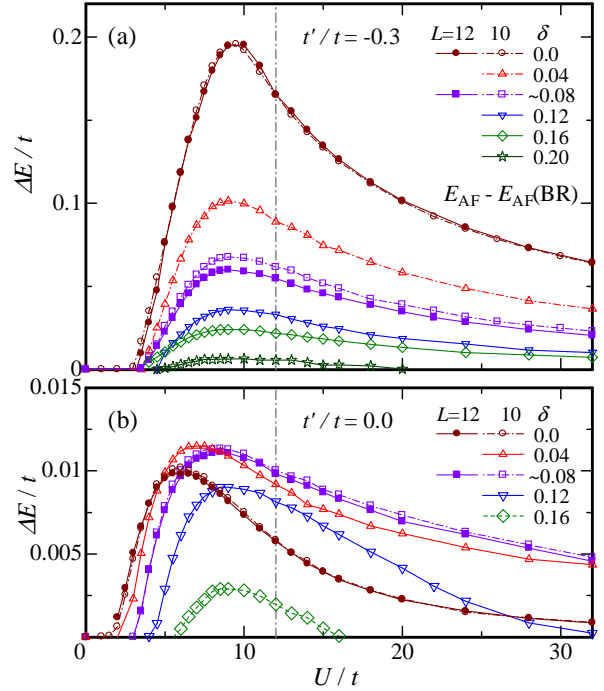} 
\end{center} 
\vskip -50mm
\caption{(Color online)
Energy gain brought about by BRE in $\Psi_{\rm AF}$ plotted as a function of $U/t$ for 
(a) $t'/t=-0.3$ and (b) $t'/t=0$. 
Note that the scale on the vertical axis in (a) is 10 times larger than that in (b).
Data for several doping rates are shown. 
}
\label{fig:DelE-vsU-jpsj}
\end{figure}
In Fig.~\ref{fig:DelE-vsU-jpsj}, we show the $U/t$ dependence of the energy reduction 
by BRE [Eq.~(\ref{eq:DelE})]. 
In contrast to the case of $d$-wave pairing, a very large energy improvement 
is brought about by BRE for $U>U_{\rm AF}$, consistent with the large BR shown 
in Fig.~\ref{fig:para-t-all-jpsj} for $t'/t=-0.3$. 
Energy improvement occurs even for $t'/t=0$ 
[Fig.~\ref{fig:DelE-vsU-jpsj}(b)] because the BRE on $t_3$ and $t_4$ are not 
small, as seen in Fig.~\ref{fig:para-t-all-vsa}, although $\Delta E/t$ is 
an order of magnitude smaller than that for $t'/t=-0.3$. 
The $U/t$ dependence of $\Delta E/t$ for $t'/t=0.3$ (not shown) is 
quantitatively similar (somewhat smaller for a large $\delta$) to the 
case of $t'/t=-0.3$. 
We find from Fig.~\ref{fig:DelE-vsU-jpsj}(a) that $\Delta E/t$ is a 
monotonically decreasing function of $\delta$ for a large $|t'/t|$ 
(actually, when $t'/t\lesssim -0.2$ and $t'/t\gtrsim 0.3$). 
This has been illustrated in Sect.~\ref{sec:normal}, where the $\delta$ dependence of 
$\Delta E/t$ in $\Psi_{\rm AF}$ was shown in 
Fig.~\ref{fig:DelE-vsn-a03-comp} for $U/t=12$ and $t'/t=-0.3$. 
$\Delta E/t$ decreases as $\delta$ increases, but the area of finite 
$\Delta E/t$ is considerably extended up to $\delta\sim 0.22$ for this 
parameter set. 
We repeat that the energy reduction by BRE in $\Psi_{\rm AF}$ is much larger than that in the normal and $d$-wave states. 
Such an improvement occurs in wide ranges of $U/t$ ($>U_{\rm AF}/t$) and 
$t'/t$ ($\lesssim -0.15$). 
\par

\begin{figure}
\begin{center}
\vskip -0mm
\hskip -0mm
\includegraphics[width=10.5cm,clip]{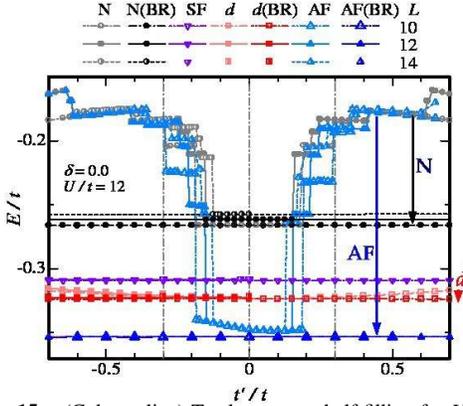} 
\end{center} 
\vskip -42mm
\caption{(Color online)
Total energy at half filling for $U/t=12$ compared among various states 
with or without BR as a function of $t'/t$. 
Dark (pale) symbols indicate cases with (without) BRE. 
Open, filled, and half-filled symbols indicate the data of $L=10$, $12$, and $14$, respectively. 
Arrows denote the energy reductions brought about by BRE for the different states. 
`SF' indicates a staggered flux state. 
}
\label{fig:evsa-comp}
\end{figure}
Now, we compare the total energy among various states in the regime of 
Mott physics ($U>U_{\rm c}$). 
In Fig.~\ref{fig:evsa-comp}, we compare the $t'/t$ dependence of $E/t$ 
at half filling among $\Psi_{\rm N}$, $\Psi_d$, and $\Psi_{\rm AF}$. 
For each, the values with BRE and without BRE are plotted. 
Without BRE, the total energy more or less depends on $t'/t$, whereas 
if BRE are introduced, $\Psi_\Lambda(t'/t)$ ($\Lambda=$N, $d$, AF) 
is optimized at $\Psi_\Lambda(0)$ for any $t'/t$.  
Consequently, $E/t$ becomes independent of $t'/t$ because the diagonal 
hopping energy vanishes: 
\begin{equation}
E_{t'}\equiv\langle{\cal H}_{t'}\rangle=0 \qquad (\mbox{for }\delta=0), 
\end{equation}
and $E_t$ and $E_U$ become constant with respect to $t'/t$. 
As a result, the energy of $\Psi_{\rm AF}$ (and $\Psi_{\rm N}$) is greatly 
reduced for large values of $|t'/t|$. 
The order of the energy becomes 
\begin{equation}
E_{\rm AF}< E_d < E_{\rm SF}< E_{\rm N} 
\label{eq:E-order}
\end{equation}
for a wide range of $|t'/t|$ (at least $<0.7$) at a fixed $U/t$ 
($>U_{\rm c}/t$).
Here, `SF' indicates a staggered flux state, which is a candidate 
pseudogap state in cuprates\cite{theory,SF} and will be taken up 
in Sect.~\ref{sec:SF}.
\par
\begin{figure}
\begin{center}
\vskip 0mm
\hskip -0mm
\includegraphics[width=10.0cm,clip]{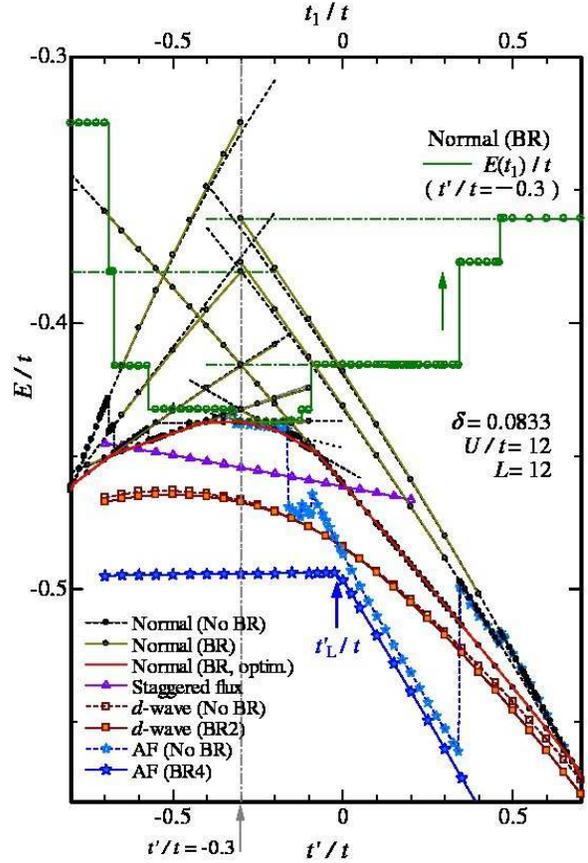} 
\end{center} 
\vskip -68mm
\caption{(Color online) 
Comparison of total energy among normal, AF, $d$-wave, and staggered 
flux\cite{SF} states with some levels of BR as a function of $t'/t$ for 
$L=12$, $\delta=0.0833$, and $U/t=12$. 
The blue arrow indicates the Lifshitz transition point of the AF state. 
In addition, we give an illustration of the procedure for obtaining the 
optimized energy with BRE for the normal state (red line) from raw data 
without BRE (black circles) for $t'/t=-0.3$. 
The green line denotes the variational energy for BRE as a function of 
$t_1/t$ (upper axis) for $t'/t=-0.3$, which corresponds to the dark-green 
line in Fig.~\ref{fig:evsa-a-a03}. 
For details, see Appendix\ref{sec:normal-A}. 
}
\label{fig:evsatilde-a03}
\end{figure}

To consider doped cases ($\delta>0$), $E/t$ for various states are 
compared in Fig.~\ref{fig:evsatilde-a03} for $\delta=0.0833$ ($L=12$). 
Similar results for other values of $\delta$ were presented in Fig.~2 in a preceding 
report.\cite{proceedings3} 
The energy reduction in $\Psi_{\rm AF}$ brought about by BRE for large $|t'/t|$ is still 
sizable, and $E/t$ exhibits different linear behaviors on opposite sides of the Lifshitz transition point $t'_{\rm L}/t$. 
In $\Psi_{\rm N}$ and $\Psi_d$, $E/t$ tends to decrease for $t'/t>0$ as 
$t'/t$ increases, and also decreases for $t'/t\lesssim -0.4$--$-0.5$ as 
$|t'/t|$ increases, mainly owing to the decrease in $E_{t'}$. 
Consequently, the order in Eq. (\ref{eq:E-order}) does not 
change for a wide range of $\delta$ except for is the SF state, where it 
rapidly becomes unstable, especially for $t'/t>0$.\cite{SF} 
Incidentally, by analyzing the charge-density structure factor, we find that 
$\Psi_{\rm AF}$ becomes metallic for $\delta<0$.\cite{proceedings1} 
At any rate, $\Psi_{\rm AF}$ with BRE has much lower energy than $\Psi_d$ 
in the whole range of $t'/t$ in Fig.~\ref{fig:evsatilde-a03}. 
This is not the case for $\Psi_{\rm AF}$ without BRE (see also 
Table \ref{table:Summary-AF}). 
\par 

\begin{figure}
\begin{center}
\vskip -0mm
\hskip -0mm
\includegraphics[width=9.0cm,clip]{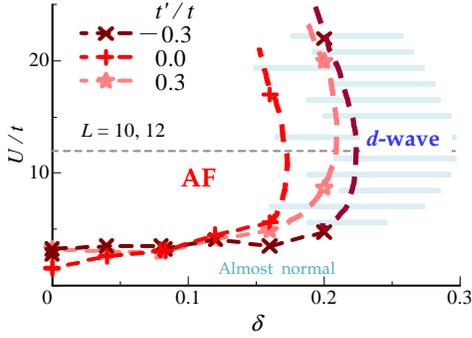} 
\end{center} 
\vskip -0mm
\caption{(Color online) 
Rough phase diagram in $U/t$-$\delta$ space constructed within 
$\Psi_{\rm AF}$ and $\Psi_d$, both with BRE. 
Because the dashed border lines indicate the locus of vanishing AF orders, 
the region of the $d$-wave may somewhat extend to the AF side. 
The region of the $d$-wave is schematic, especially, on the large-$\delta$ 
side. 
}
\label{fig:boundary-jpsj}
\end{figure}
By drawing similar figures for various values of $U/t$, $t'/t$, and $\delta$, 
we construct the phase diagram in the $U/t$-$\delta$ space shown in 
Fig.~\ref{fig:boundary-jpsj}. 
It is notable that, in contrast to previous studies, the AF area for 
$t'/t=-0.3$ becomes wider than those for $t'/t=0$ and $0.3$ and covers 
a very wide range of model parameters $U/t$, $t'/t$, and $\delta$. 
\par

\subsection{Lifshitz transition and electron-hole asymmetry
\label{sec:Lifshitz}}
%
\begin{figure}
\begin{center}
\vskip -0mm
\hskip -0mm
\includegraphics[width=10.5cm,clip]{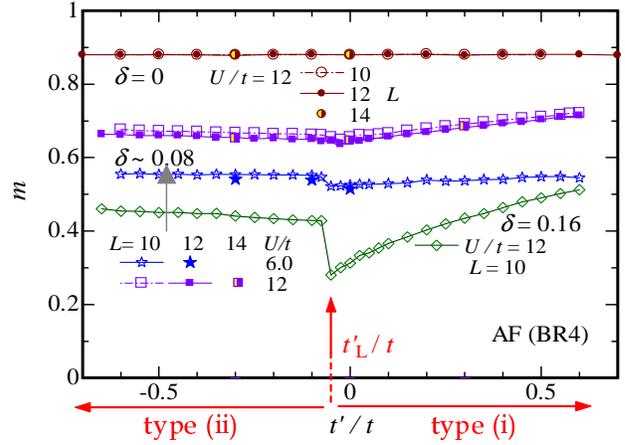} 
\end{center} 
\vskip -19mm
\caption{(Color online) 
Sublattice magnetization [Eq.~(\ref{eq:mag})] in the AF phase plotted 
as a function of $t'/t$ for some values of $\delta$ and $U/t$. 
The red arrow indicates the Lifshitz transition point. 
}
\label{fig:mag}
\end{figure}
Before discussing the Lifshitz transition, we mention the behavior of the 
staggered magnetization [Eq.~(\ref{eq:mag})] in $\Psi_{\rm AF}$. 
We find that $m$ gradually increases as $U/t$ increases for 
$U_{\rm AF}<U\lesssim 12t$ and is almost constant for $U\gtrsim 12t$, 
irrespective of $\delta$ and $t'/t$ (not shown). 
Shown in Fig.~\ref{fig:mag} is the $t'/t$ dependence of $m$ for some values 
of $\delta$ and $U/t$. 
At half filling, $m$ is constant and $\sim 0.88$ ($m$ becomes 1 for the 
N\'eel state) because $\Psi_{\rm AF}$ is invariant for $t'/t$, as mentioned 
in Sect.~\ref{sec:AF-para}. 
For $\delta>0$, an anomaly appears at $t'=t'_{\rm L}$, and the difference 
in the two areas becomes more conspicuous as $\delta$ increases. 
\par

\begin{figure*}[t!]
\begin{center}
\vskip -0mm
\hskip -3mm
\includegraphics[width=18.5cm,clip]{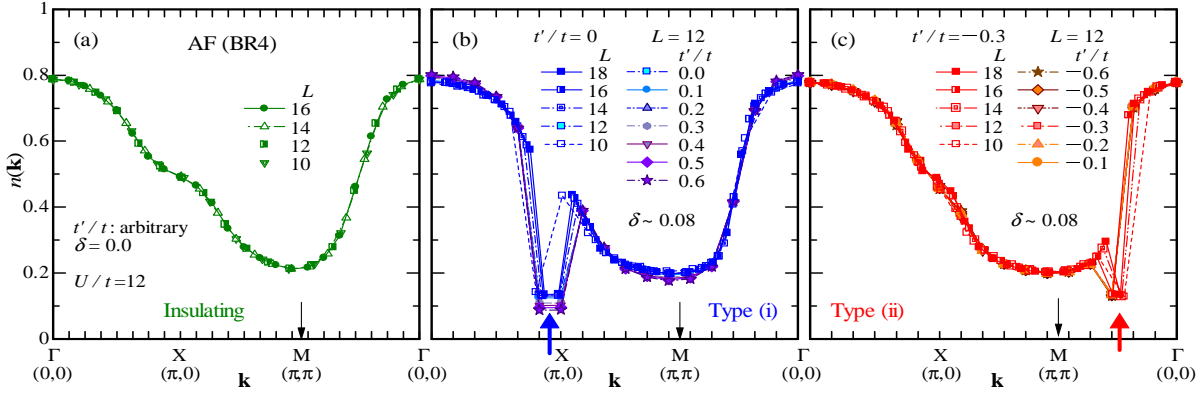} 
\end{center} 
\vskip -17mm
\caption{(Color online)
Momentum distribution function plotted along the path 
$\Gamma\rightarrow{\rm X}\rightarrow{\rm M}\rightarrow\Gamma$ for $U/t=12$ 
in three cases: 
(a) half-filled case, in which $\Psi_{\rm AF}$ becomes independent of 
$t'/t$, 
(b) doped case ($\delta\sim0.08$) with $t'>t'_{\rm L}$ [type (i)], and 
(c) doped case ($\delta\sim0.08$) with $t'<t'_{\rm L}$ [type (ii)]. 
Data for $L=10$-$18$ are plotted together. 
Pocket Fermi surfaces for the doped cases are indicated by thick arrows. 
}
\label{fig:nk-AF-jpsj}
\end{figure*}

\begin{figure*}[t!] 
\begin{center}
\vskip 0mm 
\hskip -3mm
\includegraphics[width=19.5cm,clip]{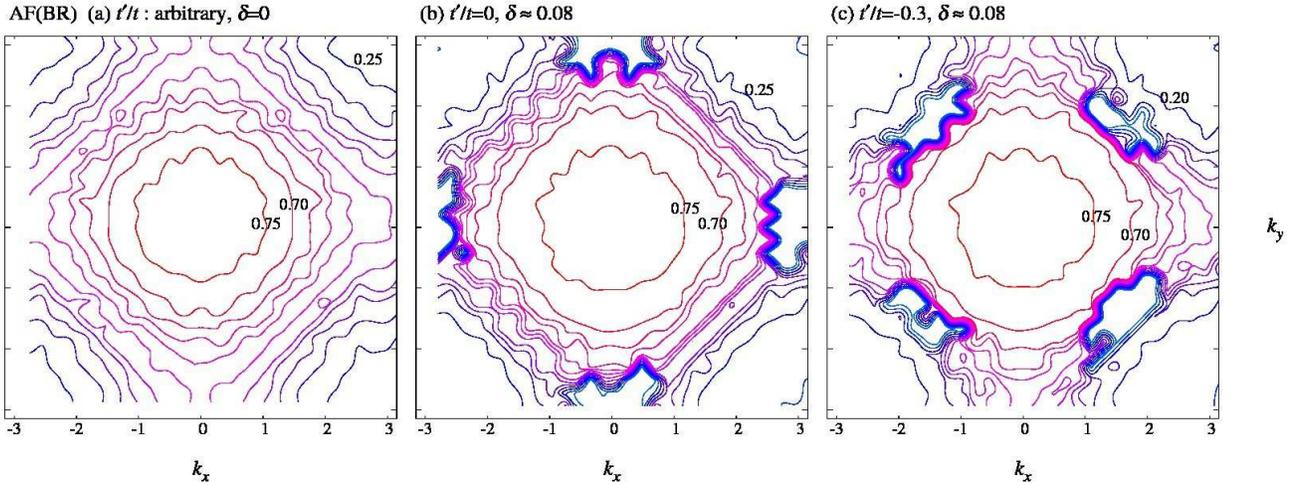} 
\end{center} 
\vskip -12mm
\caption{(Color online) 
Contour maps of momentum distribution function $n({\bf k})$ of the optimized 
pure AF state for $U/t=12$ shown for (a) $\delta=0$ with arbitrary $t'/t$ and 
for $\delta\sim 0.08$ with (b) $t'/t=0$ [type (i)] and (c) $-0.3$ [type (ii)]. 
The parameters in (a), (b), and (c) correspond to those in (a), (b), and (c) 
in Fig.~\ref{fig:nk-AF-jpsj}, respectively. 
The maps are constructed using data for $L=10$--$18$. 
In these contour maps (and similar ones displayed henceforth), the fourfold 
rotational symmetry is somewhat smeared on account of anisotropic boundary 
conditions, open-shell conditions, and functions of the graphic software used. 
}
\label{fig:AFn2k} 
\end{figure*}
To confirm that the transition arising at $t'_{\rm L}/t$ is a kind of 
Lifshitz transition, we plot in Fig.~\ref{fig:nk-AF-jpsj} the momentum 
distribution function [Eq.~(\ref{eq:nk})] in $\Psi_{\rm AF}$ 
for $U/t=12$ along the path in the original Brillouin zone mentioned in the 
caption.   
In panel (a), $n({\bf k})$ at half filling is drawn for $L=10$--$16$, 
which is smooth along the whole path, indicating that the state is insulating. 
The system-size dependence is very small. 
On the other hand, in doped cases with $\delta\sim 0.08$ shown in panels 
(b) [type (i)] and (c) [type (ii)], pocket Fermi surfaces appear and 
the state becomes metallic.\cite{note-N(q)} 
In each panel, we plot data for various values of $t'/t$ ($L=12$) and for 
various system sizes for a typical $t'/t$ ($0$ or $-0.3$) at the same time. 
In the type (i) [(ii)] regime, a pocket Fermi surface arises around 
the antinodal point $(\pi,0)$ [around $(\pi/2,\pi/2)$ in the nodal 
direction]. 
To visualize this feature, we constructed corresponding contour maps of 
$n({\bf k})$ as shown in Fig.~\ref{fig:AFn2k}. 
The location of the pocket Fermi surface suddenly jumps from $\sim(\pi,0)$ 
to $\sim(\pi/2,\pi/2)$ at $t'=t'_{\rm L}$, 
although the behavior of $n({\bf k})$ other than the Fermi surface changes 
only slightly with $t'_{\rm L}/t$. 
Note that the form of the pocket is almost preserved for a fixed $\delta$ 
as $t'/t$ is varied. 
It is notable that the pocket is narrow but very 
deep, suggesting that the advantages of half filling, such as the nesting condition, 
are well preserved by filling this narrow pocket with doped 
carriers and leaving the other parts intact. 
Anyway, this first-order transition occurs with a topological change in the 
Fermi surface. 
\par

\begin{figure}
\begin{center}
\vskip -2mm
\hskip -3mm
\includegraphics[width=11.0cm,clip]{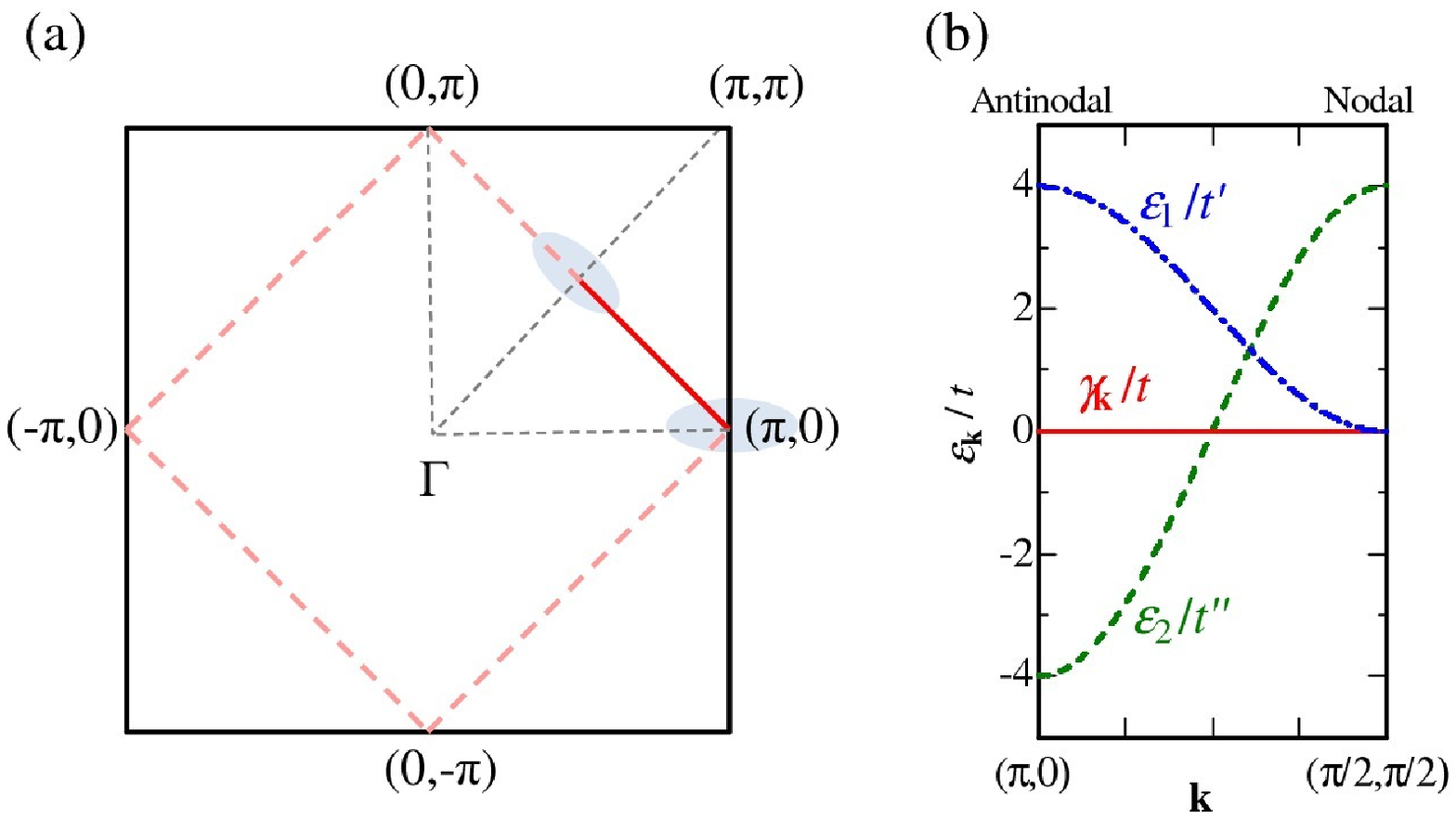} 
\end{center} 
\vskip -22mm
\caption{(Color online) 
(a) Bare Fermi surface at half filling in the tight-binding model with 
$t'/t=0$ shown with a pink dashed line in the first Brillouin zone. 
The nodal $(\pi/2,\pi/2)$ and antinodal $(\pi,0)$ areas are marked 
by shadows. 
(b) Elements of bare band dispersion relations along 
$(\pi,0)\rightarrow(\pi/2,\pi/2)$: 
$\gamma_{\bf k}/t=-2(\cos k_x+\cos k_y)$, 
$\varepsilon_1/t'=-4\cos k_x\cos k_y$, 
and $\varepsilon_2/t''=-2(\cos 2k_x+\cos 2k_y)$. 
$t''$ indicates the hopping integral to the third-neighbor sites 
($\pm 2,0$) and ($0,\pm 2$), which is disregarded in this paper. 
}
\label{Ek-br:mag}
\end{figure}
The source of this topological transition may have already arisen in the bare 
tight-binding dispersion or at the mean-field level. 
In Fig.~\ref{Ek-br:mag}(a), we show the Fermi surface at half filling for 
$t'=0$, namely, the AF Brillouin zone boundary, on which 
$\tilde\varepsilon_{\bf k}=\gamma_{\bf k}=0$ as shown in 
Fig.~\ref{Ek-br:mag}(b) in red. 
If we add an infinitesimal diagonal hopping term 
[$\varepsilon_1({\bf k})$] (blue), the degeneracy on 
$(\pi,0)$--$(\pi/2,\pi/2)$ is lifted and the band maximum appears at 
$(\pi,0)$ or $(\pi/2,\pi/2)$ according to whether $t'/t>0$ or $t'/t<0$. 
As shown in green in Fig.~\ref{Ek-br:mag}(b), the third-neighbor hopping 
term $\varepsilon_2({\bf k})$ has a similar effect, if the sign of $t''$ is 
opposite the sign of $t'$, although we do not treat it here. 
If we consider ordinary AF mean-field theory, the situation is 
similar because the quasi-particle dispersion 
\begin{equation}
E^{\rm AF}_{\bf k}=\frac{U}{2}-\sqrt{\gamma_{\bf k}^2+\Delta_{\rm AF}^2} 
\end{equation}
is degenerate in the region $(\pi,0)$--$(\pi/2,\pi/2)$. 
When we add ${\cal H}_{t'}$ as a perturbation to this framework, the leading 
difference in the dispersion relation is again 
$\langle\Phi_{\rm AF}|{\cal H}_{t'}|\Phi_{\rm AF}\rangle\propto
\varepsilon_1({\bf k})$. 
In these examples, the boundary of the topological change is at $t'/t=0$. 
Nevertheless, it is not trivial whether this topological change is connected 
to the ones in strongly correlated cases (and even with a large $\delta$), 
and, if it is connected, why $t'_{\rm L}/t$ slightly deviates to the negative 
side of $t'/t$. 
\par

A topological change equivalent to the present result was found in 
the spectral function $A({\bf k},\omega)$ for the cases in which a few carriers are doped 
in the $t$-$t'$-$J$ model and its extensions using various 
methods.\cite{Lee-Shih,Tohyama} 
In particular, Refs.~\citen{SCBA} and \citen{Nagaosa} clearly argued, 
by means of a self-consistent Born approximation and a VMC method, 
respectively, that the location of the band maximum is different between 
hole- and electron-doped cases using typical parameters of cuprates. 
Actually, angle-resolved photoemission spectroscopy (ARPES) experiments revealed that the evolution of the Fermi surface with 
doping is different in hole-doped\cite{hole-ARPES} and 
electron-doped\cite{electron-ARPES} cases in lightly doped systems, 
in accordance with the results of the above theoretical study. 
Our result for the Hubbard model directly corresponds to these results for 
slightly doped $t$-$J$-type models. 
\par 

As we will discuss in Sects.~\ref{sec:Co} and \ref{sec:SF}, this topological 
difference in the Fermi surfaces in $\Psi_{\rm AF}$ determines whether or not 
the $d$-wave SC order coexists with AF orders. 
\par

\section{BRE on Mixed State of AF and SC Orders\label{sec:mix}}
In this section, we study a mixed state of AF and $d$-SC orders in a 
strongly correlated regime ($U>U_{\rm c}$): 
\begin{equation} 
\Psi_{\rm mix}={\cal P}\Phi_{\rm mix}, 
\label{eq:psimix}
\end{equation}
where $\varepsilon_{\bf k}^{\rm AF}$ and $\varepsilon_{\bf k}^{\rm SC}$ 
are independently optimized. 
In Sect.~\ref{sec:PS}, we study the stability against phase separation (PS) 
and discuss whether charge fluctuation thereby correlates with the enhancement 
of $d$-SC.  
In Sect.~\ref{sec:Co}, we consider the mechanism for the coexistence or mutual 
exclusivity of AF and $d$-SC orders. 
In Sect.~\ref{sec:SF}, the notion treated in Sect.~\ref{sec:Co} is applied 
to the relationship between the staggered flux and $d$-SC states. 
In Sect.~\ref{sec:arc}, we refer to the relationship between the pocket Fermi 
surfaces in the type-(ii) AF state and the Fermi arcs observed in the pseudogap phase 
of cuprates. 
\par
\subsection{Stability against phase separation\label{sec:PS}}
%
\begin{table}
\caption{
Relative and intrinsic stabilities of pure AF states and mixed states of 
AF and $d$-SC gaps obtained in recent studies using the Hubbard model 
summarized according to the level of BR and to whether $t'/t\sim 0$ or $-0.3$. 
In the $U/t$ column, a typical target value is given. 
The first row denotes the range of finite AF orders. 
The second row indicates whether the system is homogeneous or phase-separated (P.~S.). 
The third row for the mixed states shows whether AF and $d$-SC orders 
are coexisting or mutually exclusive (and the dominant order) in the main 
(or small-$\delta$) area of $m>0$. 
}
\label{table:Summary-AF}
\begin{center}
\begin{tabular}{l|c|c|c|cl}
\hline
Trial states
& $U/t$ & $t'/t\sim 0$ & $t'/t\sim -0.3$ & References  
\\
\hline\hline
AF (no BR)
& $8,12$ & $\delta\lesssim 0.15$  & no AF & \citen{Y2013}
\\                       
& & P.~S. & --- & 
\\                       
\hline                   
AF (BR)                       
& $12$ & $\delta\lesssim 0.16$  & $\delta\lesssim 0.22$ & 
\citen{proceedings1,proceedings2} \&
\\                       
&& P.~S. & homogeneous &   this work 
\\
\hline\hline
Mixed (no BR) 
&    & $\delta\lesssim 0.2$ & --- &  
\\
& $10$ & --- & --- & \citen{GL}
\\
&& coexisting & --- &
\\
\hline
Mixed (BR only 
& & $\delta\lesssim 0.15 $  & $\delta\lesssim 0.15$ & 
\\                      
\qquad\qquad in SC)
& $10$ & P.~S.  & P.~S. & \citen{Koba-old,Koba-ISS14}
\\
&& coexisting  & exclusive, AF & 
\\
\hline                      
Mixed (BR in
& & $\delta\lesssim 0.16$  & $\delta\lesssim 0.25$ & \citen{proceedings3} \&
\\
\qquad~ AF \& SC)
& $12$ & P.~S.      & homogeneous  &  this work
\\
&& coexisting & exclusive, AF &
\\
\hline
Mixed (many
&    & $\delta\lesssim 0.18$  & $\delta\lesssim 0.24$ &
\\
\qquad parameters)
& $10$ & P.~S.  & homogeneous &  \citen{Misawa}
\\
&& coexisting & exclusive, AF &
\\
\hline
\end{tabular}
\end{center}
\vskip -9mm
\label{table:E-d-wave}
\end{table}
Before discussing $\Psi_{\rm mix}$, we refer to known aspects as to 
intrinsic stability of $\Psi_{\rm N}$, $\Psi_d$, and 
$\Psi_{\rm AF}$ against PS. 
Except for the limit of $\delta\rightarrow0$, at which the anomaly of the Mott 
transition appears, the normal state $\Psi_{\rm N}$ is stable against 
PS.\cite{proceedings1} 
As for $\Psi_d$, $E/t$ is a linear function of $\delta$ 
($\chi_{\rm c}\rightarrow\infty$) for a small $\delta$, as we will discuss 
later, indicating that the stability against PS is marginal. 
However, this is distinct from the apparent instability of $\Psi_{\rm AF}$ 
toward PS. 
In the second row 
of Table \ref{table:Summary-AF}, we summarize the conclusions of related VMC studies on 
the stability against PS of the AF and mixed states. 
The pure (not mixed) AF state is known to be unstable toward PS for 
$t'/t\sim 0$\cite{Y2013,proceedings1} but stable for 
$t'/t\sim\pm 0.3$.\cite{proceedings1}
A mixed state in which BRE are introduced into $\varepsilon_{\bf k}^{\rm SC}$ 
but the AF part is fixed as 
$\varepsilon_{\bf k}^{\rm AF}=\gamma_{\bf k}$\cite{Koba-ISS14} 
exhibits instability toward PS for both $t'/t=0$ and $-0.3$. 
To summarize, states with AF orders exhibit a tendency toward PS according 
to the value of $t'/t$. 
\par

\begin{figure}
\begin{center}
\vskip 4mm
\hskip -3mm
\includegraphics[width=9.0cm,clip]{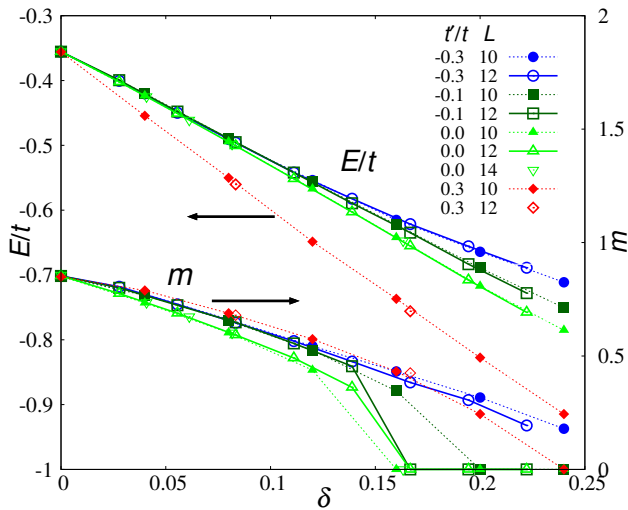} 
\end{center} 
\vskip -5mm
\caption{(Color online)
Total energy per site and staggered magnetization (right axis) obtained 
for $\Psi_{\rm mix}$ with $U/t=12$ plotted as a function of doping rate. 
Data for four values of $t'/t$ and $L=10$-$14$ are shown. 
}
\label{fig:Em}
\end{figure}
We study this property for $\Psi_{\rm mix}$ [Eq.~(\ref{eq:psimix})]. 
In Fig.~\ref{fig:Em}, the total energy and sublattice magnetization 
[Eq.~(\ref{eq:mag})] in $\Psi_{\rm mix}$ are shown as a function of the doping 
rate. 
First, we discuss the range in which the finite AF order occurs. 
As compared with the pure AF state $\Psi_{\rm AF}$,\cite{proceedings1} 
the value of $\delta$ at which $m$ vanishes ($\delta_{\rm AF}$) is almost 
unchanging for $t'/t=0$: $\delta_{\rm AF}\sim 0.16$, whereas 
$\delta_{\rm AF}$ somewhat increases for a large $|t'/t|$. 
This small change in $\delta_{\rm AF}$ stems from the small energy 
difference between $\Psi_{\rm mix}$ and $\Psi_{\rm AF}$ (or $\Psi_d$), 
as shown in Table \ref{table:E-comp}. 
\par 

\begin{table}
\caption{
Second-order coefficient $c_2$ estimated by the least-squares method 
for $E(\delta)/t$ [Eq.~(\ref{eq:Eap})] in the AF phase ($U/t=12$) of 
$\Psi_{\rm mix}$. 
For positive (negative) $c_2$, $\Psi_{\rm mix}$ is stable against 
(unstable toward) phase separation. 
Digits in round brackets indicate the error in the last digit.
} 
\label{table:c2}
\begin{center}
\begin{tabular}{r|r|r}
\hline
\multicolumn{1}{c}{$t'/t$} \vline & 
\multicolumn{1}{c}{$c_{2}$ ($L=10$)} \vline & 
\multicolumn{1}{c}{$c_{2}$ ($L=12$)} \\
\hline
$-0.4$  & $ 2.42(8)~~ $ & \multicolumn{1}{c}{---} \\
$-0.3$  & $ 1.85(9)~~ $ & $ 1.95(8)~~ $           \\
$-0.1$  & $ 0.509(6)$   & $ 0.323(8)$             \\
$ 0.0$  & $-0.551(5)$   & $-0.553(7)$             \\
$ 0.3$  & $ 0.830(4)$   & \multicolumn{1}{c}{---} \\
\hline
\end{tabular}
\end{center}
\vskip -5mm
\end{table}

We turn to the stability against PS. 
This property is often judged by the sign of the charge compressibility $\kappa$ 
$[=(1-\delta)^2\chi_{\rm c}]$ or charge susceptibility $\chi_{\rm c}$ 
[$=(\partial^2E/\partial\delta^2)^{-1}$].
For $\chi_{\rm c}>0$ ($\chi_{\rm c}<0$), the state is stable against 
(unstable toward) PS. 
Thus, we need to consider the $\delta$ dependence of $E/t$ (Fig.~\ref{fig:Em}). 
Similarly to for $\Psi_{\rm AF}$,\cite{proceedings1} we find for $\Psi_{\rm mix}$ 
that $E(\delta)/t$ is fitted well by the parabolic form 
\begin{eqnarray}
E(\delta)/t\simeq c_{0}+c_{1}\delta+c_{2}\delta^{2} 
\label{eq:Eap}
\end{eqnarray}
in the whole AF range ($\delta<\delta_{\rm AF}$); we have a unique value 
$\chi_{\rm c}=c_2^{-1}$ in the AF phase. 
The values of $c_2$ thus estimated for some values of $t'/t$ and $L$ are 
summarized in Table \ref{table:c2}. 
It reveals that $c_2$ (namely $\chi_{\rm c}$) becomes negative only for a 
narrow range near $t'/t=0$, minutely $t'_{\rm L}<t'\lesssim 0.2t$ 
(see Fig.~\ref{fig:pd2} later). 
This aspect is basically the same as that of the pure AF 
state.\cite{proceedings1} 
Thus, the instability toward charge inhomogeneity originates in the AF order 
and is not directly connected with SC, as we will discuss shortly. 
\par

\begin{figure}
\begin{center}
\vskip 4mm
\hskip -3mm
\includegraphics[width=8.5cm,clip]{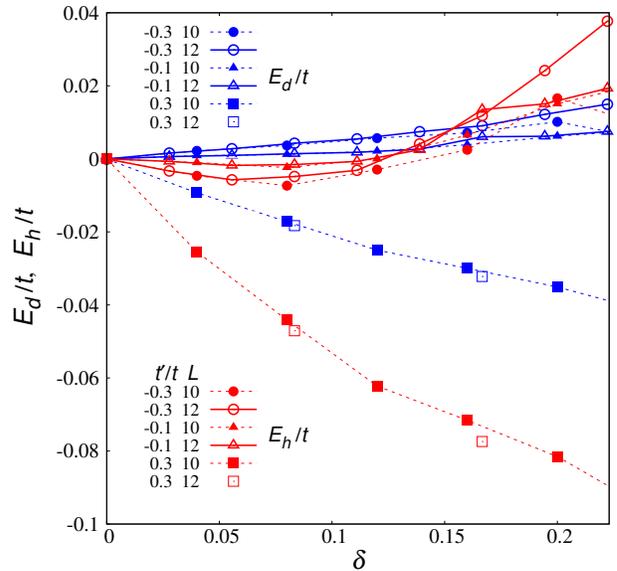} 
\end{center}  
\vskip -5mm
\caption{(Color online)
Two elements [$E_{\rm d}$ (blue) and $E_{\rm h}$ (red)] of diagonal-hopping energy $E_{t'}$ ($=E_{\rm d}+E_{\rm h}$) plotted as functions 
of doping rate for three values of $t'/t$. 
Data for $L=10$ and $12$ with $U/t=12$ are shown.
}
\label{fig:tpic2}
\vskip -5mm
\end{figure}
Now, we identify the origin of the stability against PS for large values 
of $|t'/t|$. 
First, we analyze $E/t$ by dividing it into its components $E_U/t$, $E_t/t$, and $E_{t'}/t$. 
Because $E_t/t$ ($E_U/t$) is almost linear (somewhat convex) as a function 
of $\delta$ for any value of $t'/t$ and $U>U_{\rm c}$ (not shown), these 
components do not contribute to phase stability. 
On the other hand, $E_{t'}/t$ is concave for any $t'/t$, but, of course, 
the degree of concavity diminishes as $|t'/t|$ decreases and vanishes at 
$t'/t=0$.  
We further analyze $E_{t'}$ by dividing it into the two components $E_{\rm d}$ and 
$E_{\rm h}$ ($E_{t'}=E_{\rm d}+E_{\rm h}$); $E_{\rm d}$ ($E_{\rm h}$) is 
the contribution of diagonal hopping that changes (does not change) 
the number of doublons.\cite{Y2013} 
In other words, $E_{\rm d}$ is generated by the creation or annihilation of 
D-H pairs, while $E_{\rm h}$ is generated by the hopping of doped (isolated) holes. 
In Fig.~\ref{fig:tpic2}, we show the $\delta$ dependences of $E_{\rm d}$ and 
$E_{\rm h}$ for three values of $t'/t$. 
We find that both $E_{\rm d}$ and $E_{\rm h}$ are concave but the curvature 
is much sharper for $E_{\rm h}$. 
To summarize, diagonal hopping ($t'$ term), especially that of doped holes, brings 
about intrinsic stability against PS. 
\par

\begin{figure}
\begin{center}
\vskip 4mm
\hskip 3mm
\includegraphics[width=9.0cm,clip]{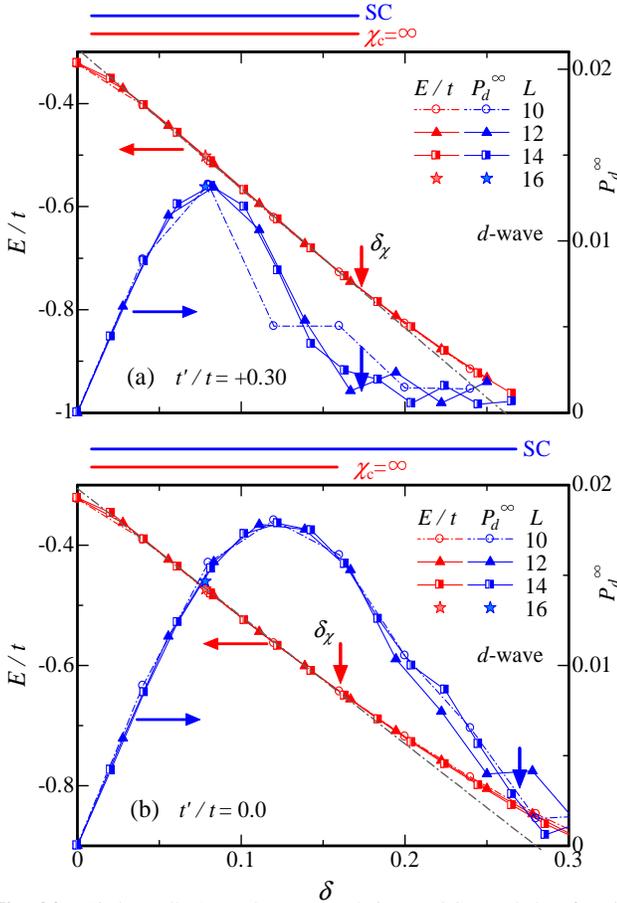} 
\end{center}  
\vskip -60mm
\caption{(Color online)
Total energy and $d$-wave SC correlation function (right axis) obtained 
for $\Psi_d$ with $U/t=12$ plotted for (a) $t'/t=0.3$ and (b) $t'/t=0$ 
as functions of doping rate. 
Data for four values of $L$ are shown. 
Above each panel, the ranges of appreciable SC and where $\chi_c=\infty$ are shown.
The straight dash-dotted line for $E/t$ is a guide for determining $\delta_\chi$. 
}
\label{fig:Etot+pd-vsn-jpsj}
\end{figure}

A recent VMC study\cite{Misawa} argued that the increase in $\chi_{\rm c}$ 
has a one-to-one correspondence with the enhancement of SC order in the 
wave function used. 
We check this point for the present $\Psi_d$ and $\Psi_{\rm mix}$. 
First, we discuss the pure SC state, $\Psi_d$, whose $\delta$ dependence of 
$E/t$ for $t'/t=0.3$ and $0$ is shown in Fig.~\ref{fig:Etot+pd-vsn-jpsj}. 
Aside from a Mott anomaly for $\delta\rightarrow 0$, $E/t$ becomes almost 
linear ($\chi_{\rm c}$ tends to diverge) for $\delta\lesssim\delta_{\rm\chi}$ 
(spinodal point), while $E/t$ becomes concave 
($\chi_{\rm c}$ remains moderate) for $\delta>\delta_{\rm\chi}$. 
Note that $\chi_{\rm c}$ does not become negative unlike the case of 
$\Psi_{\rm AF}$.
Such behavior of $E/t$ is preserved if $t'/t$ is varied, but the range of 
$\chi_{\rm c}\rightarrow\infty$ shrinks as $t'/t$ decreases; 
$\delta_{\rm\chi}\sim 0.17, 0.15$, and $0.12$ for $t'/t=0.3, 0$, and $-0.3$, 
respectively. 
On the other hand, the SC correlation function exhibits the opposite behavior. 
As shown in Fig.~\ref{fig:Etot+pd-vsn-jpsj}, $P_d^\infty$ exhibits a 
well-known dome shape and the SC order is perceptible for 
$0<\delta<\delta_{\rm SC}$. 
Because the statistical fluctuation of $P_d^\infty$ becomes large for 
$\delta\sim\delta_{\rm SC}$, we estimate $\delta_{\rm SC}$ very roughly 
by the condition that the optimized gap parameter $\Delta_d/t$ 
becomes $0.03$ ($\Delta_d/t<0.03$ for $\delta>\delta_{\rm SC}$). 
We confirmed a known tendency that $\delta_{\rm SC}$ increases as $t'/t$ 
decreases [for instance, see Fig.~25(d) in Ref.~\citen{Y2013}]; $\delta_{\rm SC}\sim 0.20, 0.27$, and $0.31$ for $t'/t=0.3, 0$, and 
$-0.3$, respectively. 
Thus, the behaviors of $\delta_{\rm\chi}$ and $\delta_{\rm SC}$ as functions of $t'/t$ 
are opposite; the increase in $\chi_{\rm c}$ rather has a negative 
correlation with the magnitude of SC in $\Psi_d$. 
\par

\begin{figure}
\begin{center}
\vskip 4mm
\hskip -7mm
\includegraphics[width=9.0cm,clip]{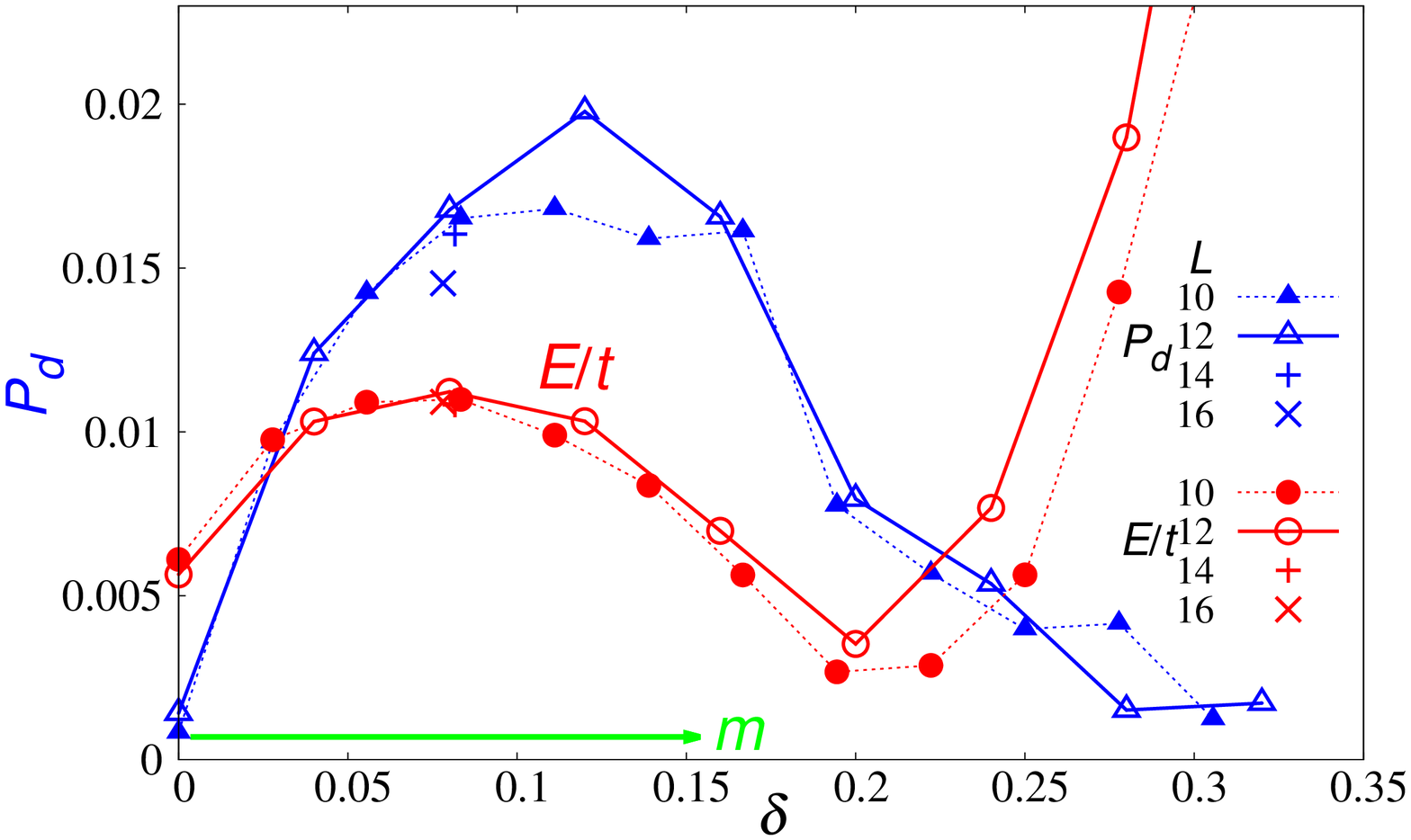} 
\end{center}  
\vskip -5mm
\caption{(Color online)
Total energy (arbitrary unit) and $d$-wave SC correlation function obtained 
for $\Psi_{\rm mix}$ with $U/t=12$ and $t'/t=0$ plotted as a function 
of doping rate. 
Instead of $E/t$, we plot $(E-\tilde{c_0}-\tilde{c_1}\delta)/t$
with $\tilde{c_0}$ and $\tilde{c_1}$ appropriately adjusted to emphasize 
the curvature of $E/t$. 
The area of finite staggered magnetization is shown by a green arrow. 
}
\label{fig:kai}
\end{figure}
Next, we consider the case of $\Psi_{\rm mix}$. 
As mentioned above, the range of $\chi_{\rm c}<0$ is included in the regime 
of type-(i) AF, $-0.05\lesssim t'/t\lesssim 0.2$. 
As an example, we show in Fig.~\ref{fig:kai} the $\delta$ dependence of $E/t$ 
for $t'/t=0$. 
We repeat that the area where $E/t$ is convex precisely coincides with that 
of finite $m$ ($\delta<\delta_{\rm AF}$) indicated by a green arrow. 
For $\delta>\delta_{\rm AF}$, where the state is SC, $E/t$ is concave. 
Furthermore, $P_d$ is smooth at $\delta=\delta_{\rm AF}$ and not specially 
enhanced in the area of $\chi_{\rm c}<0$. 
Anyway, as $\delta$ increases, after the AF order (or instability toward PS) vanishes 
at $\delta_{\rm AF}\sim 0.16$, SC survives up to $\delta_{\rm SC}\sim 0.27$ 
for $t'/t=0$. 
In contrast, for $t'/t=0.1$ [as in Fig.~\ref{fig:pdm2}(a)], the SC first becomes 
weak at $\delta\sim 0.12$, but the area of $\chi_{\rm c}<0$ (and AF order) 
continues up to $\delta\sim 0.18$. 
The extents of $\delta$ where $\chi_{\rm c}<0$ and $P_d>0$ are reversed 
as $t'/t$ varies. 
\par

Through the above analyses, we can conclude that the instability toward PS 
does not directly correlate with $d$-SC, although the ranges of $t'/t$ where 
SC and PS arise are similar as seen in Fig.~\ref{fig:pd2}. 
As discussed in Refs.~\citen{ZGRS} and \citen{Y2013}, we consider that 
the AF spin correlation and the suppression of charge fluctuation owing to 
the Mott physics are responsible for the behavior of the $d$-wave SC. 
We will return to this topic in Sect.~\ref{sec:Co}. 
\par

Finally, we emphasize the importance of BRE again. 
As seen in Table \ref{table:E-d-wave}, a mixed state in which BRE are 
considered only in $\varepsilon_{\bf k}^{\rm SC}$ exhibits instability 
toward PS even for $t'/t=-0.3$.\cite{Koba-ISS14} 
In this mixed state, $\varepsilon_{\bf k}^{\rm AF}$ is fixed 
at $\gamma_{\bf k}$ [Eq.~(\ref{eq:gamma})], which resembles the optimized 
$\varepsilon_{\bf k}^{\rm AF}$ for $t'/t=0$ 
($t_1$, $t_2\sim 0$, see Fig.~\ref{fig:para-t-all-vsa} for instance) 
belonging to the PS area. 
This means that the BRE on $\varepsilon_{\bf k}^{\rm AF}$ (independent of 
the BRE on $\varepsilon_{\bf k}^{\rm SC}$) are crucial for this property. 
\par

\subsection{Coexistence or mutual exclusivity of AF and $d$-SC orders\label{sec:Co}}
Previous studies using various mixed states with 
BRE\cite{Shih-mix,Koba-mix,Koba-ISS14,Misawa} and a recent study using 
density matrix embedding theory (DMET)\cite{DMET} argued that the orders of AF and $d$-SC are coexisting 
or mutually exclusive according to whether $t'/t\sim 0$ or $t'/t\lesssim -0.1$. 
Here, we systematically study this point for $\Psi_{\rm mix}$ and deduce 
the origin of the coexistence of the two orders, which is closely related to 
the mechanism of $d$-SC. 
\par

\begin{figure}[htb] 
\begin{center} 
\vskip 4mm
\hskip -3mm
\includegraphics[width=8.5cm,clip]{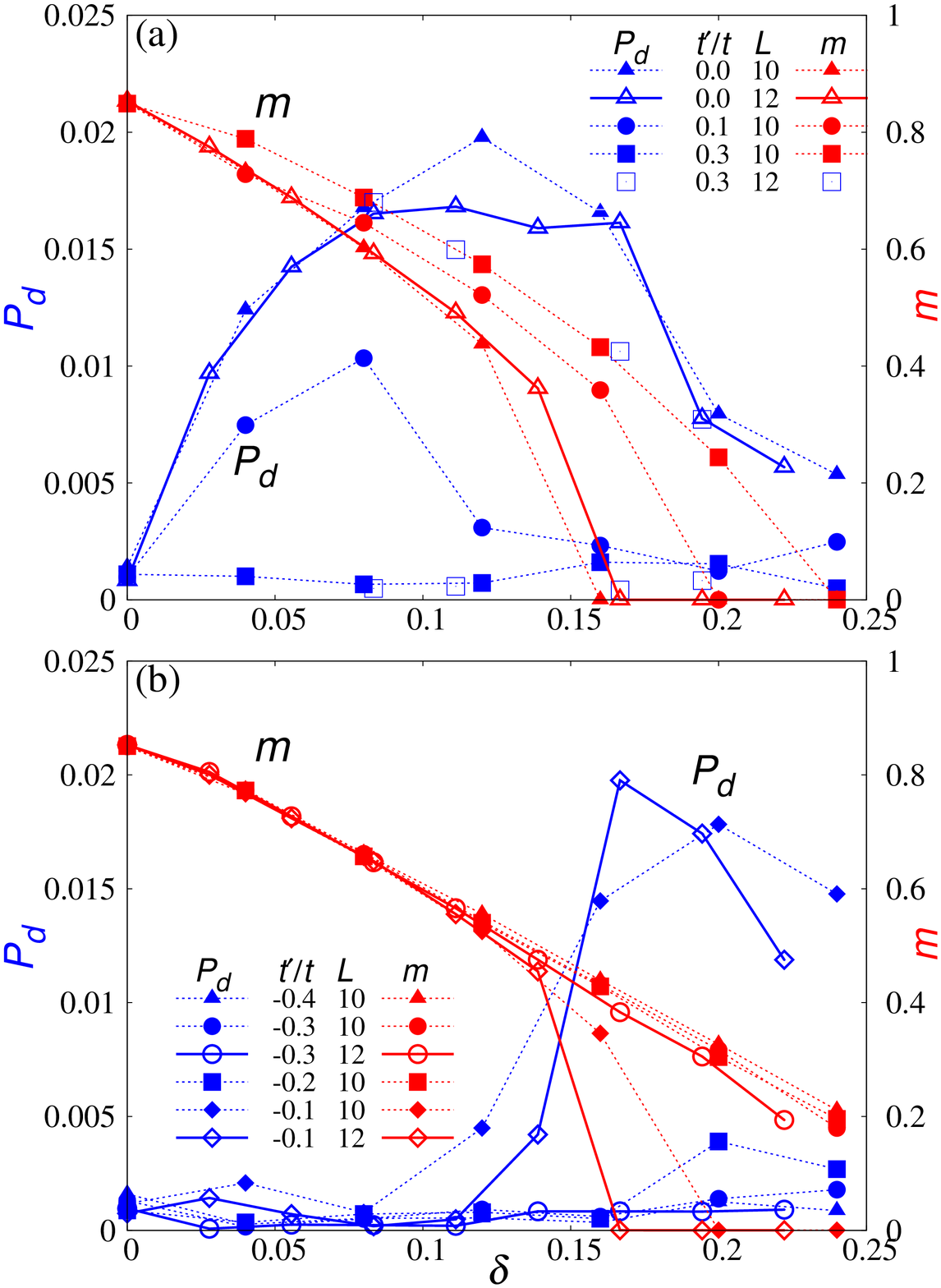} 
\end{center} 
\vskip -5mm
\caption{(Color online) 
Doping-rate dependence of $d$-wave SC correlation function $P_d$ (left axis) 
and staggered magnetization $m$ (right axis) shown for $U/t=12$. 
(a) Cases of $t'>t'_{\rm L}$ [type-(i) regime] and 
(b) those of $t'<t'_{\rm L}$ [type-(ii) regime] for both 
$L=10$ and $12$. 
The Lifshitz transition point of the pure AF state is 
$t'_{\rm L}/t\sim -0.05$. 
}
\label{fig:pdm2}
\end{figure}

In Fig.~\ref{fig:pdm2}, we show the $\delta$ dependence of the $d$-SC 
correlation function and staggered magnetization [Eq.~(\ref{eq:mag})]
measured in $\Psi_{\rm mix}$. 
For $\Psi_{\rm mix}$, we represent the $d$-SC correlation function by 
$P_d\equiv P_d({\bf R})$ [Eq.~(\ref{eq:pd})] with ${\bf R}$ being the vector 
connecting the distant points in the system used [For instance, 
${\bf R}=(5,5)$ for a system of $L=10$], because we focus on a strongly 
correlated regime (See Appendix C in Ref.~\citen{Y2013}). 
We show the results separately for the type-(i) AF and type-(ii) 
AF regimes in panels (a) and (b), respectively, because the features are 
distinct in the two regimes. 
In the type-(i) regime [panel (a)], the SC order ($P_d$) arises or vanishes 
regardless of whether the AF order ($m$) is present or absent. 
For example, for $t'/t\sim 0$, AF and SC long-range orders coexist for 
$\delta<\delta_{\rm AF}$ and SC remains for 
$\delta_{\rm AF}<\delta<\delta_{\rm SC}$ as a pure SC order. 
On the other hand, in the type-(ii) regime [panel (b)], $P_d$ is almost zero 
for $\delta<\delta_{\rm AF}$ and grows after the AF order vanishes 
($\delta>\delta_{\rm AF}$). 
Thus, the two orders are mutually exclusive. 
More accurately, in panel (b), a narrow range of coexistence exists 
near the boundary $\delta=\delta_{\rm AF}$ for small $|t'/t|$, typically 
for $t'/t=-0.1$. 
Anyway, the boundary between coexistence and mutual exclusivity is situated 
at $t'=t_{\rm L}\sim -0.05t$, which is consistent with the previous 
results.\cite{Shih-mix,Koba-mix,Koba-ISS14,Misawa,DMET}  
In the present results, it seems that the AF state is always more robust than 
the $d$-SC state and that the features of the underlying AF state control whether 
$d$-SC appears or not. 
We will return to these points shortly. 
\par

\begin{figure}[htb] 
\begin{center}
\vskip 4mm
\hskip -3mm
\includegraphics[width=7.50cm,clip]{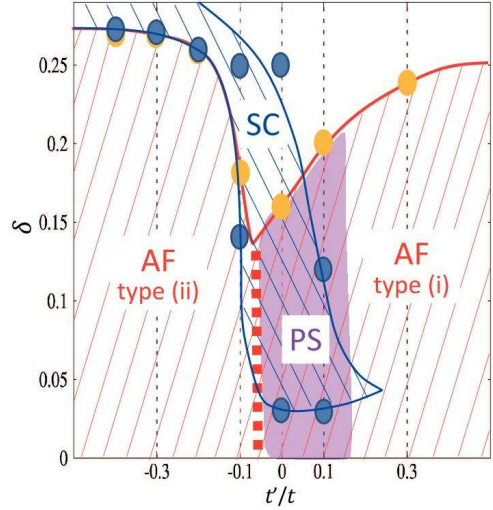} 
\end{center} 
\vskip -4mm
\caption{(Color online) 
Phase diagram in $\delta$-$t'$ space for $U/t=12$ constructed for mixed 
state $\Psi_{\rm mix}$. 
The purple shaded area indicates the regime unstable toward phase separation, 
which is limited to within the type-(i) AF phase. 
The bold red dotted line indicates the Lifshitz transition boundary 
$t'_{\rm L}/t$.  
}
\label{fig:pd2} 
\end{figure}
%
On the basis of the results for $\Psi_{\rm mix}$ above, we constructed the 
phase diagram in the $\delta$-$t'$ space shown in Fig.~\ref{fig:pd2}. 
In accordance with Fig.~\ref{fig:boundary-jpsj} for the pure states, the AF 
state occupies a wide area. 
Except for the range of $-0.1\lesssim t'/t\lesssim 0.2$, SC does not appear 
for low doping rates ($\delta\lesssim 0.2$). 
Furthermore, as mentioned, $\chi_{\rm c}$ becomes negative for 
$t'_{\rm L}<t'\lesssim 0.2t$. 
The state phase separates into an AF state at half filling and a state in 
the overdoped regime ($\delta\gtrsim 0.15$). 
Therefore, homogeneous SC does not appear in the underdoped regime for any 
value of $t'/t$. 
This result greatly modifies the results of previous VMC studies without 
BRE, in which $d$-SC widely prevails for $t'/t<0$, but is consistent with 
recent results of studies aapplying many-parameter VMC methods to Hubbard-type 
models\cite{Misawa} and a $d$-$p$ model\cite{Tamura} and a study employing DMET.\cite{DMET}  
Such predominance of the long-range AF phase is inconsistent with the results of experiments 
on hole-doped cuprates as well as recent advanced studies on 
electron-doped cuprates.\cite{Naito,Adachi,Horio} 
We will discuss this point in Sect.~\ref{sec:summary}. 
\par

\begin{figure*}[t!] 
\begin{center}
\vskip 0mm 
\hskip -0mm
\includegraphics[width=19.5cm,clip]{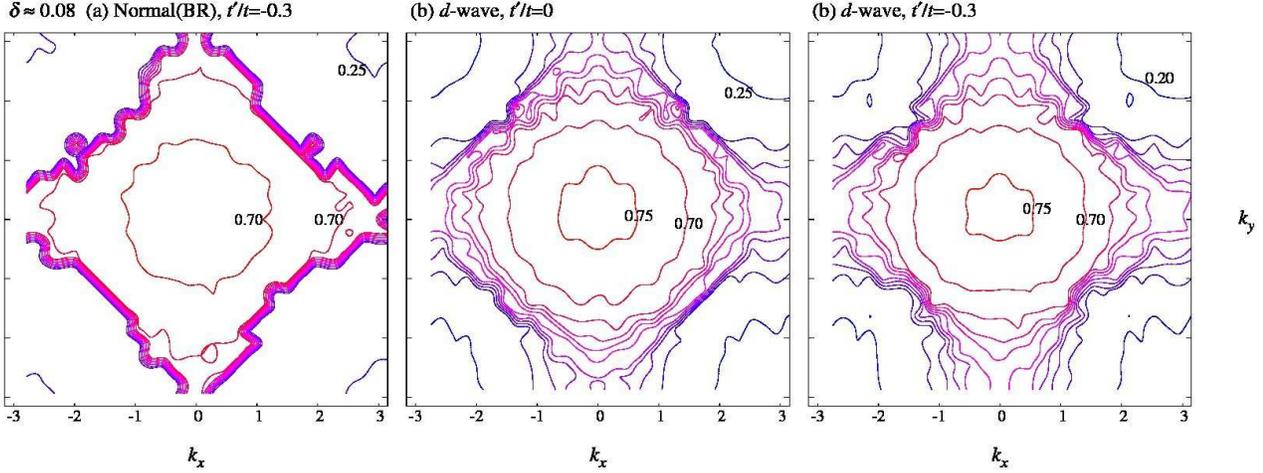} 
\end{center} 
\vskip -12mm
\caption{(Color online) 
Contour maps of $n({\bf k})$ at $U/t=12$ and $\delta\sim 0.08$ shown for 
(a)optimized normal (paramagnetic) state $\Psi_{\rm N}$ with $t'/t=-0.3$ 
($L=10$--$18$) and for 
(b) and (c) optimized pure $d$-wave pairing state $\Psi_d$ with 
$t'/t=0$ (b) and $-0.3$ (c) ($L=10$--$16$). 
}
\label{fig:dn2k}
\end{figure*}
\begin{figure}[htb] 
\begin{center}
\vskip -0mm
\hskip -3mm
\includegraphics[width=7.5cm,clip]{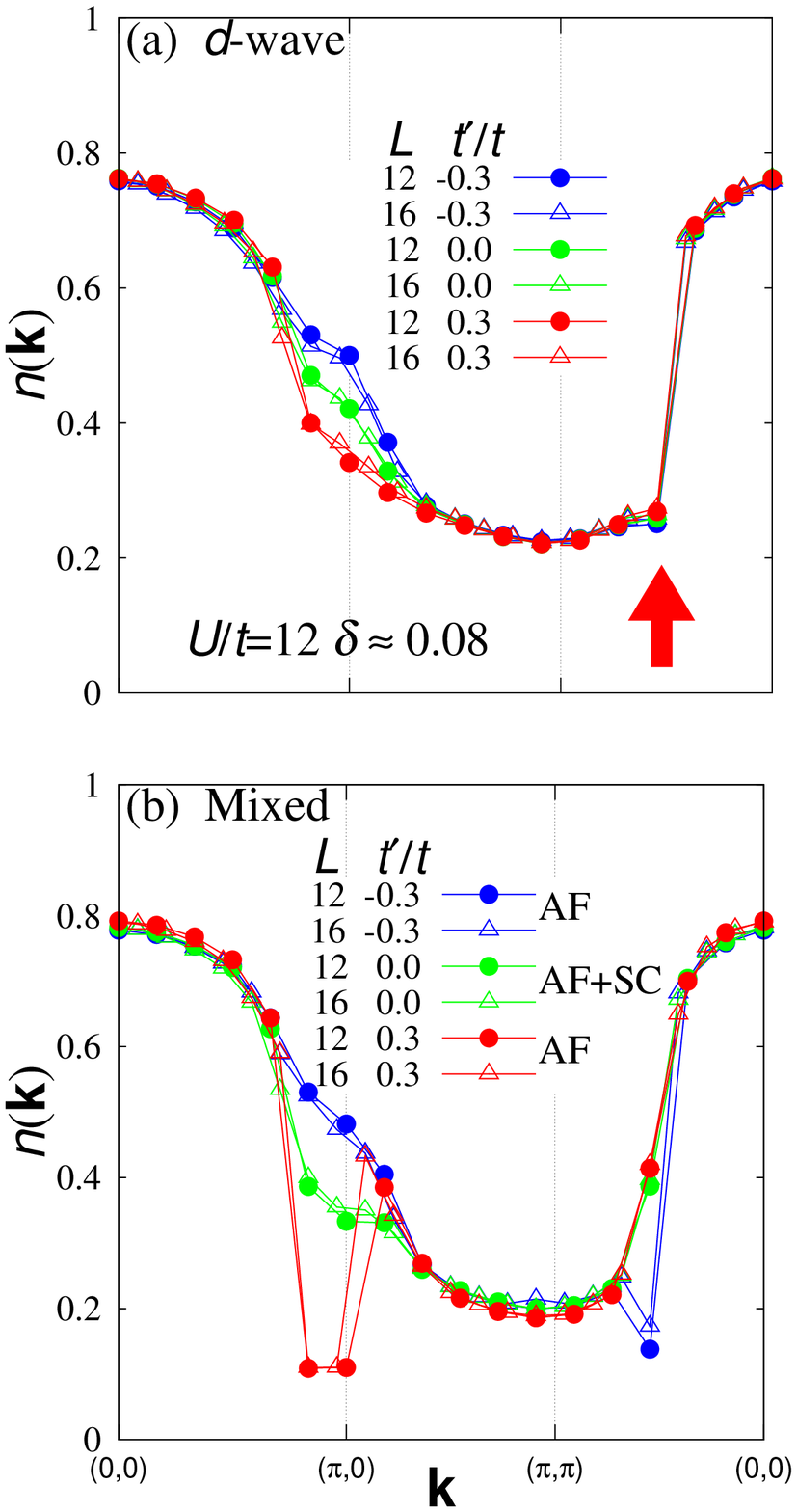} 
\end{center} 
\vskip -8mm
\caption{(Color online) 
Momentum distribution functions for (a) $d$-SC state and 
(b) mixed state for $U/t=12$ and $\delta\sim 0.08$ compared among 
three values of $t'/t$. 
In (a), a nodal Fermi surface is indicated by an arrow. 
Data for $L=12$ and $16$ are plotted. 
}
\label{fig:nk-mix}
\end{figure}
\begin{figure*}[t!] 
\begin{center}
\vskip 0mm 
\hskip -3mm
\includegraphics[width=19.5cm,clip]{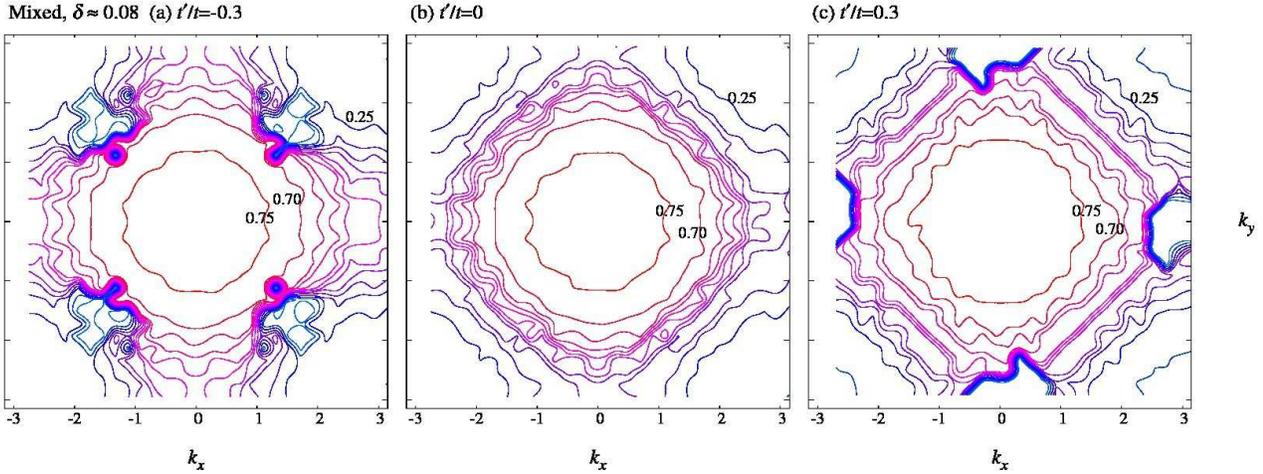} 
\end{center} 
\vskip -12mm
\caption{(Color online) 
Contour maps of $n({\bf k})$ for the optimized mixed state at $U/t=12$ and 
$\delta\sim 0.08$ shown for (a) $t'/t=-0.3$, (b) $0$, and (c) $0.3$. 
The data in these maps include the data used in Fig.~\ref{fig:nk-mix}(b).
Systems with $L=10$--$16$ are used. 
}
\label{fig:Mn2kx}
\end{figure*}
\begin{table}
\caption{
Locations of the centers of local Fermi surfaces in the state realized 
for a small $\delta$ (leftmost state for the mixed state) and $U/t=12$ 
summarized for the AF, $d$-SC, and mixed states.
For the mixed state, the evolution of the realized states as $\delta$ increases 
is shown for $\delta\lesssim 0.3$. 
`Co' (`N') indicates a coexisting state with AF and $d$-SC orders (normal 
state). 
}
\label{table:mixda}
\begin{center}
\begin{tabular}{r|l|l|l|l}
\hline
\multicolumn{1}{c}{$t'/t$}\vline & \multicolumn{1}{c}{AF}\vline & 
\multicolumn{1}{c}{$d$-SC}\vline & \multicolumn{2}{c}{Mixed}     \\
\cline{4-4}
& & & \multicolumn{1}{c}{Evolution of state}\vline & \\
\hline
$-0.3$ & ($\pi/2$, $\pi/2$) &                    & AF(ii) $\rightarrow$ SC 
            & ($\pi/2$, $\pi/2$) \\
$-0.1$ & ($\pi/2$, $\pi/2$) & \ \ Always         & AF(ii) $\rightarrow$ (Co) $\rightarrow$ SC & ($\pi/2$, $\pi/2$) \\
$ 0.0$ & ($\pi$, $0$)       & ($\pi/2$, $\pi/2$) & Co $\rightarrow$ SC $\rightarrow$ N      & No                 \\
$ 0.1$ & ($\pi$, $0$)       &                    & Co $\rightarrow$ AF(i) $\rightarrow$ N   & No                 \\
$ 0.3$ & ($\pi$, $0$)       &                    & AF(i) $\rightarrow$ N 
            & ($\pi$, $0$)       \\
\hline
\end{tabular}
\end{center}
\vskip -8mm
\end{table}
Now, we consider why a $d$-SC order can coexist with a type-(i) AF 
order but is incompatible with a type-(ii) AF order. 
We can deduce the reason by considering the location of the Fermi surface 
in the underlying pure AF state. 
First, we review relevant properties of the $d$-SC state. 
In Figs.~\ref{fig:dn2k}(b) and \ref{fig:dn2k}(c), we show contour maps of 
$n({\bf k})$ for $\Psi_d$ with $t'/t=0$ and $t'/t=-0.3$, respectively. 
The steep slope of $n({\bf k})$, indicative of a Fermi surface, exists only 
near $(\pm\pi/2, \pm\pi/2)$, and the gentle slopes around $(\pm\pi,0)$ and 
$(0,\pm\pi)$ indicate gaps, in contrast with the feature of the normal 
state shown in Figs.~\ref{fig:dn2k}(a), which clearly exhibits a Fermi surface 
in any direction. 
In Fig.~\ref{fig:nk-mix}(a), $n({\bf k})$ in $\Psi_d$ [corresponding to 
Figs.~\ref{fig:dn2k}(b) and \ref{fig:dn2k}(c)] is shown along the same path 
as in Fig.~\ref{fig:nk-AF-jpsj} for three values of $t'/t$. 
As $t'/t$ varies, $n({\bf k})$ around $(\pi,0)$ greatly varies but the nodal 
Fermi surface near $(\pi/2,\pi/2)$\cite{Himeda-BR} is almost 
unchanging.\cite{Y2013} 
This indicates that the electronic states near $(\pi,0)$ are closely related to 
SC, because properties associated with SC such as $P_d$ greatly change with $t'/t$. 
Actually, antinodal Fermi surfaces have the following advantages for 
$d$-SC on the square lattice: 
\par
(i) The density of states diverges at $(\pi,0)$ owing to a van Hove singularity 
for $|t'/t|\le 0.5$. 
\par
(ii) A $d$-SC gap $\Delta_{\bf k}$ with a similar form to Eq.~(\ref{eq:gap}) 
has a large magnitude at $(\pi,0)$. 
\par
(iii) The scattering of ${\bf q}=(\pi,\pi)$, which is induced by the AF exchange 
correlation between nearest-neighbor sites, is possible by connecting two 
antinodal points with opposite signs of $\Delta_{\bf k}$. 
\par

In Fig.~\ref{fig:nk-mix}(b), we plot $n({\bf k})$ obtained in 
$\Psi_{\rm mix}$ for the same parameter sets as in Fig.~\ref{fig:nk-mix}(a). 
Corresponding contour maps are displayed in Fig.~\ref{fig:Mn2kx}. 
For $t'/t=\pm 0.3$, the results for $\Psi_{\rm mix}$ are almost the same as 
those for $\Psi_{\rm AF}$ shown in Figs.~\ref{fig:nk-AF-jpsj}(b) and 
\ref{fig:nk-AF-jpsj}(c) because the SC order does not appear. 
The results are also similar in Figs.~\ref{fig:AFn2k}(c) and \ref{fig:Mn2kx}(a) for 
$t'/t=-0.3$. 
However, for $t'/t=0$, where SC appears, the pocket Fermi surfaces at 
the antinodes in $\Psi_{\rm AF}$ in Fig.~\ref{fig:nk-AF-jpsj}(b) are replaced 
with gap behavior (green) similar to the decreasing slope in $\Psi_d$ in 
Fig.~\ref{fig:nk-mix}(a). 
It is clearer to compare Fig.~\ref{fig:AFn2k}(b) for $\Psi_{\rm AF}$ 
with Fig.~\ref{fig:Mn2kx}(b) for $\Psi_{\rm mix}$. 
This reveals that for the $d$-SC order, Fermi surfaces in the nodal directions 
are not necessary but gap formation in the antinodes is vital. 
Incidentally, the resultant SC in the coexisting state, if any, does not 
have a feature of cuprate SCs, namely, nodal Fermi surfaces 
[Fig.~\ref{fig:nk-mix}(b)]; it is smeared out by an AF gap. 
To provide an overview of this topic, we summarize in Table~\ref{table:mixda} 
the locations of the local Fermi surface centers of the three states for typical 
values of $t'/t$. 
On the basis of this table with the above discussion, we may derive two 
requisites for $d$-SC in the mixed state: 
\par
(I) In the underlying pure AF (or normal) state, Fermi surfaces exist in the 
antinodes [around $(\pi, 0)$ and equivalent points]. 
\par
(II) The hot spots determined by $\varepsilon_{\bf k}^{\rm SC}$ (see 
Sect.~\ref{sec:d-BRE}) are situated in the Fermi surface area mentioned 
in (I). 
\par

\begin{figure}[htb] 
\begin{center}
\vskip -10mm
\hskip -25mm
\includegraphics[width=11.0cm,clip]{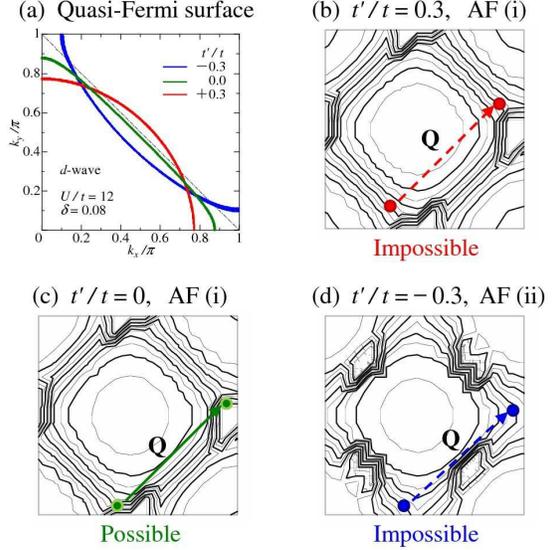} 
\end{center} 
\vskip -50mm
\caption{(Color online) 
(a) Quasi-Fermi surface in $\Psi_d$ obtained using 
$\varepsilon_{\bf k}^{\rm SC}$ in the first quadrant of the Brillouin zone 
for $U/t=12$ and $\delta=0.08$ for three typical values of $t'/t$. 
The gray dash-dotted line indicates the AF Brillouin zone boundary. 
(b)-(d) Possibility of scattering of ${\bf q}={\bf Q}$ in $\Psi_{\rm mix}$ 
in the full Brillouin zone.  
The hot spots in the SC part are indicated by circles. 
Data for $\Psi_{\rm AF}$ with $U/t=12$ and $\delta=0.0816$ ($L=14$) are used to 
draw the contours. 
}
\label{fig:schematic}
\end{figure}
On the basis of these conditions, we can explain the evolution of the states 
realized in $\Psi_{\rm mix}$ mentioned in Table~\ref{table:mixda}. 
We show the main point schematically in Fig.~\ref{fig:schematic}. 
For $t'<t'_{\rm L}$, item (I) is not satisfied for a small $\delta$, 
and $d$-SC does not emerge as shown in Fig.~\ref{fig:schematic}(d). 
However, as $\delta$ approaches $\delta_{\rm AF}$, the edge of the Fermi 
surface centered at $(\pi/2, \pi/2)$ extends to the antinodes, as will be 
shown in Fig.~\ref{fig:n2t-03AF}(c). 
The scattering therein possibly yields a narrow window of coexistence, for 
example, for $t'/t=-0.1$ [$\delta\sim 0.12$ and $\sim 0.139$ for $L=10$ 
and $12$, respectively] in Fig.~\ref{fig:pdm2}(b). 
Regarding item (II), the hot spots stay near the antinodes in this range 
of $t'/t$ [Fig.~\ref{fig:schematic}(a)]. 
On the other hand, for $t'>t'_{\rm L}$, item (I) is satisfied. 
For a small $|t'/t|$, item (II) is also satisfied 
[Fig.~\ref{fig:schematic}(c)], so that a coexisting state appears as in 
Fig.~\ref{fig:pdm2}(a). 
However, as $t'/t$ increases, the hot spots shift toward the nodal area 
[red in Fig.~\ref{fig:schematic}(a)] and deviate from the Fermi surface 
range in the antinodes [Fig.~\ref{fig:schematic}(b)], which is relatively 
narrow as shown later in Fig.~\ref{fig:n2t+00AF}. 
Consequently, $d$-SC does not appear appreciably for $t'/t=0.3$, as seen 
in Fig.~\ref{fig:pdm2}(a). 
This behavior contrasts with that of the pure $d$-SC state 
(Fig.~\ref{fig:Etot+pd-vsn-jpsj}), in which $d$-SC becomes weak more 
slowly because the hot spots are always situated at the Fermi surface 
of the underlying state $\Psi_{\rm N}$, and the scattering intensity becomes 
weak as the hot spots move away from the antinodes. 
\par

To summarize, because the AF state underlies the $d$-SC order, substantial 
$d$-SC arises only when the scattering of ${\bf Q}$ in the antinodes is 
compatible with the AF behavior. 
The requisites for this are given by (I) and (II) above. 
\par

\subsection{Coexistence of $d$-wave SC and staggered flux orders\label{sec:SF}}
\begin{figure}[htb] 
\begin{center}
\hskip -0mm
\includegraphics[width=9.0cm,clip]{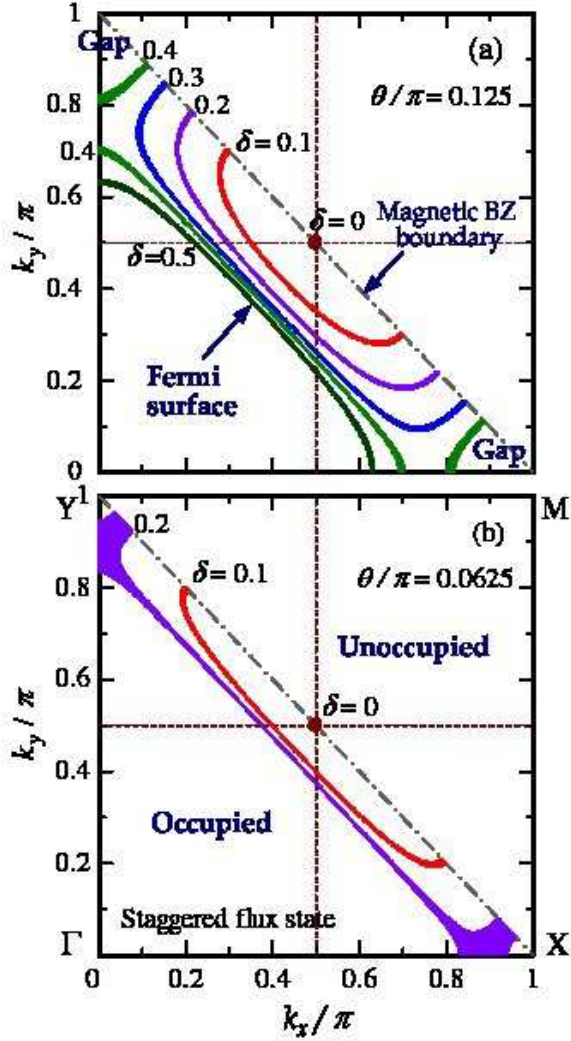} 
\end{center} 
\vskip -60mm
\caption{(Color online) 
Fermi surfaces of a staggered flux state for two values of $\theta$ and 
some doping rates drawn in the first quadrant of the Brillouin zone: 
$\Gamma=(0,0)$, ${\rm X}=(\pi,0)$, ${\rm M}=(\pi,\pi)$, and ${\rm Y}=(0,\pi)$. 
The thickness of the Fermi lines [e.g. for $\delta=0.2$ in (b)] indicates that 
$\varepsilon_{\bf k}^{\rm SF}$ is relatively flat. 
}
\label{fig:fsth1-8+1-16-jpsj} 
\end{figure}
To highlight the importance of Fermi surfaces in the antinodes for inducing 
a $d$-wave SC order, we consider the bare dispersion of a staggered flux 
(or $d$-density wave) state.\cite{theory,SF} 
Although this state has been extensively studied as a candidate for the 
pseudogap state as well as the ground state of cuprates, here we avoid 
referring to various interesting aspects of this state and focus on its 
bare dispersion: 
\begin{equation}
\varepsilon_{\bf k}^{\rm SF}
=-2t\sqrt{\cos^2{k_x}+2\cos{2\theta}\cos{k_x}\cos{k_y}+\cos^2{k_y}}, 
\label{eq:E-SF}
\end{equation}
where $\theta$ corresponds to a quarter of the magnetic flux penetrating each plaquette 
of the square lattice and is treated as a variational parameter here. 
For $\theta=0$, $\varepsilon_{\bf k}^{\rm SF}$ is reduced to 
$\gamma_{\bf k}$ [Eq.~(\ref{eq:gamma})]; for $\theta=\pi/4$ ($\pi$-flux 
state), $\varepsilon_{\bf k}^{\rm SF}$ at half filling yields a Dirac cone 
with a linear dispersion with apices at $(\pm\pi/2, \pm\pi/2)$. 
In Fig.~\ref{fig:fsth1-8+1-16-jpsj}, we show the Fermi surfaces generated by 
$\varepsilon_{\bf k}^{\rm SF}$ in the first quadrant of the Brillouin zone for 
two values of $\theta$ and some values of $\delta$ for each $\theta$. 
At half filling, the Fermi surface is the apex of an elongated Dirac cone at 
$(\pi/2, \pi/2)$. 
For $\delta>0$, a Fermi surface appears as a slice of an elongated Dirac cone 
around the nodal point $(\pi/2,\pi/2)$. 
Gaps open in the antinodes around ($\pi,0$) and ($0,\pi$). 
The form of the pocket Fermi surfaces and the antinodal gaps resembles the 
features in the pseudogap phase of cuprates. 
Note that the pocket Fermi surface becomes slender and its edge approaches 
the antinodes as $\theta$ decreases and/or $\delta$ increases. 
\par

\begin{figure}[htb] 
\begin{center} 
\vskip -0mm
\hskip -0mm
\includegraphics[width=9.5cm,clip]{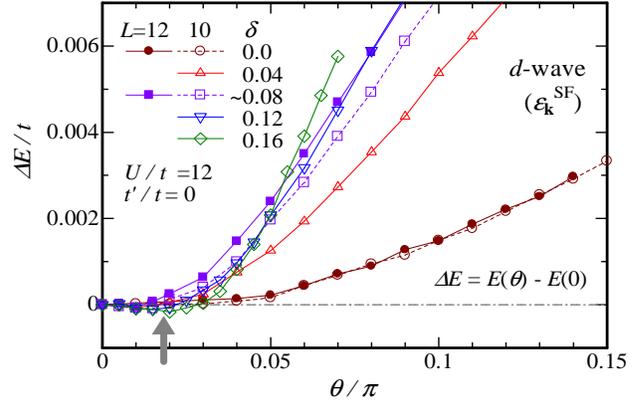} 
\end{center} 
\vskip -20mm
\caption{(Color online) 
Difference in total energy upon introducing a staggered flux $4\theta$ in 
$\varepsilon^{\rm SF}_{\bf k}$ [Eq.~(\ref{eq:E-SF})] for $U/t=12$ and 
$t'/t=0$. 
The thick arrow indicates the minimum (at $\theta/\pi\sim0.02$) for 
$\delta=0.16$.
}
\label{fig:evsth-n}
\end{figure}
Here, we study how the energy in $\Psi_d$ [Eq.~(\ref{eq:Phi_d})] varies 
when we use $\varepsilon_{\bf k}^{\rm SF}$ instead of $\gamma_{\bf k}$ 
as $\varepsilon_{\bf k}^{\rm SC}$. 
If the coexistence of staggered flux and $d$-wave SC orders is favored, 
the energy in $\Psi_d$ may be reduced by a finite value of $\theta$. 
In Fig.~\ref{fig:evsth-n}, we show the increment in energy per site $\Delta E$ 
as compared with that in $\Psi_d$ with $\gamma_{\bf k}$ as a function of 
$\theta$ for $t'/t=0$ and $U/t=12$. 
For large values of $\theta$ ($\gtrsim 0.05\pi$), the energy markedly 
increases regardless of $\delta$. 
On the other hand, for a small $\theta$ and large $\delta$, $\Delta E$ is 
small or slightly negative, as indicated by the arrow, meaning that the two 
orders possibly coexist. 
In these cases, the Fermi surfaces reach the antinodes. 
This is consistent with the notion that the gap in the antinodes in 
$\varepsilon_{\bf k}$ for the underlying state is unfavorable to the $d$-SC 
order. 
To summarize, a robust staggered flux order and a $d$-wave SC order are 
unlikely to coexist, although, to ensure this conclusion, we should investigate 
an appropriate mixed state of the two orders. 
\par

\subsection{Possible relation with pseudogap\label{sec:arc}}
One of the anomalous features arising in the pseudogap phase ($T_{\rm c}<T<T^*$)
of underdoped cuprates is the Fermi arcs\cite{Fermi-arc} observed in ARPES 
spectra, namely, unclosed Fermi surfaces whose centers are situated 
in the nodal directions near $(\pm\pi/2, \pm\pi/2)$, and similarly 
Fermi surface pockets\cite{YBCO,Hg}. 
If $T$ is fixed, a Fermi arc becomes longer as $\delta$ increases and 
becomes connected to other arcs in adjacent quadrants of the Brillouin zone 
to form an ordinary closed Fermi surface at the phase boundary ($T=T^*$). 
The origin of the pseudogap has not yet been elucidated. 
First, as a possible candidate for the Fermi arc, we consider the pocket 
Fermi surface of a doped $\Psi_{\rm AF}$.  
\par

\begin{figure*}[t!] 
\begin{center}
\hskip -0mm
\includegraphics[width=19.5cm,clip]{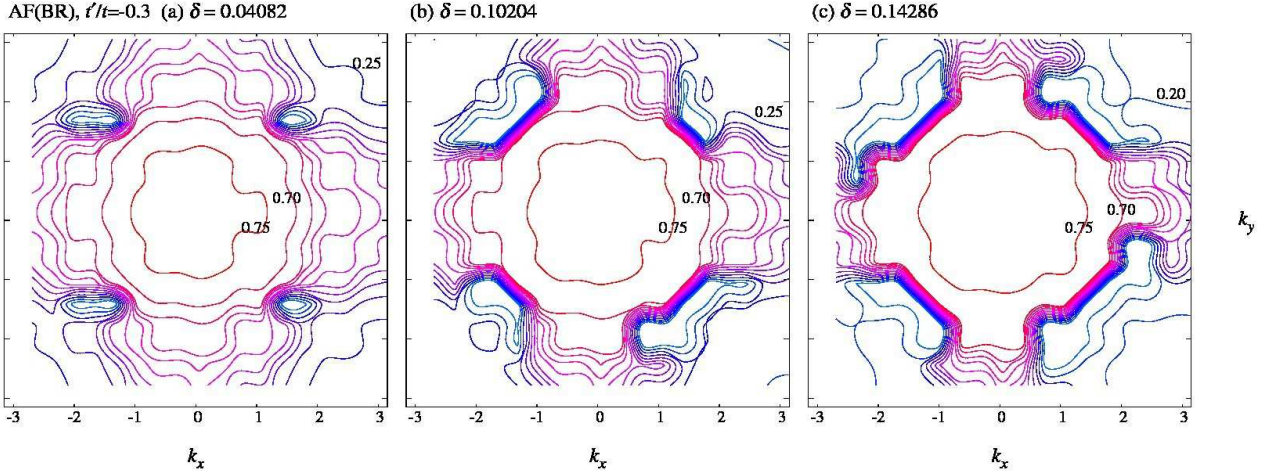} 
\end{center} 
\vskip -12mm
\caption{(Color online) 
Contour maps showing the evolution of $n({\bf k})$ as the doping rate increases 
in the type-(ii) AF state for $t'/t=-0.3$ ($U/t=12$).  
Three typical doping rates are selected. 
Data for $L=14$ are used. 
}
\label{fig:n2t-03AF} 
\end{figure*}
As shown in Fig.~\ref{fig:AFn2k}(c), a pocket Fermi surface of a type-(ii) AF 
state is formed around $(\pi/2, \pi/2)$ and is similar to the Fermi arc 
observed by ARPES. 
$\Psi_{\rm AF}$ has energy gaps around the antinodes in the sense that 
$n({\bf k})$ is smooth with a finite $|\nabla n({\bf k})|$. 
In Fig.~\ref{fig:n2t-03AF}, we show contour maps of $n({\bf k})$ for 
different doping rates, where the 
other conditions are the same ($t'/t=-0.3$, $U/t=12$). 
This figure reveals how the pocket Fermi surface evolves as $\delta$ increases;   
a small pocket Fermi surface appears around $(\pi/2, \pi/2)$ for very light 
doping, the arc length becomes long along the AF Brillouin zone boundary 
$(\pi,0)$--$(0,\pi)$, finally forming a connected Fermi surface centered 
at $\Gamma=(0,0)$ for $\delta=0.2245$ (not shown), where the AF order 
vanishes. 
This behavior is consistent with that of the Fermi arc of cuprates. 
For the appearance of such behavior at a finite temperature, it is also important that 
the type-(ii) $\Psi_{\rm AF}$ has a very low energy and is stable against 
phase separation. 
Furthermore, the type-(ii) $\Psi_{\rm AF}$ does not coexist with $d$-SC except for 
$\delta\sim\delta_{\rm AF}$. 
Although this result cannot be directly applied to the pseudogap phase of 
cuprates because an AF long-range order has not been observed, it is 
intriguing that short-range AF orders of $20$-$30$ lattice constants were 
observed up to high temperatures.\cite{AF-SRO}
\par

\begin{figure*}[t!] 
\begin{center}
\hskip -0mm
\includegraphics[width=19.5cm,clip]{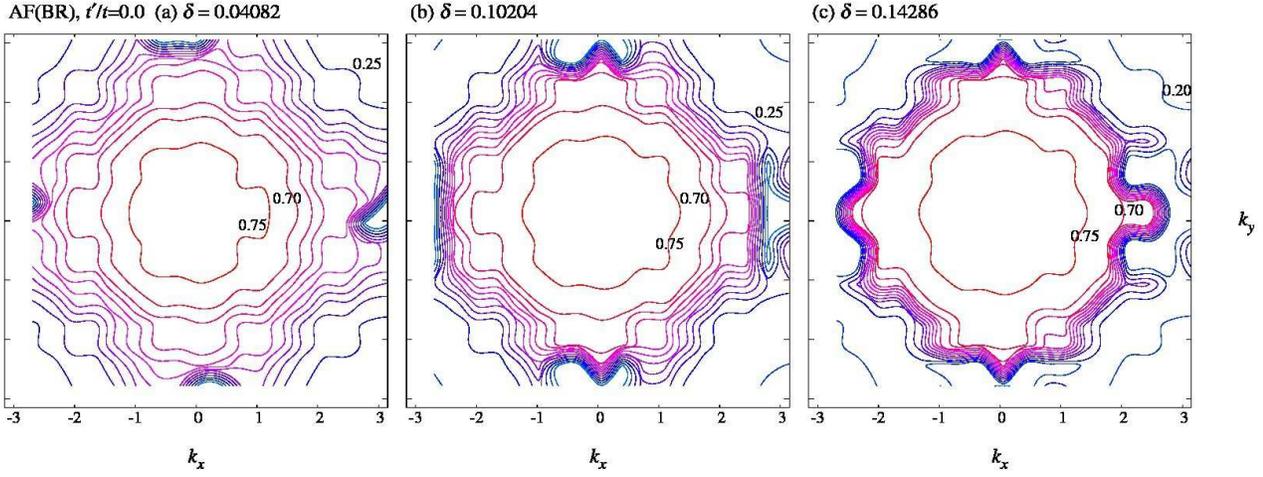} 
\end{center} 
\vskip -12mm
\caption{(Color online) 
Contour maps showing the evolution of $n({\bf k})$ as the doping rate increases 
in the type-(i) AF state for $t'/t=-0.3$ ($U/t=12$).  
Data for the same doping rates as in Fig.~\ref{fig:n2t-03AF} are displayed for 
comparison. 
Systems with $L=14$ are used. 
}
\label{fig:n2t+00AF} 
\end{figure*}
In Fig.~\ref{fig:n2t+00AF}, we show the evolution of contour maps of $n({\bf k})$ 
as $\delta$ increases in the type-(i) AF state ($t'/t=0$). 
In contrast to the type-(ii) AF state, a pocket Fermi surface grows from 
the antinodes in the nodal directions and finally forms a closed Fermi 
surface centered at $\Gamma=(0,0)$ for $\delta=0.1633$ (not shown). 
Because energy gaps open for $\delta<\delta_{\rm AF}$ in the nodal 
directions, the type-(i) AF state, which corresponds to electron-doped cuprates, 
is not directly related to the Fermi arc phenomena. 
\par

\section{Summary and Discussion\label{sec:summary}}
In this paper, we studied band renormalization effects (BRE) owing to electron 
correlation on a mixed state of $d_{x^2-y^2}$-wave pairing ($d$-SC) and 
antiferromagnetic (AF) orders, as well as normal (paramagnetic), pure 
$d$-SC, and pure AF states, by applying a variational Monte Carlo (VMC) 
method to the Hubbard ($t$-$t'$-$U$) model. 
For the mixed state, BRE were introduced into the AF and $d$-SC parts 
independently; BRE on AF orders, previously not investigated,\cite{note-Misawa} 
markedly change the previous knowledge of the Hubbard model. 
By searching widely in the model-parameter space with wave functions on 
various levels, we obtained systematic insights, particularly into the 
following subjects: 
(A) Ground-state phase diagrams in the space of $U/t$, $t'/t$, and $\delta$. 
(B) In what regime and through what mechanisms does the coexistence of AF and 
$d$-SC arise? 
(C) In what regime and from what cause does instability toward inhomogeneous 
phases occur? 
First, we itemize the main results in this work: 
\par

(1) In the $d$-SC state, the effective band $\varepsilon_{\bf k}^{\rm SC}$ is 
markedly renormalized for the model parameters of $U/t\gtrsim 6$, 
a large $|t'/t|$, and a small $\delta$ $(\lesssim 0.1)$ 
(Figs.~\ref{fig:DelE-br-jpsj} and \ref{fig:DelE-br-vsn}), as known previously. 
We found, however, owing to BRE, not only is the improvement in energy much 
smaller than those in the normal and AF states, but also quantities related 
to SC [$P_d$, $S({\bf q})$, $n({\bf k})$] are modified only very slightly
(Figs.~\ref{fig:pd-jpsj}-\ref{fig:nk-jpsj}). 
\par

(2) In the normal state, BRE apply $U\gtrsim U_{\rm c}$, 
$\delta\lesssim 0.05$, and $|t'/t|\gtrsim 0.1$ with $U_{\rm c}/t$ being the 
Mott transition point (Fig.~\ref{fig:DelE-N-jpsj}).  
The improvement in energy is an order of magnitude larger than that of the 
$d$-SC state but an order of magnitude smaller than that of the AF state 
(Fig.~\ref{fig:DelE-vsn-a03-comp}). 
\par

(3) In all the states studied, band renormalization takes place to reduce 
the kinetic energy ($E_t$) at the cost of the interaction energy ($E_U$), 
which corresponds to the tendency of a strongly correlated state to undergo 
a phase transition to reduce the kinetic energy.\cite{YTOT,Y2013} 
In the resultant renormalized band, the nesting condition tends to be 
restored ($t_1/t\rightarrow 0$).  
\par

(4) For the AF state, BRE are useful in reducing the energy, especially for 
$t'/t<0$ (Fig.~\ref{fig:DelE-vsU-jpsj}); the qualitative features are almost 
independent of $U/t$ for $U>U_{\rm AF}$. 
As a result, the AF state occupies a wide area ($\delta\lesssim 0.2$) in 
the phase diagrams (Figs.~\ref{fig:boundary-jpsj} and \ref{fig:pd2}). 
The AF area is considerably wider for $t'/t=-0.3$ than for $t'/t=0$, which 
contrasts with the results without BRE. 
In a doped metallic AF state, as $t'/t$ is varied, a kind of first-order 
Lifshitz transition takes place at $t'=t_{\rm L}\sim -0.05t$ regardless 
of the values of $U/t$ and $\delta$. 
In the type-(i) [(ii)] AF regime ($t'>t_{\rm L}$) [($t'<t_{\rm L}$)], 
local pocket Fermi surfaces arise around $(\pi, 0)$ [$(\pi/2, \pi/2)$] and 
equivalent points (Figs.~\ref{fig:nk-AF-jpsj} and \ref{fig:AFn2k}). 
This difference plays a critical role in inducing the $d$-SC mentioned in (6) before.  
The Fermi surface in the type-(ii) AF is possibly related to the Fermi arcs 
found in cuprates. 
\par

(5) In the mixed state, the range of instability toward phase separation 
(PS) is found to be $t'_{\rm L}/t<t'/t\lesssim 0.2$, similarly to in the AF 
states.\cite{Y2013,proceedings1} 
The AF order is responsible for this instability, which does not directly 
correlate with $d$-SC. 
Elsewhere, the state is stable against PS. 
This stability is mainly due to the diagonal hopping of 
doped carriers. 
\par

(6)The coexistence or mutual exclusivity of AF and $d$-SC orders was studied 
in the mixed state (Fig.~\ref{fig:pdm2}). 
The AF order has preferentially exhibits this property because the AF part 
greatly reduces the energy compared with the SC part 
(Fig.~\ref{fig:evsatilde-a03}). 
By checking various cases, we found two requisites for the $d$-SC 
order to arise (Fig.~\ref{fig:schematic}): 
(i) In the underlying pure AF (or normal) state, Fermi surfaces exist 
in the antinodes [near $(\pi, 0)$ and equivalent area]. 
(ii) The hot spots determined by $\varepsilon_{\bf k}^{\rm SC}$ are situated 
in the Fermi surface area. 
These requisites indicate that the scattering of ${\bf q}=(\pi, \pi)$ in the 
antinodes is vital for $d$-SC. 
Thus, the coexistence basically occurs in the type-(i) AF regime. 
The range of $t'/t$ in which coexistence occurs is similar to that for the instability toward 
PS (Fig.~\ref{fig:pd2}), but this similarity is accidental. 
These requisites seem to apply to the coexistence of $d$-SC and staggered 
flux orders. 
\par
The present results are quantitatively consistent with recent studies with 
advanced techniques,\cite{Misawa,Otsuki,DMET} and make it possible to 
reasonably interpret individual features of previous studies. 
\par

Finally, we discuss the relationship with cuprates.  
The present results that the AF order is predominant for a wide range of model 
parameters ($U/t\gtrsim 6$, $\delta\lesssim 0.2$, most $t'/t$) and that uniform 
$d$-SC disappears in the underdoped regime are consistent 
with those of recent VMC,\cite{Misawa} DMFT,\cite{Otsuki} and 
DMET\cite{DMET} studies based on the Hubbard model. 
Furthermore, recent VMC studies on the $t$-$J$\cite{Sato} and 
$d$-$p$\cite{Tamura} models display the same tendency. 
Nevertheless, these results are inconsistent with properties common to 
hole-doped cuprate SCs: the AF long-range order is broken by less than 5\% doping 
with carriers and high-$T_{\rm c}$ $d$-SC appears in the 
underdoped regime. 
In addition, it was recently shown that well-annealed electron-doped samples 
with small doping rates (5-10\%) exhibit no AF long-range orders but metallic 
or SC behavior\cite{Naito,Adachi} with entirely closed Fermi 
surfaces.\cite{Horio} 
Assuming that the AF order is excluded for some reason, most properties of 
the remaining $d$-SC derived by theories so far are basically consistent with 
those of cuprates. 
Thus, it is important to clarify why AF long-range order is robust in 
the theory. 
It seems that the approximations applied are not responsible for the 
predominant AF orders, but the models are lacking in certain important 
factors that destabilize AF orders. 
They are possibly disorders or impurities inherent in cuprate SCs. 
It seems that theoretical research on cuprate SCs may proceed to this direction. 
\par
After the submission of this paper, we noticed that BRE on AF states were already considered in a VMC study of Watanabe, 
Shirakawa and Yunoki for three-band as well as single-band Hubbard models.\cite{65} 
They used the optimization method mentioned as `an alternative approach' in Appendix B. 
Their results are basically consistent with ours.
\par

\begin{acknowledgment}
We thank Kenji Kobayashi, Masao Ogata, Shun Tamura, Junya Otsuki, Yuta Toga, Hiroshi Watanabe, 
Kentaro Sato, and Masaki Fujita for useful discussions and information. 
This work was supported in part by Grants-in-Aid from the Ministry of 
Education, Culture, Sports, Science and Technology, Japan. 
\par
\end{acknowledgment}

\appendix
\section{Details of Optimization in Normal State\label{sec:normal-A}}
In this Appendix, we explain how to actually deal with the BR of the normal 
(paramagnetic) state $\Psi_{\rm N}$ [Eq.~(\ref{eq:normal})]
for finite-size systems. 
As mentioned in Sect.~\ref{sec:form}, $\Psi_{\rm N}$ depends only on the 
choice of $\{{\bf k}\}_{\rm occ}$ (Fermi surface) and not explicitly 
on $\varepsilon_{\bf k}$. 
In the thermodynamic limit ($L=\infty$), where ${\bf k}$ is a continuous 
variable, $\{{\bf k}\}_{\rm occ}$ continuously changes as 
$\varepsilon_{\bf k}$ or the band parameters therein ($t_1/t$, etc.) gradually 
vary. 
This means that $\{{\bf k}\}_{\rm occ}$ directly depends on the band parameters. 
On the other hand, in the finite systems we treat here, for which the available 
${\bf k}$ are discrete, $\{{\bf k}\}_{\rm occ}$ (namely $\Psi_{\rm N}$) 
is invariable in a certain range of band parameters (or 
$\varepsilon_{\bf k}$) and discontinuously changes at the edges of 
the range. 
Generally, this range becomes wider for a smaller $L$. 
To begin with, we illustrate this point assuming that the effective band is 
given by $\varepsilon_{\bf k}^{\rm N}$ in Eq.~(\ref{eq:disp-N}). 
Even for this simple form of $\varepsilon_{\bf k}$, we believe that full 
BRE are achieved in most cases. 
\par

\begin{figure}
\begin{center}
\vskip -0mm
\hskip -0mm
\includegraphics[width=9.0cm,clip]{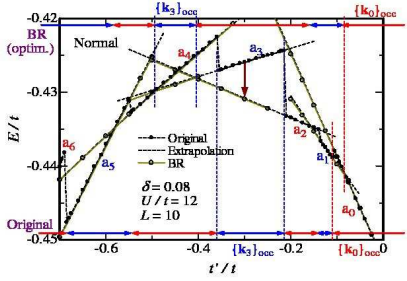} 
\end{center} 
\vskip -25mm
\caption{(Color online)
Illustration of how to obtain the band-renormalized energy in $\Psi_{\rm N}$ 
within a single variational band parameter $t_1/t$ for a specific model 
parameter set ($L=10$, $\delta=0.08$, $U/t=12$). 
The total energy per site is plotted as a function of $t'/t$ ($<0$). 
For details, see text. 
}
\label{fig:evsa-a03-jpsj}
\end{figure}
To avoid confusion between $t'/t$ in ${\cal H}$ (model parameter) and the 
variational band parameter $t_1/t$ in $\Psi_{\rm N}$, we start with the 
noninteracting case ($U=0$). 
In this case, the exact ground state is given by Eq.~(\ref{eq:FS}), in which 
$\{{\bf k}\}_{\rm occ}$ is determined by the bare band dispersion 
$\tilde\varepsilon_{\bf k}$ in Eq.~(\ref{eq:bareband}), indicating that $t_1=t'$ 
and $\varepsilon_{\bf k}=\tilde\varepsilon_{\bf k}$ for $U=0$ in variation 
theory. 
If we decrease the sole band parameter $t'/t$ in ${\cal H}$ from zero, 
$\{{\bf k}\}_{\rm occ}$ is switched from one configuration to another at certain discrete 
values of $t'/t$. 
In Fig.~\ref{fig:evsa-a03-jpsj}, we show such evolution of 
$\{{\bf k}\}_{\rm occ}$, for $L=10$ and $\delta=0.08$ as an example, with 
alternate red and blue arrows near the lower horizontal axis; 
$\{{\bf k}\}_{\rm occ}$ is switched as 
\begin{equation}
\{{\bf k}_{\rm 0}\}_{\rm occ}\rightarrow \{{\bf k}_{\rm 1}\}_{\rm occ}
\rightarrow
\{{\bf k}_{\rm 2}\}_{\rm occ}\rightarrow \{{\bf k}_{\rm 3}\}_{\rm occ}
\rightarrow\cdots
\label{eq:a_ell}
\end{equation}
at $t'/t\sim -0.107$, $-0.137$, $-0.213$, $-0.357$, $\cdots$. 
Let A$_\ell$ ($\ell$: integer) denote the area of $t'/t$ where 
$\{{\bf k}\}_{\rm occ}=\{{\bf k}_\ell\}_{\rm occ}$ as shown 
in Fig.~\ref{fig:evsa-a03-jpsj}, for example, A$_2=[-0.213,-0.137]$. 
Note that within each A$_\ell$, the ground-state wave function 
$\Psi_{\rm N}$ (=$\Phi_{\rm N}$) is unchanging but the energy changes 
with $t'/t$ owing to the diagonal hopping term. 
\par

\begin{figure}
\begin{center}
\vskip -5mm
\hskip -0mm
\includegraphics[width=9.0cm,clip]{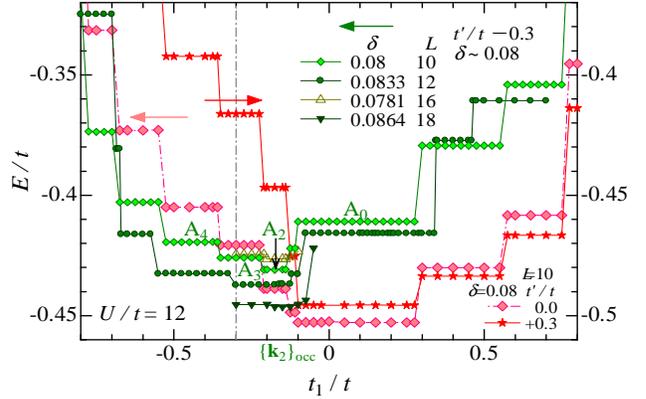} 
\end{center} 
\vskip -22mm
\caption{(Color online) 
Energy expectation values of the normal state $\Psi_{\rm N}$ for some 
model-parameter sets ($L$, $\delta$, $t'/t$, $U/t=12$) plotted as 
functions of the band parameter $t_1/t$. 
The data with light-green diamonds and lettering A$_\ell$ correspond 
to the set discussed in the text ($10$, $0.08$, $-0.3$, $12$). 
Those plotted as dark circles are identical to those shown 
in Fig.~\ref{fig:evsatilde-a03}.
}
\label{fig:evsa-a-a03}
\end{figure}
Next, we consider interacting cases ($U>0$). 
Let the model parameters be fixed, for example, at $L=10$, $\delta=0.08$, 
$t'/t=-0.3$, and $U/t=12$. 
For such a parameter set, we need to optimize $\Psi_{\rm N}$ by adjusting the 
band parameter $t_1/t$ independently of $t'/t$ together with the correlation 
parameters. 
Because $\Phi_{\rm N}$ in Eq.~(\ref{eq:FS}) depends only on 
$\{{\bf k}_\ell\}_{\rm occ}$ but not directly on $t_1/t$, $\Psi_{\rm N}$ 
should exhibit completely flat energy as a function of $t_1/t$ in A$_\ell$ 
and discontinuities at the edges of A$_\ell$. 
Actually, in Fig.~\ref{fig:evsa-a-a03}, we show the $t_1/t$ dependence of the total 
energy for the above parameter set with light-green diamonds, along with 
the same quantity for other parameter sets. 
Because the effective band dispersion $\varepsilon^{\rm N}_{\bf k}$ 
[Eq.~(\ref{eq:disp-N})] in $\Psi_{\rm N}$ is assumed to be the same form 
as the bare band dispersion $\tilde\varepsilon_{\bf k}$ 
[Eq.~(\ref{eq:bareband})],\cite{refined-e} the division of the areas 
(A$_\ell$) for $t'/t$ discussed above directly corresponds to the division of 
$t_1/t$, as also marked by A$_\ell$ in Fig.~\ref{fig:evsa-a-a03}. 
The energy minimum for the above model parameter set ($t'/t=-0.3$) is 
obtained not in A$_3$ (including $t_1/t=-0.3$) but in A$_2$, meaning that BRE 
manifest themselves. 
\par

Owing to this locally flat behavior of $E/t$, ordinary optimization 
techniques that need information about gradients of $E/t$ are 
inapplicable to $\Psi_{\rm N}$. 
Here, we use another way of optimization. 
Below, we describe its outline with an illustration in 
Fig.~\ref{fig:evsa-a03-jpsj} for a model-parameter set 
($L=10$, $\delta=0.08$, $U/t=12$) as an example. 
(i) Calculate the total energy $E/t$ densely as a function of $t'/t$ for a fixed 
set of the other parameters ($L$, $\delta$, $U/t$) without introducing 
BRE, namely by putting $t_1=t'$. 
In Fig.~\ref{fig:evsa-a03-jpsj}, the $E/t$ thus obtained are plotted with small 
solid circles with a thick dashed line. 
We find that $E/t$ is described by a distinct nearly straight curve for 
each A$_\ell$. 
(ii) Each segmented curve (say in A$_\ell$) can be well extrapolated 
using a first- or second-order least-squares method: 
\begin{equation}
E_\ell(t'/t)=c^{(\ell)}_0+c^{(\ell)}_1(t'/t)+c^{(\ell)}_2(t'/t)^2. 
\end{equation}
The extrapolated curves are shown with thin dashed lines 
in Fig.~\ref{fig:evsa-a03-jpsj} and practically coincide with the values 
of $E/t$ actually calculated with $\{{\bf k}_\ell\}_{\rm occ}$ (BRE) outside 
A$_\ell$, whose values are shown with open circles joined by thin dull-green 
curves. 
Therefore, we may substitute such extrapolated values for the results of 
actual BRE calculations to save labor. 
(iii) The optimized energy allowing for BRE for a fixed value of $t'/t$ 
is given by the lowest extrapolated value among the all the A$_\ell$. 
For $t'/t=-0.3$, for example, the lowest energy is given by 
$\{{\bf k}_2\}_{\rm occ}$, and the improvement in energy owing to BRE 
($\Delta E/t$) is indicated by a brown arrow. 
We actually estimated the optimized BRE energies of $\Psi_{\rm N}$ for most 
model parameter sets through this procedure. 
To obtain other quantities, however, calculations using the optimized 
parameters are necessary. 
\par
Under the upper horizontal axis in Fig.~\ref{fig:evsa-a03-jpsj}, we show 
the areas of $\{{\bf k}_\ell\}_{\rm occ}$ which yield the optimized 
$E/t$ with red and blue arrows. 
It reveals that these areas of $\{{\bf k}_\ell\}_{\rm occ}$ with BRE often 
deviate from the areas of $\{{\bf k}_\ell\}_{\rm occ}$ for bare cases 
shown near the lower horizontal axis. 
Thus, in this model parameter set, the energy reduction owing to BRE is 
brought about discontinuously as a function of $t'/t$ 
[see Fig.~\ref{fig:DelE-N-jpsj}(b)]. 
In Fig.~\ref{fig:evsatilde-a03}, we actually illustrate the above process of optimization associated with BRE 
for $\Psi_{\rm N}$ with $L=12$, $\delta=0.0833$, and $U/t=12$. 
The red line indicates the optimized line for $\Psi_{\rm N}$. 
In this parameter, BRE are ineffective for or small 
$-0.573\lesssim t'/t\lesssim 0.343$. 
\par

\section{Details of Optimizing AF and Mixed States\label{sec:AF-A}}
In optimizing $\Psi_{\rm AF}$ and $\Psi_{\rm mix}$, a similar difficulty exists in the case of $\Psi_{\rm N}$. 
Namely, if we determine $\{\bf k\}_{\rm occ}^{\rm AF}$ according to 
$\varepsilon_{\bf k}^{\rm AF}$, as $t_\eta$ is gradually varied, total energy $E/t$ discontinuously changes 
at a value where 
$\{\bf k\}_{\rm occ}^{\rm AF}$ is switched to another configuration. 
In contrast to $\Psi_{\rm N}$, we have to optimize 
$\varepsilon_{\bf k}^{\rm AF}$ in addition to $\{\bf k\}_{\rm occ}^{\rm AF}$ 
for $\Psi_{\rm AF}$ and $\Psi_{\rm mix}$, as shown in Table \ref{table:BR}. 
What is worse, $E/t$ depends on $t_\eta$ only very weakly. 
For this reason, the stochastic reconfiguration method and quasi-Newton 
methods did not work effectively, and we returned to a primitive linear 
optimization method in most cases.
\par

\begin{figure}
\begin{center}
\vskip 2mm
\hskip -0mm
\includegraphics[width=9.5cm,clip]{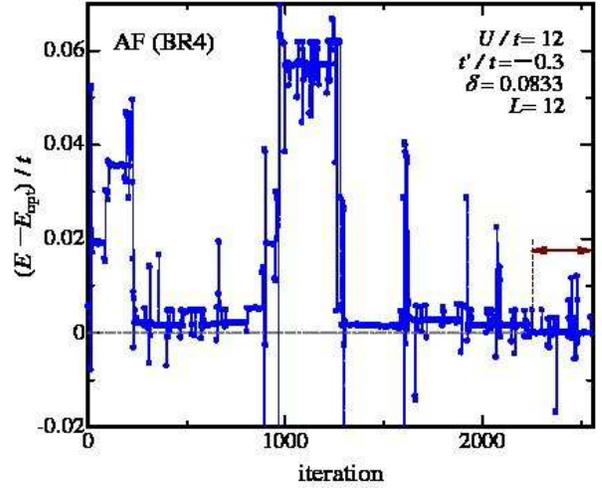} 
\end{center} 
\vskip -25mm
\caption{(Color online) 
Evolution of the energy expectation value in $\Psi_{\rm AF}$ obtained by VMC process 
using a simple linear optimization method. 
The results of eight calculations successively performed are plotted in sequence, in each 
of which 320 linear optimizations were carried out. 
The initial parameter values in each calculation were set to those that yielded 
the lowest plateau energy in the previous calculations. 
We estimated the optimized energy, in this case, by averaging the final 
results indicated by the arrow. 
In averaging, we exclude scattered data that are more than twice the standard deviation from the mean. 
}
\label{fig:a03it-jpsj} 
\end{figure}
We show an example of optimizing $\Psi_{\rm AF}$ 
in Fig.~\ref{fig:a03it-jpsj}, where the expectation value of $E/t$ obtained 
in each linear optimization of the parameters is plotted for the specified 
model parameter set. 
Typically, $2.5\times 10^5$ samples are used for the linear optimization. 
The expectation value of $E$ does not monotonically decrease to the 
optimized value $E_{\rm opt}$ but irregularly fluctuates, exhibiting wide 
and narrow plateaus and irrelevant spikes. 
A given configuration $\{\bf k\}_{\rm occ}^{\rm AF}$ yields a plateau or plateaus with the same 
$E$. 
We determined $E_{\rm opt}$ by averaging $E$ in the lowest plateau and 
checking that the estimated value is smoothly connected to those of other 
model parameter sets. 
For $t'\sim t'_{\rm L}$, the statistical fluctuations become very large because 
multiple $\{\bf k\}_{\rm occ}^{\rm AF}$ have a value of $E$ comparable to 
$E_{\rm opt}$. 
Therefore, in this regime, we carried out up to fifty calculations for a single 
model-parameter set, especially for $\Psi_{\rm mix}$. 
\par

As an alternative approach, we may optimize $\Psi_{\rm AF}$ and 
$\Psi_{\rm mix}$ with a fixed $\{\bf k\}_{\rm occ}^{\rm AF}$ using the 
stochastic reconfiguration method. 
By carrying out such operations for various values of 
$\{\bf k\}_{\rm occ}^{\rm AF}$, we can single out the $\Psi$ with the 
lowest $E/t$. 
Because the number of choices of $\{\bf k\}_{\rm occ}^{\rm AF}$ rapidly 
increases as $L$ increases, we may adopt the way of choosing 
$\{\bf k\}_{\rm occ}$ used for $\Psi_{\rm N}$ 
in Appendix\ref{sec:normal-A}.
Anyway, the task of optimizing $\Psi_{\rm AF}$ and $\Psi_{\rm mix}$ is 
much more burdensome than that for $\Psi_d$.
\par




\begin{thebibliography}{99}
\bibitem{note-Misawa} In a recent VMC study with a different 
formulation (Ref.~\citen{Misawa}), BRE on AF parts seem to be implicitly 
introduced.
\bibitem{note-HF} The usual Hartree-Fock approximation for the completely 
nested case is not suited to the case of $t'/t\ne 0$, and the 
quasiparticle band dispersion 
$$
E_{\bf k}=\frac{U}{2}\pm\sqrt{\varepsilon_{\bf k}^2+\Delta_{\rm AF}^2}
$$
is not appropriate for determining ${\bf k}_{\rm F}$. 
However, the form of Eq.~(\ref{eq:AF}) is useful for the AF part of 
the variational wave function even for $t'/t\ne 0$.
\bibitem{note-hotspot} To consider hot spots more seriously, it is necessary 
to optimize the $d$-wave gap form $\Delta_d$ appropriately, as carried out 
in Refs.~\citen{Wata-t-J} and \citen{Hirashima}.
In this sense, the discussion here is rather broad.
\bibitem{note-N(q)} This change from an insulator to a metal is also 
confirmed by the small-$|{\bf q}|$ behavior of the charge density structure 
factor $N({\bf q})$.
\end{thebibliography}
\end{document}